\documentclass[twoside,openright,a4paper,11pt]{book}
\usepackage{amssymb,amsfonts,amsmath}   
\usepackage{graphicx,subcaption,float}   
\usepackage{palatino}                   
\usepackage{tikz}
\usepackage{physics}
\usepackage{dsfont}                     
\usepackage{mathrsfs}                   
\usepackage{tcolorbox}                  
\usepackage[normalem]{ulem}             
\usepackage{todonotes}					
\usepackage{xcolor}
\usepackage[toc,page]{appendix}
\usepackage[square,sort,comma,numbers,sort&compress]{natbib}
\usepackage[pagebackref=false,pdftex,pdfborderstyle={/S/U/W 1},hyperfootnotes=true,colorlinks,
linkcolor={darkred},
citecolor={darkviolet},
urlcolor={darkred}]{hyperref}
\usepackage{setspace}
\usepackage{wrapfig,sidecap,,fullpage,morefloats}
\usepackage{grffile,float,xspace}
\usepackage[sort&compress]{natbib}     
\newcounter{daggerfootnote}

\usepackage{caption}                     
\definecolor{darkred}{rgb}{0.55, 0.0, 0.0}
\definecolor{darkviolet}{rgb}{0.58, 0.0, 0.83}	
\definecolor{midnightblue}{rgb}{0.1, 0.1, 0.44}	
\definecolor{oxfordblue}{rgb}{0.0, 0.13, 0.28}
\definecolor{prussianblue}{rgb}{0.0, 0.19, 0.33}
\definecolor{goldenrod}{rgb}{0.85, 0.65, 0.13}
\definecolor{darkgreen}{rgb}{0.0, 0.5, 0.0}
\definecolor{Ultramarine}{rgb}{18,10,143}

\newcommand{\thesistitle}{Numerical Investigations of Phase Transitions in Lattice Field Theories}
 
\newcommand{\degree}{Doctor of Philosophy}
\newcommand{\candidate}{Vamika}

\newcommand{\department}{\href{https://web.iisermohali.ac.in/dept/physics/}{Department of Physical Sciences}}
\newcommand{\university}{\href{https://www.iisermohali.ac.in/}{Indian Institute of Science Education \& Research (IISER) Mohali,\\ Sector 81 SAS Nagar, Manauli PO 140306 Punjab, India}}

\begin{document}
\pagenumbering{roman}
\begin{titlepage}
	
	\vfill
	\begin{center}
	\vfill
	{\huge {\sc \textbf{\thesistitle}} \par}\vspace{0.01cm}
	\vfill
	
	{\LARGE \sc \candidate \vskip 0.3cm} 
	
	\vfill
	\textit{\large A thesis submitted for the partial fulfillment of }\\[0.1cm] 
	\textit{\large the degree of \degree}\\[0.4cm]
	\vfill
	
	\centering
	\includegraphics[width=3.5cm]{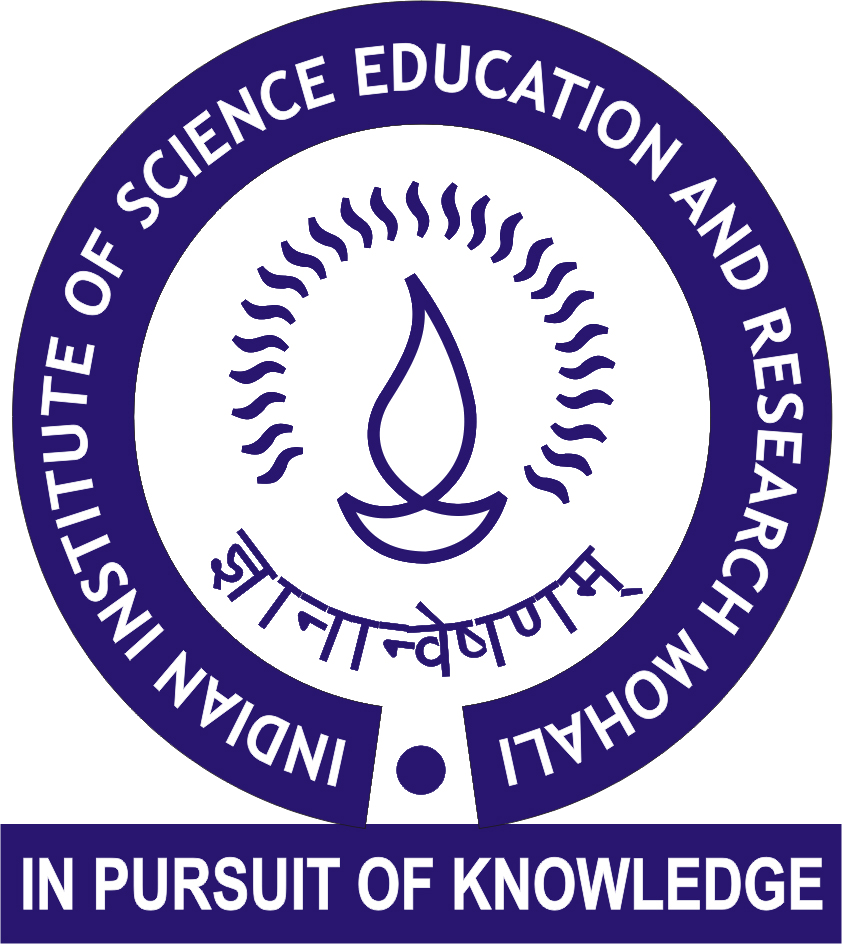}
	\vfill
	\sc \department
	\\
	\vspace{0.5cm}
	\university
	\vfill
	\end{center}
	\vfill
	
\end{titlepage}

\clearpage\mbox{}\clearpage
\begin{center}
	{\Large  \textbf{Abstract}}
    
\end{center}

\vspace{6mm}

The study of phase transitions plays an important role in understanding qualitative changes in the behaviour of physical systems at criticality. Despite decades of progress, there is still a strong demand for high-precision numerical tools capable of resolving subtle critical phenomena. Motivated by this need, in this thesis, we present two complementary numerical investigations of phase transitions in lattice systems. The first uses real-space tensor renormalization group techniques to study the two-dimensional generalized XY (gXY) model. The second develops and benchmarks a configurational temperature estimator, constructed from gradients and Hessians of the Euclidean lattice action, in compact U(1) lattice gauge theories.

In the first part of the thesis, we apply the higher-order tensor renormalization group (HOTRG) to the two-dimensional gXY model. This model interpolates between three distinct phases through a deformation parameter $\Delta$. The phases are a ferromagnetic phase, characterized by integer vortices; a nematic phase, characterized by half-integer vortices, and a paramagnetic (disordered) phase. With the help of character expansion and GPU-accelerated HOTRG with controlled truncation, we compute the partition function of the model in the thermodynamic limit. We extract thermodynamic observables, including specific heat, conventional and nematic magnetic susceptibilities, and magnetizations, using impurity-tensor techniques. These measurements allow us to clearly distinguish the three phases of the model and map their corresponding phase boundaries. Our results improve upon earlier Monte Carlo and matrix-product-state studies. They demonstrate that carefully controlled, GPU-accelerated HOTRG provides a highly accurate and efficient framework for resolving complex phase structures.

The second part of the thesis focuses on a configurational temperature estimator based on Rugh's geometric formulation of statistical mechanics. This estimator determines the temperature using purely configurational information, and it is derived from the gradient and Hessian of the action, without reference to momenta. We extend this estimator for Eulidean quantum field theories. There, it serves as a diagnostic tool for detecting sampling inconsistencies, slow thermalization, and numerical errors in Monte Carlo simulations. The estimator can handle gauge theories as well since it is formulated in a gauge-invariant way. Such diagnostics are particularly valuable in lattice gauge theories, where discretization effects and algorithmic artifacts can distort thermodynamic observables and critical temperatures. We test the estimator in compact U(1) lattice gauge theories in one, two, and four Euclidean dimensions, comparing the measured configurational temperature against the known value.

Our studies address two complementary challenges in numerical investigations of critical phenomena. On the one hand, tensor network methods can accurately capture rich phase structures when truncation and finite-bond effects are adequately controlled. On the other hand, the configurational temperature estimator provides an independent, low-overhead means of validating thermal sampling across different algorithms and models. The estimator can also be deployed as a runtime diagnostic to identify sampling pathologies before large-scale production runs.

The thesis concludes by outlining the future directions, including the application of HOTRG to generalized XY models with higher fractionalization $q \geq 3$ or to systems with coupled gauge degrees of freedom. Also, the integration of configurational temperature monitoring into large-scale Hybrid Monte Carlo simulations of non-Abelian gauge theories and finite-temperature QCD. These approaches are not limited to lattice field theories. They are expected to be equally valuable in other domains, such as condensed matter systems, where precise numerical control and reliable diagnostics of phase transitions are important for large-scale computations. 


\clearpage\mbox{}\clearpage
\chapter*{List of Publications and Proceedings}
\begin{enumerate}

\item \noindent{N. S. Dhindsa, Anosh Joseph and \textbf{Vamika Longia}, 
``Gradient and Hessian-Based Temperature Estimator in Lattice Gauge Theories: A Diagnostic Tool for Stability and Consistency in Numerical Simulations'' 
\href{https://doi.org/10.1007/JHEP10(2025)015}{JHEP \textbf{10}, 015 (2025)}, 
\href{https://doi.org/10.48550/arXiv.2508.05595}{arXiv:2508.05595 [hep-lat]}}.

\item \noindent{Abhishek Samlodia, \textbf{Vamika Longia}, Raghav G. Jha and Anosh Joseph, 
``Phase diagram of generalized XY model using tensor renormalization group'' 
\href{https://link.aps.org/doi/10.1103/PhysRevD.110.034504}{Phys. Rev. D \textbf{110}, 034504 (2024)}}.

\item \noindent{\textbf{Vamika Longia}, N. S. Dhindsa and Anosh Joseph, 
``Configurational Thermometer for Lattice Gauge Theories'' 
\href{https://doi.org/10.22323/1.518.0044}{PoS(LATTICE2025)044}, \href{https://arxiv.org/abs/2601.17436v1}{arXiv:2601.17436 [hep-lat]}}.

\item \noindent{\textbf{Vamika Longia}, Abhishek Samlodia, Raghav G. Jha and Anosh Joseph, 
``Investigating the Two-Dimensional Generalized XY Model using Tensor Networks'' 
XXV DAE-BRNS High Energy Physics Symposium (2022), 
\href{https://doi.org/10.48550/arXiv.2307.13593}{arXiv:2307.13593}}.

\end{enumerate}
\clearpage\mbox{}\clearpage
\listoffigures
\clearpage\mbox{}\clearpage
\listoftables
\clearpage\mbox{}\clearpage
\chapter*{List of Abbreviations}
\addcontentsline{toc}{chapter}{List of Abbreviations}

\begin{tabular}{ll}
\textbf{Abbreviation} & \textbf{Full Form} \\[0.3cm]

BKT  & Berezinskii--Kosterlitz--Thouless \\[0.15cm]
gXY  & Generalized XY \\[0.15cm]
HMC  & Hybrid Monte Carlo \\[0.15cm]
HOSVD & Higher-Order Singular Value Decomposition \\[0.15cm]
HOTRG & Higher-Order Tensor Renormalization Group \\[0.15cm]
LGT  & Lattice Gauge Theory \\[0.15cm]
MCMC & Markov Chain Monte Carlo \\[0.15cm]
MPS  & Matrix Product State \\[0.15cm]
RG   & Renormalization Group \\[0.15cm]
SVD  & Singular Value Decomposition \\[0.15cm]
TN   & Tensor Network \\[0.15cm]
TRG  & Tensor Renormalization Group \\[0.15cm]
U(1) & Unitary Group of Degree 1 \\[0.15cm]
SU(2) & Special Unitary Group of Degree 2 \\[0.15cm]
SU(3) & Special Unitary Group of Degree 3 \\

\end{tabular}
\clearpage\mbox{}\clearpage
\chapter*{List of Symbols}
\addcontentsline{toc}{chapter}{List of Symbols}

\begin{tabular}{ll}
\textbf{Symbol} & \textbf{Meaning} \\[0.3cm]

$D$ 
& Bond dimension in tensor renormalization methods \\[0.15cm]

$D_{\mathrm{cut}}$ 
& Truncation bond dimension used in HOTRG \\[0.15cm]

$\chi_{\mathrm{MPS}}$ 
& Bond dimension of a Matrix Product State (MPS) \\[0.15cm]

$\Delta$ 
& Deformation parameter of the generalized XY model \\[0.15cm]

$T$ 
& Temperature \\[0.15cm]

$T_c$ 
& Critical temperature \\[0.15cm]

$\beta$ 
& Inverse temperature \\[0.15cm]

$\beta_M$ 
& Configurational inverse-temperature estimator \\[0.15cm]

$h$ 
& External magnetic field \\[0.15cm]

$h_1$ 
& Nematic symmetry-breaking field \\[0.15cm]

$Z$ 
& Partition function \\[0.15cm]

$C_v$ 
& Specific heat \\[0.15cm]

$\chi$ 
& Magnetic susceptibility \\[0.15cm]

$M$ 
& Magnetization \\[0.15cm]

$M_1$ 
& Nematic magnetization \\[0.15cm]

$V$ 
& Lattice volume \\[0.15cm]

$\sigma_i$ 
& Singular values obtained from the SVD \\[0.15cm]

$\epsilon_{\mathrm{disc}}$ 
& Discarded weight in tensor truncation \\

\end{tabular}
\clearpage\mbox{}\clearpage
\tableofcontents
\clearpage\mbox{}\clearpage
\pagenumbering{arabic}
	
\chapter{Introduction}
\label{ch:intro}

In many body physics, phase transitions play a fundamental role. They mark qualitative changes in macroscopic behavior when external parameters, such as temperature or coupling, are varied. 
A canonical example is the liquid-vapor transition in water. At low pressures, there is a first-order coexistence line. It terminates at a critical point, beyond which the distinction between liquid and vapor disappears. 

At a continuous (second-order) critical point, thermodynamic observables exhibit singular behavior. For example, near the Curie point, in a ferromagnet, the heat capacity $C$, magnetization $M$, and susceptibility $\chi$ follow power laws:
\[
C \sim |T - T_c|^{-\alpha}, \quad M \sim (T_c - T)^{\beta}, \quad \chi \sim |T - T_c|^{-\gamma}, \quad T \to T_c.
\]
Strikingly, the exponents $\alpha$, $\beta$, $\gamma$, $\ldots$ that characterize these divergences are largely universal. 
They take the same values across seemingly different systems. 
This universality, explained theoretically by the renormalization group (RG), underlies why diverse systems such as magnetic materials, liquid-vapor mixtures, and lattice spin models can exhibit quantitatively similar critical behavior despite microscopic differences. 
Understanding these critical phenomena, and their classification into universality classes, has been a major success of statistical physics and quantum field theory, both conceptually and through advanced computational methods. 

In the past century, the theory of critical phenomena evolved from Landau's mean-field symmetry-breaking framework to the seminal work of Kosterlitz, Thouless, Wilson, Fisher, and many others \cite{Landau1937, Kosterlitz1973, Wilson1974}. 
The RG approach of Wilson in the 1970s explained why microscopically different models could share the same critical exponents by showing that long-distance physics near criticality is governed by a fixed point of scale transformations. 
This RG insight also explains why continuous symmetries in two dimensions cannot spontaneously break (the Mermin--Wagner theorem) and why special infinite-order transitions, such as the Berezinskii--Kosterlitz--Thouless (BKT) phenomenon, can occur. 
For example, in the two-dimensional XY model (with an O(2) spin symmetry), there is no conventional long-range order at any nonzero temperature. 
However, there is a topological BKT transition driven by vortex--antivortex unbinding. 
The XY model has been extensively studied in contexts ranging from superfluid films to two-dimensional melting. 

Modern investigations of critical behavior rely heavily on numerical methods. 
In lattice field theory and statistical mechanics, Monte Carlo simulations of spin and gauge models have long been a workhorse for locating phase transitions and measuring critical exponents. 
However, Monte Carlo methods can suffer from long autocorrelation times (critical slowing down), finite-size effects, and potential systematic biases \cite{Binder1987, Sokal1997}. 
In lattice gauge theories, for instance, one must carefully calibrate algorithms such as Hybrid Monte Carlo (HMC) to avoid subtle sampling errors that can distort thermodynamic observables. 
These challenges motivate the development of independent, high-precision numerical tools and diagnostics to complement traditional approaches. 
Recent years have seen significant progress in tensor network methods --- real-space renormalization schemes where the partition function of a lattice model is expressed as a contraction of multi-legged tensors. 
Algorithms such as the higher-order tensor renormalization group (HOTRG) allow the evaluation of observables in the thermodynamic limit with controlled approximations. 
These tensor renormalization group (TRG) methods can capture complex critical behavior (including both BKT-type and discrete-symmetry transitions) by coarse-graining the lattice directly in position space, providing an alternative to Monte Carlo.

In the first part of this thesis, we study a specific spin model -- the generalized XY (gXY) model in two dimensions -- using a GPU-accelerated HOTRG implementation. 
The gXY model is a deformation of the classical 2D XY model proposed by Korshunov, Lee, and Grinstein. 
It introduces an additional ``nematic'' coupling that allows both integer and half-integer vortices, leading to a rich phase diagram with three distinct phases: a ferromagnetic (integer-vortex) phase, a nematic (half-integer-vortex) phase, and a high-temperature disordered (paramagnetic) phase. 
These phases are separated by two Berezinskii--Kosterlitz--Thouless (BKT) transitions (one for integer vortices, one for half-integer vortices) and an intervening Ising-like transition, which are expected to meet at a multicritical point. 

The three phases are separated by transition lines of different universality classes. 
In particular, the nematic term in the Hamiltonian allows half-integer vortices that carry $\pi$ winding, in addition to the usual $2\pi$ vortices of the XY model. 
Previous studies (using Monte Carlo and matrix-product-state methods) have established the existence of these phases. However, a precise mapping of the phase boundaries and the nature of the multi-critical region has remained challenging. 
In this work, we apply HOTRG to the gXY model to compute its partition function directly in the thermodynamic limit. 
We construct a tensor network by expanding the Boltzmann weight in character (dual) variables, and then coarse-grain this tensor using HOTRG. 
By inserting impurity tensors that encode perturbations (such as an external magnetic field), we can extract observables. For example, we compute the (spin and nematic) magnetizations and susceptibilities from appropriate impurity contractions. 
From these quantities, we identify the critical temperatures of the integer-BKT, half-BKT, and Ising transitions as a function of the coupling parameter.


The second thrust of this thesis is not a new phase diagram for a spin model, but rather the development of a general diagnostic for thermodynamic consistency in lattice field theory simulations. 
We build on a configurational-temperature estimator originally introduced by Rugh. 
In Rugh's geometric formulation of statistical mechanics, the temperature of a microcanonical ensemble can be expressed in terms of derivatives of the Hamiltonian rather than kinetic energies. 
Specializing to lattice (Euclidean) field theories, one obtains a purely configuration-based estimator: it depends only on the gradient and Hessian of the action with respect to the fields, without any reference to momenta. 
Crucially, this estimator can be made gauge-invariant by using covariant derivatives, so it applies even to compact gauge theories. 

We test the configurational temperature estimator on compact U(1) lattice gauge theory in one, two, and four dimensions. 
These cases serve as a controlled suite of testbeds. 
In one and two dimensions, U(1) gauge theory is analytic or trivial (no phase transition), so one can directly check against exact results. 
In four dimensions, compact U(1) theory is known to have a weak first-order confining--Coulomb phase transition around $\beta \approx 1.0$, providing a nontrivial benchmark. 

Overall, our studies showcase two complementary advances in numerical studies of critical phenomena. 
On one hand, tensor network renormalization (HOTRG) can accurately capture complex phase diagrams. 
On the other hand, the configurational temperature provides a gauge-invariant diagnostic tool to validate Monte Carlo sampling. 

The thesis is organized as follows. 
Chapter \ref{ch:tn} presents an overview of tensor network methods for quantum many-body systems, including an introduction to tensor networks (TNs), singular value decomposition (SVD), and the higher-order tensor renormalization group (HOTRG) method. 
In Chapter \ref{ch:gxy}, we study the generalized XY model and investigate the multicritical region using the HOTRG approach. 
Chapter \ref{ch:tmp_est} is devoted to the development and formulation of the configurational temperature-estimator. 
In Chapter \ref{ch:u1}, we benchmark the configurational temperature estimator on U(1) lattice gauge theory. 
Finally, Chapter \ref{ch:conclusion} concludes the thesis with a summary of results and a discussion of future directions.

\clearpage\mbox{}\clearpage
\chapter{Tensor Network Methods for Quantum Many-Body Systems}
\label{ch:tn}

The formulation of quantum mechanics in the early twentieth century provided a fundamental description of microscopic systems \cite{Dirac1930, Vonneumann1932, Sakurai1994}. 
However, it simultaneously exposed one of the most formidable challenges in theoretical physics: the quantum many-body problem \cite{Anderson1972}. 
For a system of $N$ interacting particles, the dimension of the Hilbert space typically scales exponentially with $N$ \cite{Nielsen2000}. 
This exponential growth renders traditional numerical approaches, such as exact diagonalization, computationally infeasible even for modest system sizes \cite{Feynman1982, Schollwock2011-dmrg, Verstraete2008-mps}. 

A decisive breakthrough occurred in 1993 when White introduced the Density Matrix Renormalization Group (DMRG) \cite{White1993}. 
DMRG demonstrated remarkable efficiency in treating one-dimensional (1D) strongly correlated systems by systematically retaining the most relevant eigenstates of reduced density matrices. 
This success indicated that physically relevant low-energy states occupy only a restricted, highly structured region of the full Hilbert space. 

Subsequent developments led to the formalization of the tensor network (TN) framework \cite{Verstraete2008, Schollwock2011-dmrg, Orus2014}. 
In this paradigm, quantum many-body wave functions are represented as networks of interconnected tensors whose structure encodes entanglement and locality. 
A central insight underpinning tensor networks is the \textit{area law} for entanglement entropy. 
For gapped local Hamiltonians, the bipartite entanglement entropy $S_A$ of a subsystem $A$ scales as \cite{Hastings2007, Eisert2010}
\begin{equation}
S_A = - \mathrm{Tr}(\rho_A \log \rho_A) \sim |\partial A|,
\end{equation}
where $\rho_A$ is the reduced density matrix and $|\partial A|$ denotes the boundary of subsystem $A$, rather than its volume. 
This restricted entanglement growth explains why low-rank tensor decompositions efficiently approximate ground states. 

By explicitly encoding locality and entanglement structure, tensor networks reduce exponential complexity to polynomial scaling in physically relevant regimes \cite{Verstraete2008, Orus2014}. 
This renders strongly correlated quantum systems, once considered intractable, manageable to controlled numerical investigation. 

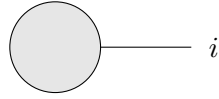
\begin{figure}[ht]
\centering
\begin{tikzpicture}[scale=1.5]
    \draw[fill=gray!20] (0,0) circle (0.4);
    \draw (0.4,0) -- (1.2,0);
    \node at (1.4,0) {$i$};
\end{tikzpicture}
\caption{Penrose graphical representation of a tensor node with one index.}
\label{fig:tensor-node}
\end{figure}

Tensor networks (TNs)\footnote{For an easy and accessible understanding of tensor networks, the reader is encouraged to refer to the following webpages: \url{https://tensornetwork.org/} and \url{https://www.tensors.net/}.} offer an efficient framework for representing and computing with the exponentially large wave functions characteristic of many-body quantum systems. 
A tensor is defined as a multidimensional array of complex numbers, where the rank corresponds to the number of indices. 
For example, a $rank-0$ tensor is a scalar, a $rank-1$ tensor is a vector, and a $rank-2$ tensor is a matrix \cite{Orus2014}. 
In tensor network theory, diagrammatic notation is commonly employed, specifically \textit{Penrose graphical notation} \cite{Penrose1971}. 
Each tensor is depicted as a node, and each index is represented as a line that extends from the node.
In Fig. \ref{fig:tensor-node} we show the Penrose graphical representation of a tensor node with one index.

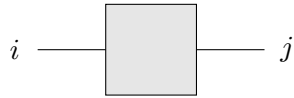
\begin{figure}[ht]
\centering
\begin{tikzpicture}[scale=1.5]
    \draw[fill=gray!20] (-0.4,-0.4) rectangle (0.4,0.4);
    \draw (-1,0) -- (-0.4,0);
    \draw (0.4,0) -- (1,0);

    \node at (-1.2,0) {$i$};
    \node at (1.2,0) {$j$};
\end{tikzpicture}
\caption{Graphical representation of a rank-2 tensor (matrix).}
\label{fig:matrix-tensor}
\end{figure}

For example, a matrix is illustrated as a box with two lines corresponding to its two indices.
See Fig. \ref{fig:matrix-tensor}.

A simple three-index tensor is illustrated below in Fig. \ref{fig:rank3tensor}.

\begin{figure}[ht]
\centering
\begin{tikzpicture}[scale=1.5]
    \draw[fill=gray!20] (0,0) circle (0.5);

    \draw (-1,0) -- (-0.5,0);
    \draw (1,0) -- (0.5,0);
    \draw (0,1) -- (0,0.5);

    \node at (-1.2,0) {$i$};
    \node at (1.2,0) {$j$};
    \node at (0,1.2) {$k$};
\end{tikzpicture}
\caption{Graphical representation of a rank-3 tensor $T_{ijk}$.}
\label{fig:rank3tensor}
\end{figure}
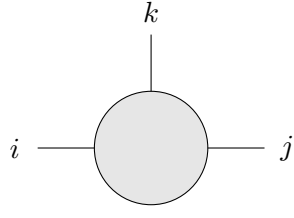

A tensor network consists of multiple tensors with certain indices contracted, which means summed pairwise. 
Graphically, contractions are shown by connecting the corresponding lines of two tensors. 
In algebraic terms, if two tensors $A_{ij}$ and $B_{jk}$ share an index $j$, their contraction yields
\begin{equation}
C_{ik} = \sum_j A_{ij} B_{jk},
\end{equation}
which is analogous to matrix multiplication.

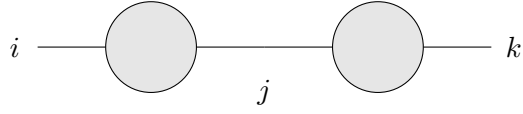
\begin{figure}[ht]
\centering
\begin{tikzpicture}[scale=1.5]

    \draw[fill=gray!20] (-1,0) circle (0.4);
    \draw (-1.4,0) -- (-2,0);
    \draw (-0.6,0) -- (0,0);
    
    \draw[fill=gray!20] (1,0) circle (0.4);
    \draw (0,0) -- (0.6,0);
    \draw (1.4,0) -- (2,0);
    
    \node at (-2.2,0) {$i$};
    \node at (2.2,0) {$k$};
    \node at (0,-0.4) {$j$};

\end{tikzpicture}
\caption{Tensor contraction between two tensors sharing index $j$.}
\label{fig:tensor-contraction}
\end{figure}

\begin{figure}[ht]
\centering
\begin{tikzpicture}[scale=1.5]

    \draw[fill=gray!20] (-2,0) circle (0.4);
    \draw (-2.4,0) -- (-3.5,0);     
    \draw (1.6,0) -- (-1.6,0); 
    
    \draw[fill=gray!20] (0,0.8) circle (0.4);
    \draw (-0.4,0.8) -- (-1.72,0.3); 
    \draw (0.4,0.8) -- (1.72,0.3);   
    \draw (0,1.2) -- (0,2);         
    
    \draw[fill=gray!20] (2,0) circle (0.4);
    \draw (2.4,0) -- (3.5,0);         
    
    \node at (-3.7,0) {$j$};
    \node at (0,2.2) {$i$};
    \node at (3.7,0) {$k$};
    
    \node at (-1,0.8) {$l$};
    \node at (0,-0.6) {$M_T$};
    \node at (1,0.8) {$n$};

\end{tikzpicture}
\caption{Tensor network representation of the contraction.}
\label{fig:tn-chain-corrected}
\end{figure}
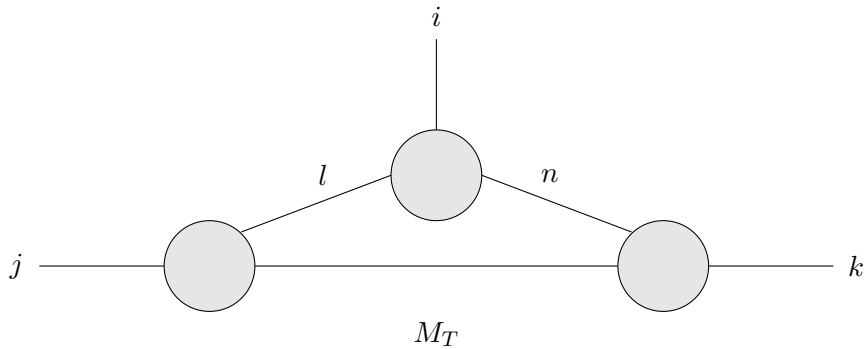

Summing over a shared index of two tensors results in a new tensor or a scalar if no free indices remain. 
Tensor networks can encode complex algebraic structures graphically, where open lines represent the free indices of the resulting object. 
An open chain of contracted tensors produces a final tensor with the remaining free indices.
For example, we have
\begin{equation}
D_{ijk} = \sum_{l,m,n} A_{ljm} B_{iln} C_{nmk}.
\end{equation}
In Fig. \ref{fig:tn-chain-corrected} we show the tensor network representation of the contraction given above. 

This diagrammatic formalism provides an intuitive and computationally powerful language for manipulating many-body wave functions. 

\section{Representation of Many-Body States}
\label{sec:mbs}

A principal application of tensor networks is the representation of many-body quantum states. 
A general state of $N$ particles, each with local dimension $p$, is described by amplitude coefficients forming an order-$N$ tensor with $p^N$ entries, written as
\[
|\Psi\rangle = \sum_{i_1, i_2, \ldots, i_N = 1}^{p} C_{i_1 i_2 \ldots i_N} |i_1 i_2 \ldots i_N\rangle,
\]
where $C_{i_1 i_2 \ldots i_N}$ is an order-$N$tensor with $p^N$ coefficients. 
As $N$ increases, this number grows exponentially, making explicit storage intractable. 
Tensor networks address this challenge by decomposing the large tensor into a network of smaller tensors connected by contracted indices. 
Each smaller tensor contains significantly fewer parameters, and the network structure encodes the entanglement properties of the many-body system. 
TN methods were specifically developed to exploit the structure of physically relevant quantum states. 
Mathematically, this approach rewires the full-amplitude tensor into a set of local tensors whose contraction reconstructs the original state. 
When the quantum state exhibits limited entanglement, as is typical for ground states of local Hamiltonians, this decomposition reduces the number of parameters from exponential to polynomial in system size. 
Thus, enabling substantial dimensional reduction by focusing on the physically relevant subspace of Hilbert space where low-energy states reside.

\subsection{One-Dimensional Representation}

In one dimension, the natural tensor network representation is the Matrix Product State (MPS). 
A general quantum state\footnote{A clear pedagogical introduction to MPS can be found at \url{https://tensornetwork.org/mps/}.}
\begin{equation}
|\psi\rangle = \sum_{i_1,i_2,\dots,i_N} C_{i_1 i_2 \dots i_N} |i_1 i_2 \dots i_N\rangle
\end{equation}
contains exponentially many coefficients $C_{i_1 i_2 \dots i_N}$. 
The MPS ansatz factorizes this coefficient tensor into a product of local tensors:
\begin{equation}
C_{i_1 i_2 \dots i_N} = \sum_{\{\alpha\}} A^{[1] i_1}_{\alpha_1} A^{[2] i_2}_{\alpha_1 \alpha_2} \cdots A^{[N] i_N}_{\alpha_{N-1}},
\end{equation}
where $\alpha_k$ are auxiliary (bond) indices of dimension $\chi_{MPS}$. 
The bond dimension $\chi_{MPS}$ controls the amount of entanglement captured by the ansatz. 
For gapped 1D systems satisfying an area law, relatively small $\chi_{MPS}$ suffices for accurate approximations.

\begin{figure}[ht]
\begin{center}
\begin{tikzpicture}[scale=1.5]

    \draw[rounded corners=6pt, fill=gray!30] (-3.6,0) rectangle (-1.2,-0.8);
    
    \foreach \x in {-3.4,-3.0,-2.6,-2.2,-1.8,-1.4}
    {
    \draw (\x,-0.8) -- (\x,-1.4);
    }
    
    \node at (-2.35,0.3) {$C$};
    
    \node at (-0.4,-0.4) {$=$};
    
    
    \foreach \x in {0.5,1.6,2.7,3.8,4.9,6.0}
    {
    \draw[fill=pink!80] (\x,-0.4) circle (0.25);
    }
    
    \draw (0.75,-0.4) -- (1.35,-0.4);
    \draw (1.85,-0.4) -- (2.45,-0.4);
    \draw (2.95,-0.4) -- (3.55,-0.4);
    \draw (4.05,-0.4) -- (4.65,-0.4);
    \draw (5.15,-0.4) -- (5.75,-0.4);
    
    \foreach \x in {0.5,1.6,2.7,3.8,4.9,6.0}
    {
    \draw (\x,-0.65) -- (\x,-1.3);
    }
\end{tikzpicture}
\end{center}
\caption{Matrix Product State (MPS) / Tensor Train decomposition of a tensor $C$. 
The original tensor with six indices (shown on the left) is factorized into a chain of tensors connected by virtual bond indices.}
\end{figure}
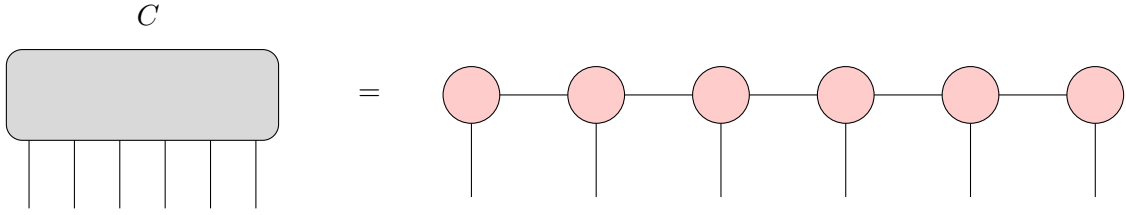

\subsection{Higher-Dimensional Generalizations}

We can extend the tensor network framework to higher dimensions in a natural way. 

We have the following in higher dimensions:
\begin{itemize}
\item \textbf{Projected Entangled Pair States (PEPS):} A two-dimensional generalization of MPS suitable for lattice systems \cite{Verstraete2004}. 
\item \textbf{Multi-scale Entanglement Renormalization Ansatz (MERA):} A hierarchical tensor network incorporating scale transformations and efficiently describing critical systems \cite{Vidal2007}.
\end{itemize}

While MPS-based methods dominate in one dimension, higher-dimensional tensor networks face increased computational complexity due to bond-dimension growth. 
An important alternative approach, inspired by real-space renormalization ideas, is the Tensor Renormalization Group (TRG) discussed in the later section. 
TRG\footnote{The method has been studied on Ising Model \cite{Cook2015} and beautifully represented here in a pictorial manner: \url{https://tensornetwork.org/trg/}} performs iterative coarse-graining of tensor networks through singular value decompositions (SVD), retaining dominant entanglement contributions while avoiding full network contraction. 
This enables efficient simulations of two-dimensional classical and quantum systems, particularly at finite temperature and near criticality.

\subsection{Tensor Decompositions}

At the core of tensor network algorithms lie fundamental tensor decompositions such as the \emph{Singular Value Decomposition} (SVD), \emph{QR decomposition}, and \emph{eigen-decomposition} \cite{NielsenChuang2000}. 
The SVD, for instance, allows one to factorize a reshaped tensor $M_T$ into
\[
M_T = U \Sigma V^\dagger,
\]
where $U$ and $V$ are unitary matrices and $\Sigma$ contains singular values $\{\sigma_i\}$. 
Truncating small singular values provides a compressed approximation with controlled error. 
This is a key step in algorithms like the Density-Matrix Renormalization Group (DMRG) and Time-Evolving Block Decimation (TEBD), where truncation retains only the most relevant entanglement components. 

Efficient contraction of large networks is achieved by optimizing the sequence of pairwise contractions to reduce computational cost. 
Many tensor network algorithms iteratively contract, decompose, and update tensors to approximate ground states or simulate time evolution. 
This formalism has become the standard framework for describing quantum lattice systems, enabling direct manipulation of the entanglement structure of quantum matter.

\section{Tensor Renormalization Group}
\label{sec:trg}

The need for reliable and scalable numerical renormalization group tools for two-dimensional classical systems led to a clear and influential proposal by Levin and Nave \cite{Cardy1996, Levin2007}. 
They reformulated the classical partition function as a contraction of a local tensor network \cite{Steinhaus2020}. 
They also introduced a deterministic coarse-graining method that compresses these tensors while retaining the degrees of freedom that matter most for long-distance physics. 
Levin and Nave built on ideas that had worked well for one-dimensional quantum systems. 
Their method uses consecutive decompositions to identify and remove the least important virtual bonds, instead of sampling configurations like in Monte Carlo. 
This creates a controlled approximation, and its accuracy is managed by an adjustable bond dimension. 

An important advancement of this approach is its clear use of quantum information concepts, especially the idea of entanglement and the distribution of singular values, to guide truncation. 
This perspective makes the algorithm well suited to capture the short-range correlations that dominate non-universal behavior. 
At the same time, it preserves the large-scale structure that determines universality classes. 
In practical terms, TRG avoids the stochastic noise and critical slowing down that can occur with Monte Carlo methods near phase transitions \cite{Cook2015}. 
It also does not have a sign problem, allowing it to be applied to models with complex or non-positive weights without the sampling failures common in probabilistic methods 

TRG is completely isotropic in terms of its methods. Coarse-graining steps treat all lattice directions the same, which helps keep rotational invariances and makes it easier to see universal, long-wavelength observables. The adjustable bond dimension makes it easy to see the trade-off between accuracy and cost, and it lets one estimate errors by looking at different truncation levels. The original TRG has led to many improvements since then, such as higher-order decompositions, better truncation criteria, and better contraction sequences. However, its main idea of using tensor networks to represent classical partition sums and linear algebraic compression based on entanglement is still an important step in linking classical renormalization and quantum information theory. 

\section{Singular Value Decomposition}
\label{sec:svd}



When Kadanoff, and later Wilson, looked at the Ising model near the critical temperature, they introduced a simple yet powerful coarse-graining idea \cite{Kadanoff1966, Wilson1974}. They suggested dividing the lattice into blocks that are large on a microscopic scale but small compared to the correlation length. Each block is replaced by its average magnetization, turning these block variables into the degrees of freedom for a new, coarser lattice. This blocking process can be repeated, leading to an effective theory that gradually filters out short-range fluctuations. Microscopic details are averaged out, while the long-range collective structure that influences critical behavior remains intact. In Wilson's framework, this observation is made systematic. Many microscopic modes do not matter for the infrared physics and can be integrated out, creating a flow in theory space that relates to just a few relevant operators. 


In this context, renormalization is the process of intentionally blurring fine details to focus on the variables that define large-scale behavior. The goal is not to capture every microscopic configuration but to create a simplified description whose degrees of freedom reflect the same macroscopic response. This concept helps explain why universal features like critical exponents, scaling functions, and fixed points are not influenced by microscopic details but depend instead on a few global properties such as symmetry and dimensionality. 



Tensor-network methods make this idea of renormalization into a mathematical algorithm that works for many-body systems. A quantum state or partition function is reworked as a network of local tensors. The RG step is then a way to group those tensors together: tensors that are close to each other are combined and replaced with a lower-rank approximation, which keeps the network manageable. In practice, this process of merging, approximating, and rescaling is done over and over until the tensors reach a fixed-point form that shows the physics of long distances.This idea is put into action by Tensor Renormalization Group (TRG) methods, which do three main things: they reconnect local tensor indices (rewiring), they use singular-value decomposition to compress the enlarged index spaces (truncation), and they contract the network to make the next coarse lattice (decimation). The SVD-based truncation is a controlled approximation that keeps only the most important correlations \cite{Aydin2006}. This is similar to how blocking gets rid of small, unimportant fluctuations while keeping the important collective degrees of freedom.

The singular-value decomposition (SVD) is the linear algebra tool that makes the physical idea of coarse-graining into a controlled numerical operation. In a real-space renormalization step, we combine nearby degrees of freedom to make a larger object. Then, we compress that object so that the coarse-grained description is still easy to work with. The SVD does exactly this kind of compression: it finds the best low-rank approximation of a matrix (in the 2-norm), and when applied to tensors that have been reshaped properly, it picks the dominant correlated subspace that holds long-range information. When we look at the squared singular values as Schmidt weights, we can see that getting rid of small singular values is like getting rid of weak Schmidt components (short-range noise) and keeping the most important correlations. This linear-algebraic truncation is the numerical equivalent of Kadanoff blocking and is the basis for tensor renormalization group (TRG) algorithms.


\subsection{Mathematical Formulation}

\begin{figure}[ht]
\centering
\begin{tikzpicture}[scale=1.0, every node/.style={scale=1.0}]
    \draw[thick] (0,0) rectangle (1.5,1.5);
    \node at (0.75,0.75) {$M_T$};

    \node at (2.2,0.75) {$=$};

    \draw[thick] (2.8,0) rectangle (4.3,1.5);
    \node at (3.55,0.75) {$U$};

    \draw[thick] (4.8,0) rectangle (6.3,1.5);
    \node at (5.55,0.75) {$\Sigma$};

    \draw[thick] (6.8,0) rectangle (8.3,1.5);
    \node at (7.55,0.75) {$V^\dagger$};
\end{tikzpicture}
\caption{Graphical representation of the singular value decomposition $M_T = U \Sigma V^\dagger$.}
\label{fig:graphical_rep}
\end{figure}
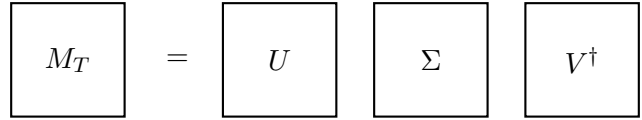

Let $M_T \in \mathbb{C}^{m\times n}$. 
Then, the (full) singular-value decomposition has the form \cite{WolframSVD}
\begin{equation}
\label{eq:svd-full}
M_T = U \,\Sigma \,V^\dagger,
\end{equation}
where $U \in \mathbb{C}^{m\times m}$ and $V \in \mathbb{C}^{n \times n}$ are unitary and $\Sigma$ is an $m \times n$ diagonal matrix with non-negative entries $\sigma_1 \ge \sigma_2 \ge \cdots \ge 0$. 
The values $\{\sigma_i\}$ are the singular values of $M_T$.
In Fig. \ref{fig:graphical_rep}, we show the graphical representation of the singular value decomposition $M_T = U \Sigma V^\dagger$.

\begin{figure}[ht]
\centering
\begin{tikzpicture}
    \foreach \i in {0,...,5} {
        \draw[fill=gray!30] (0,\i*0.3) rectangle (2,\i*0.3+0.2);
    }
    \node at (1,2.2) {All $\sigma_i$};

    \draw[->, thick] (2.5,1) -- (4,1);

    \foreach \i in {0,...,2} {
        \draw[fill=blue!40] (4.5,\i*0.3) rectangle (6.5,\i*0.3+0.2);
    }
    \foreach \i in {3,...,5} {
        \draw[fill=red!30] (4.5,\i*0.3) rectangle (6.5,\i*0.3+0.2);
    }

    \node at (5.5,2.2) {Keep $D$, discard rest};
\end{tikzpicture}
\caption{Truncation of singular values: largest $D$ retained, smaller ones discarded.}
\label{fig:truncation}
\end{figure}
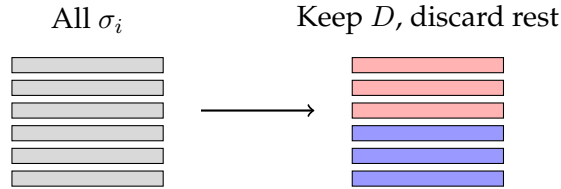

If we retain only the largest $D$ singular values, as shown schematically in Fig. \ref{fig:truncation}, 
then we would obtain the $rank-D$ approximation
\begin{equation}
\label{eq:svd-trunc}
M_T \approx \widetilde M_T = U_D \,\Sigma_D \, V_D^\dagger,
\end{equation}
where $U_D$ and $V_D$ contain the first $D$ columns of $U$ and $V$, respectively, and $\Sigma_D$ is the $D \times D$ diagonal matrix of the kept singular values. 
In Fig. \ref{fig:truncated_sing_val}, we show a graphical representation of the truncated singular value decomposition $\widetilde M_T = U_D \Sigma_D V_D^\dagger$.
Sometimes $D$ is also referred to as the {\it bond dimension}.

\begin{figure}[ht]
\centering
\begin{tikzpicture}[scale=1.0, every node/.style={scale=1.0}]
    \draw[thick] (0,0) rectangle (1.5,1.5);
    \node at (0.75,0.75) {$\widetilde M_T$};

    \node at (2.2,0.75) {$=$};

    \draw[thick] (2.8,0) rectangle (3.6,1.5);
    \node at (3.2,0.75) {$U_D$};

    \draw[thick] (4.2,0.7) rectangle (5.0,1.5);
    \node at (4.6,1.14) {$\Sigma_D$};

    \draw[thick] (5.6,0.7) rectangle (7.1,1.5);
    \node at (6.35,1.14) {$V_D^\dagger$};
\end{tikzpicture}
\caption{Graphical representation of the truncated singular value decomposition $\widetilde M_T = U_D \Sigma_D V_D^\dagger$.}
\label{fig:truncated_sing_val}
\end{figure}

By the Eckart-Young theorem this $\widetilde M_T$ is optimal in the spectral norm:
\begin{equation}
\| M_T - \widetilde M_T \|^2 = \sigma_{D+1}.
\end{equation}
A convenient global error measure is the \emph{discarded weight}
\begin{equation}
\label{eq:discarded-weight}
\varepsilon_{\rm disc} = \sum_{i > D} \sigma_i^2,
\end{equation}
which equals the Frobenius-norm squared error $\| M_T - \widetilde M_T \|_F^2$. 

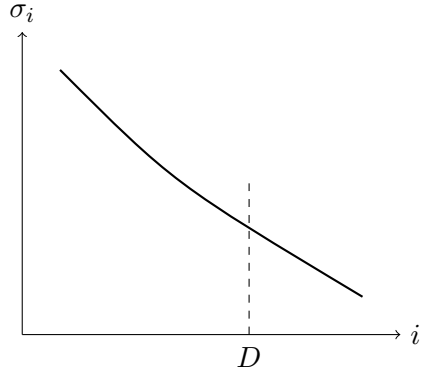
\begin{figure}[ht]
\centering
\begin{tikzpicture}
    \draw[->] (0,0) -- (5,0) node[right] {$i$};
    \draw[->] (0,0) -- (0,4) node[above] {$\sigma_i$};

    \draw[thick] (0.5,3.5) .. controls (2,2) .. (4.5,0.5);

    \draw[dashed] (3,0) -- (3,2);
    \node at (3,-0.3) {$D$};
\end{tikzpicture}
\caption{Typical decay of singular values and truncation at bond dimension $D$.}
\label{fig:typical_decay}
\end{figure}

Because the truncated SVD is optimal in the spectral norm, the discarded weight $\varepsilon_{\rm disc}$ in Eq.~\eqref{eq:discarded-weight} is a direct quantitative diagnostic of how much information is lost at a truncation. 
Practically, one should monitor the decay profile of singular values $\{\sigma_i\}$ after each truncation, the discarded weight $\varepsilon_{\rm disc}$, and the dependence of measured observables (e.g., free energy, specific heat, magnetization) on $D$.
In Fig. \ref{fig:typical_decay}, we show the typical decay of singular values and truncation at bond dimension $D$.

For critical systems, it is common practice to compute observables at several values of $D$ and extrapolate to $D \to \infty$ (or to zero discarded weight) to estimate systematic errors. 
If the model has global symmetries (e.g., $\mathbb{Z}_2$, U(1)), performing \emph{block-sparse} SVD in symmetry sectors both increases accuracy and reduces computational cost.


\subsection{Computational Cost}

The computational cost of a full SVD on a $m \times n$ matrix scales as $O(\min(mn^2, m^2n))$. 
There exist truncated SVD algorithms; they compute only the leading $D$ singular vectors (Lanczos, Arnoldi, randomized SVD).
They typically cost $O(m n D)$, up to constants. 
In TRG, the effective sizes $m, n$ grow with the bond dimension and the number of combined indices, so the cost per RG step grows rapidly with $D$.
Thus, controlling the computational budget is a central concern \cite{Levin2007}. 

Memory requirements scale similarly with intermediate tensor sizes. 
They scale typically as a power of $D$. 
Numerical stability is aided by normalization of tensors after each RG step, QR preconditioning before SVD when matrices are ill-conditioned, and the use of high-quality SVD routines (LAPACK, cuSOLVER on GPUs). 
Exploiting block sparsity coming from conserved quantum numbers is one of the most effective ways to reduce both CPU and memory cost in production runs.

\section{Higher-Order Tensor Renormalization Group Method}
\label{sec:hotrg}

The original TRG procedure, while elegant, relies on pairwise SVD decompositions and bond truncations during coarse-graining. 
However, this becomes inefficient or inaccurate in systems with complex entanglement structures, especially in higher dimensions. 
To save the day, we have higher-order tensor renormalization group (HOTRG) method. 
HOTRG addresses these challenges by employing a higher-order decomposition that considers the full multi-leg structure of local tensors, allowing for more accurate truncation and improved numerical stability. 

The HOTRG method is a powerful extension of the original TRG. 
It was introduced to address several limitations inherent in the basic TRG formulation. 
Particularly those related to truncation accuracy and computational efficiency in higher-dimensional systems. 
Building upon the foundations of tensor network coarse-graining and higher-order singular value decomposition (HOSVD), HOTRG provides a systematic and controlled framework to renormalize tensor networks representing classical partition functions or quantum many-body wave functions in two and higher dimensions. 

The key idea behind HOTRG is to generalize the SVD-based truncation scheme of TRG to a higher-order setting using HOSVD. 
In TRG, at each renormalization step, a tensor network is coarse-grained by contracting pairs of tensors and performing SVD along a single bond direction to truncate the bond dimension. 
This procedure works efficiently in two dimensions for small bond dimensions but suffers from the accumulation of truncation errors because it neglects correlations among different bond directions. 

In contrast, HOTRG uses HOSVD to simultaneously consider multiple bond directions when truncating leading to a more globally optimal projection that preserves dominant subspaces in all relevant directions. 
As a result, HOTRG achieves higher accuracy for the same bond dimension compared to TRG, and it scales more naturally to higher dimensions. 

\subsection{HOTRG Algorithm: Step-by-Step Formulation}

\begin{figure}[h]
\centering

\begin{subfigure}{0.45\textwidth}
\centering
\begin{tikzpicture}[scale=1.2]

\tikzstyle{tensor}=[circle, draw=black, fill=purple!30, text=white, minimum size=0.9cm]

\node[tensor] (T11) at (0,2) {$T^{(0)}$};
\node[tensor] (T12) at (2,2) {$T^{(0)}$};
\node[tensor] (T13) at (4,2) {$T^{(0)}$};

\node[tensor] (T21) at (0,0) {$T^{(0)}$};
\node[tensor] (T22) at (2,0) {$T^{(0)}$};
\node[tensor] (T23) at (4,0) {$T^{(0)}$};

\draw (T11) -- (T12);
\draw (T12) -- (T13);
\draw (T21) -- (T22);
\draw (T22) -- (T23);

\draw (T11) -- (T21);
\draw (T12) -- (T22);
\draw (T13) -- (T23);

\draw (T11.north) -- ++(0,0.6);
\draw (T12.north) -- ++(0,0.6);
\draw (T13.north) -- ++(0,0.6);

\draw (T21.south) -- ++(0,-0.6);
\draw (T22.south) -- ++(0,-0.6);
\draw (T23.south) -- ++(0,-0.6);

\draw (T11.west) -- ++(-0.6,0);
\draw (T21.west) -- ++(-0.6,0);

\draw (T13.east) -- ++(0.6,0);
\draw (T23.east) -- ++(0.6,0);

\end{tikzpicture}
\caption*{(a)}
\end{subfigure}
\hfill
\begin{subfigure}{0.45\textwidth}
\centering
\begin{tikzpicture}[scale=1.3]

\tikzstyle{tensor}=[circle, draw=black, fill=purple!30, text=white, minimum size=1cm]

\node[tensor] (T) at (0,0) {$T^{(0)}$};

\draw (T.north) -- ++(0,1) node[above] {$u$};
\draw (T.south) -- ++(0,-1) node[below] {$d$};
\draw (T.west) -- ++(-1,0) node[left] {$l$};
\draw (T.east) -- ++(1,0) node[right] {$r$};

\end{tikzpicture}
\caption*{(b)}
\end{subfigure}

\caption{(a) Tensor network representation. 
(b) A single rank-4 tensor $T^{(0)}_{lrud}$ with its indices.}
\label{fig:tensor0}

\end{figure}
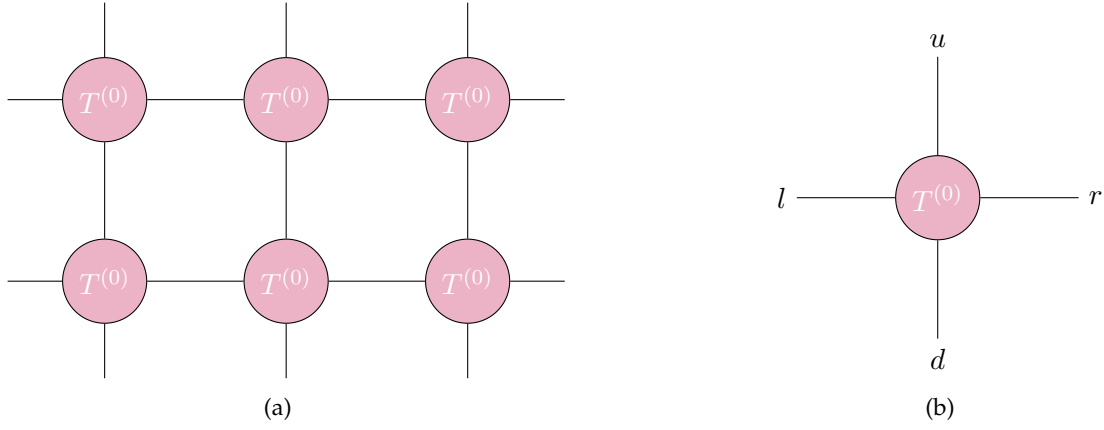

\begin{itemize}
\item[] The content of this section is from the paper: \textit{Coarse-graining renormalization by higher-order singular value decomposition} \cite{Xie2012} \textbf{\textcolor{red!70!black}{[Phys. Rev. B 86, 045139]}}
\end{itemize}
Consider a classical lattice model represented as a tensor network. (Here, we focus on the two-dimensional case, but the same can be extended to higher dimensions.) Suppose the initial network structure has tensors $T^{(0)}_{lrud}$ as shown in Fig. \ref{fig:tensor0}, with bond dimension $D$. The partition function $Z$ is represented as the full contraction of this tensor network:
\begin{equation}
    Z = \mathrm{tTr}\left[ \bigotimes_{i} T^{(0)}_{l_ir_iu_id_i} \right],
\end{equation}
where $\mathrm{tTr}$ denotes tensor trace, i.e., contraction over all internal indices.

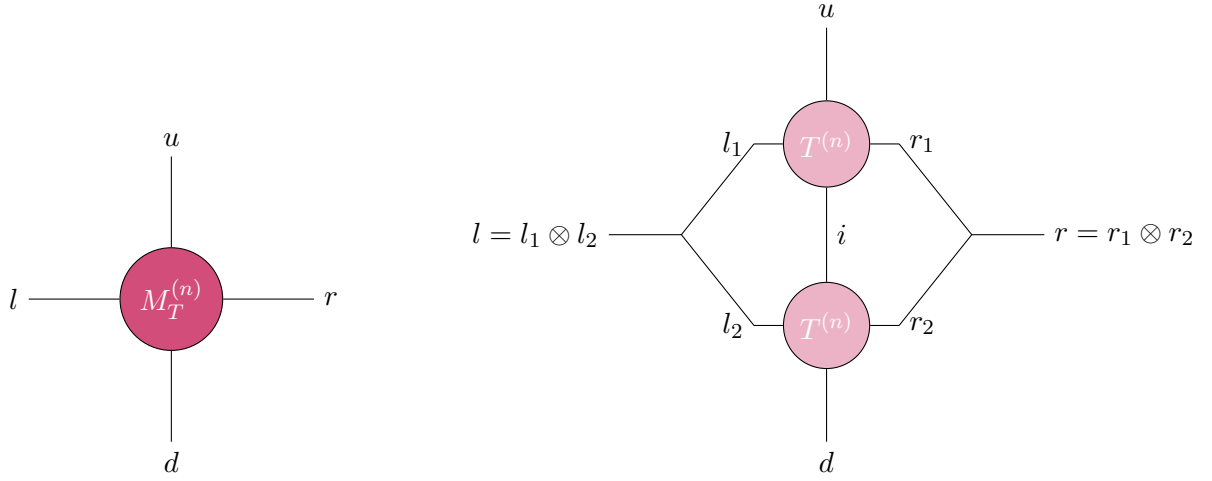
\begin{figure}[h]
\centering

\begin{subfigure}{0.32\textwidth}
\centering
\begin{tikzpicture}[scale=1.2]

\tikzstyle{tensor}=[circle, draw=black, fill=purple!70, text=white, minimum size=1cm]

\node[tensor] (M) at (0,0) {$M^{(n)}_{T}$};

\draw (M.north) -- ++(0,1) node[above] {$u$};
\draw (M.south) -- ++(0,-1) node[below] {$d$};
\draw (M.west) -- ++(-1,0) node[left] {$l$};
\draw (M.east) -- ++(1,0) node[right] {$r$};

\end{tikzpicture}
\caption*{}
\end{subfigure}
\hfill
\begin{subfigure}{0.6\textwidth}
\centering
\begin{tikzpicture}[scale=1.2]

\tikzstyle{tensor}=[circle, draw=black, fill=purple!30, text=white, minimum size=0.9cm]

\node[tensor] (T1) at (0,1) {$T^{(n)}$};
\node[tensor] (T2) at (0,-1) {$T^{(n)}$};

\draw (T1.south) -- node[right] {$i$} (T2.north);

\draw (T1.north) -- ++(0,0.8) node[above] {$u$};
\draw (T2.south) -- ++(0,-0.8) node[below] {$d$};

\draw (T1.west) -- ++(-0.32,0) node[left] {$l_1$};
\draw (T2.west) -- ++(-0.32,0) node[left] {$l_2$};

\draw (-0.8,1) -- (-1.6,0) -- (-0.8,-1);
\draw (-1.6,0) -- ++(-0.8,0) node[left] {$l = l_1 \otimes l_2$};

\draw (T1.east) -- ++(0.32,0) node[right] {$r_1$};
\draw (T2.east) -- ++(0.32,0) node[right] {$r_2$};

\draw (0.8,1) -- (1.6,0) -- (0.8,-1);
\draw (1.6,0) -- ++(0.8,0) node[right] {$r = r_1 \otimes r_2$};

\end{tikzpicture}
\caption*{}
\end{subfigure}

\caption{Resulting tensor $M^{(n)}_T$ obtained by contracting two tensors $T^{(n)}$ along the internal index $i$.}
\label{fig:Mtensor_final}

\end{figure}

The HOTRG coarse-graining procedure proceeds iteratively along different lattice directions. Consider coarse-graining along the horizontal direction by contracting neighbouring tensors along the vertical-axis to form a new composite tensor $M_T$:
\begin{equation}
    M^{(n)}_{T_{lrud}} = \sum_{i} T^{(n)}_{l_1r_1ui} \, T^{(n)}_{l_2r_2id}.
    \label{eq:Mtensor}
\end{equation}
where $l = l_1 \otimes l_2$ and $r = r_1 \otimes r_2$ label the external horizontal bonds, while $u, d, i$ label the vertical ones, and $n$ corresponds to the $n^{th}$ iteration, schematically represented in Fig. \ref{fig:Mtensor_final}. The resulting tensor $M_T$ has enlarged horizontal bond dimension $D^{2}$.

To truncate the enlarged bond space, HOTRG applies HOSVD to the tensor $M_T$. HOSVD decomposes $M_T$ as
\begin{equation}
    M^{(n)}_{T_{lrud}} = \sum_{ijkq} S_{ijkq} \, U_{li} \, V_{rj} \, W_{uk} \, X_{dq},
    \label{eq:hosvd}
\end{equation}
where $S$ is a core tensor and $U, V, W, X$ are unitary matrices associated with the $n^{th}$ mode of the tensor $M_T$ shown in Fig. \ref{fig:hosvd_small}. The decomposition is computed by performing SVD on each matricized mode of $M_T$ and collecting the leading singular vectors. 

\begin{figure}[h]
\centering
\begin{tikzpicture}[scale=0.9, transform shape]

\tikzstyle{tensor}=[circle, draw=black, fill=purple!70, text=white, minimum size=0.75cm]
\tikzstyle{matrix}=[rectangle, draw=black, fill=blue!20, minimum width=0.7cm, minimum height=0.7cm]

\node[tensor] (S) at (0,0) {$S$};

\node[matrix] (U) at (-1.6,0) {$U$};
\node[matrix] (V) at (1.6,0) {$V$};
\node[matrix] (W) at (0,1.6) {$W$};
\node[matrix] (X) at (0,-1.6) {$X$};

\draw (U.east) -- node[above] {\small $i$} (S.west);
\draw (V.west) -- node[above] {\small $j$} (S.east);
\draw (W.south) -- node[right] {\small $k$} (S.north);
\draw (X.north) -- node[right] {\small $q$} (S.south);

\draw (U.west) -- ++(-0.8,0) node[left] {\small $l$};
\draw (V.east) -- ++(0.8,0) node[right] {\small $r$};
\draw (W.north) -- ++(0,0.8) node[above] {\small $u$};
\draw (X.south) -- ++(0,-0.8) node[below] {\small $d$};

\end{tikzpicture}

\caption{HOSVD of $M^{(n)}$}
\label{fig:hosvd_small}

\end{figure}

The $S$ tensor has the following properties for any index (say k):
\begin{enumerate}
    \item orthogonality
    \begin{equation}
        \braket{S_{:,:,k,:}|S_{:,:,k',:}} = 0 \hspace{3mm} \rm{if} \hspace{2mm} k \neq k',
    \end{equation}
    \item pseodo-diagonal
    \begin{equation}
        |S_{:,:,k,:}|\geq |S_{:,:,k',:}| \hspace{3mm} \rm{if} \hspace{2mm} k < k'.
    \end{equation}
\end{enumerate}
Here, $\braket{S_{:,:,k,:}|S_{:,:,k',:}}$ is the inner product of the two sub-tensors, and $|S_{:,:,k,:}|$ is the norm of the sub-tensor which is the square root of all elements' square sum and is similar to the singular values of a matrix.
The truncation step involves selecting the leading $D_{\mathrm{cut}}$ singular vectors in each mode and constructing projectors $P^{(n+1)}$.

We can always apply the truncation on one bond and it will automatically truncate the other bond lying in the same axis as the bond of one tensor is directly linked to the bond of the other identical tensor on the neighbouring site. The choice of which bond to truncate to provide a simple and optimal approximation to minimise the truncation error can be decided on the following factor:
\begin{itemize}
    \item We compare $\varepsilon_1 = \sum_{i>D_{\mathrm{cut}}} |S_{i,:,:,:}|^2$ and $\varepsilon_2 = \sum_{j>D_{\mathrm{cut}}} |S_{:,j,:,:}|^2$.
    \item If $\varepsilon_1 < \varepsilon_2$, we truncate the $i$ to $D_{\mathrm{cut}}$ otherwise we truncate $j$ to $D_{\mathrm{cut}}$. 
\end{itemize}
In practical the vertical bonds do not need to be renormalized so we do not determine $W$ and $X$. Based on the above, the projector is defined as:
\begin{equation}
    P^{(n+1)} = U^{(n)}_{[:,1:D_{\mathrm{cut}}]} \hspace{3mm} \rm{if} \hspace{2mm} \varepsilon_1 < \varepsilon_2 \hspace{2mm} \rm{or} \hspace{1mm} \rm{otherwise}.
\end{equation}

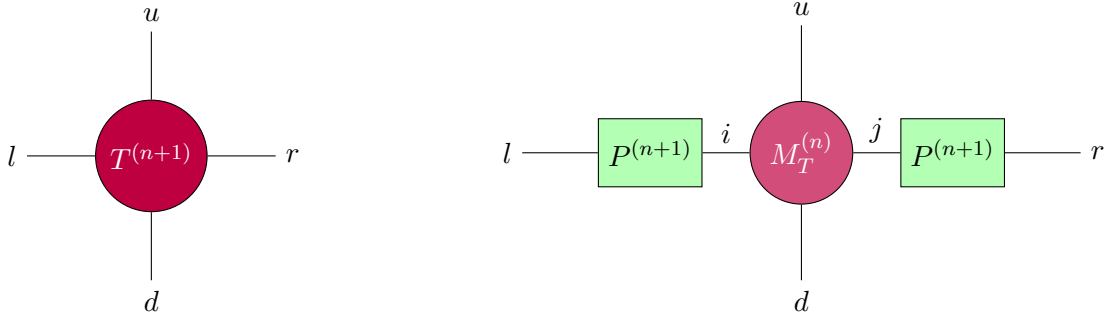
\begin{figure}[h]
\centering

\begin{subfigure}{0.32\textwidth}
\centering
\begin{tikzpicture}[scale=1.0, transform shape]

\tikzstyle{tensor}=[circle, draw=black, fill=purple!100, text=white, minimum size=0.7cm]

\node[tensor] (T) at (0,0) {$T^{(n+1)}$};

\draw (T.north) -- ++(0,0.9) node[above] {$u$};
\draw (T.south) -- ++(0,-0.9) node[below] {$d$};
\draw (T.west) -- ++(-0.9,0) node[left] {$l$};
\draw (T.east) -- ++(0.9,0) node[right] {$r$};

\end{tikzpicture}
\caption*{}
\end{subfigure}
\hfill
\begin{subfigure}{0.6\textwidth}
\centering
\begin{tikzpicture}[scale=1.0, transform shape]

\tikzstyle{tensor}=[circle, draw=black, fill=purple!70, text=white, minimum size=0.8cm]
\tikzstyle{proj}=[rectangle, draw=black, fill=green!30, minimum width=0.7cm, minimum height=0.9cm]

\node[tensor] (M) at (0,0) {$M^{(n)}_T$};

\node[proj] (P1) at (-2,0) {$P^{(n+1)}$};
\node[proj] (P2) at (2,0) {$P^{(n+1)}$};

\draw (P1.east) -- node[above] {$i$} (M.west);
\draw (M.east) -- node[above] {$j$} (P2.west);

\draw (P1.west) -- ++(-1,0) node[left] {$l$};
\draw (P2.east) -- ++(1,0) node[right] {$r$};

\draw (M.north) -- ++(0,1) node[above] {$u$};
\draw (M.south) -- ++(0,-1) node[below] {$d$};

\end{tikzpicture}
\caption*{}
\end{subfigure}

\caption{Coarse-grained tensor $T^{(n+1)}$ construction via projection.}
\label{fig:renorm_updated}

\end{figure}

Using these projectors, the coarse-grained tensor $T^{(n+1)}$ is obtained by contracting $M_T$ with the projectors along the truncated modes, as illustrated in Fig. \ref{fig:renorm_updated}:
\begin{equation}
    T^{(n+1)}_{lrud} = \sum_{ij} P^{(n+1)}_{li} \, M^{(n)}_{T_{ijud}} \, P^{(n+1)}_{jr}.
    \label{eq:renormalizedT}
\end{equation}
This defines a new tensor living on a coarse-grained lattice with reduced linear size by a factor of 2 in the horizontal direction. By alternating coarse-graining steps along different directions (e.g., $x$ then $y$ in 2D), the entire network can be systematically renormalized.

\subsection{Computational Complexity and Numerical Stability}


For a two-dimensional case, the computational cost of HOTRG scales as $\mathcal{O}(D^{7})$, compared to $\mathcal{O}(D^{6})$ for TRG. 
Although slightly more expensive, HOTRG achieves much better accuracy for the same $D$. 
In three dimensions, the scaling is $\mathcal{O}(D^{11})$, which is still manageable for moderate $D$. 

HOTRG is numerically stable because HOSVD truncation preserves the orthogonality of the projection subspaces. 
However, like all tensor network renormalization methods, errors can accumulate with the number of renormalization steps, especially near criticality. 
Careful control of $D_{\mathrm{cut}}$ and gauge fixing are often required for stable results.

\subsection{Concluding Remarks}

HOTRG has been successfully applied to a wide range of classical and quantum lattice models. 
In classical statistical mechanics, it has been used to compute partition functions and critical properties of models such as the 2D and 3D Ising models, Potts models, and clock models with high accuracy. 
In quantum many-body physics, HOTRG can be used to evaluate path-integral representations of quantum systems via Trotter decomposition, enabling finite-temperature calculations. 

For example, applying HOTRG to the 2D classical Ising model allows accurate computation of the free energy and magnetization near criticality. 
The method can capture critical exponents and phase transitions with remarkable precision even for moderate bond dimensions, illustrating the power of HOSVD-based truncation. 

Several extensions of TRG other than HOTRG have been proposed, such as second renormalization group (SRG), higher-order second renormalization group (HOSRG). 
In principle, SRG can achieve better accuracy than HOTRG by optimizing truncations globally. 
However, SRG involves solving large eigenvalue problems and is more computationally expensive. 
HOTRG, by contrast, achieves a good balance between accuracy and efficiency, particularly in higher dimensions where SRG becomes prohibitively expensive. 
The HOSRG incorporates environmental effects similarly to SRG but within the HOTRG framework. 
Other improvements include adaptive truncation strategies, randomized SVD accelerations, and anisotropic coarse-graining schemes. 

HOTRG remains a central tool in the tensor network renormalization landscape, balancing accuracy, stability, and computational cost. 
Its ability to handle higher-dimensional systems makes it an indispensable method for modern studies in statistical and condensed matter physics. 
In the subsequent chapter, we use this method to study the phase transition in a generalized version of the 2D XY model.

\clearpage\mbox{}\clearpage
\chapter{Tensor Renormalization Study of the Generalized XY Model}
\label{ch:gxy}

\textit{This chapter is based on the paper:}
\begin{itemize}
    \item[] \textit{A. Samlodia, V. Longia, R. G. Jha and A. Joseph,  
        ``Phase diagram of generalized XY model using the tensor renormalization group,''} (Published) 
    \item[] \href{https://doi.org/10.1103/PhysRevD.110.034504}{Phys. Rev. D 110, 034504 (2024)}, \href{https://doi.org/10.48550/arXiv.2404.17504}{arXiv:2404.17504 [hep-lat]}. 
\end{itemize} 

Spin models in two Euclidean dimensions with discrete or continuous global symmetries can have a wide range of interesting properties. 
Due to the famous no-go theorem known as the Hohenberg-Mermin-Wagner-Coleman (HMWC) theorem, in two-dimensional models, a continuous symmetry cannot break spontaneously. Thus, a phase transition from a disordered to an ordered phase is not allowed. 
The simplest model in two dimensions with continuous symmetry is the classical XY model, also known as the O(2) model. 
In this model, the spins take values on the circle $S^1$. 
This model has been the subject of several investigations over the past fifty years \cite{Berezinskii:1970fr, Kosterlitz:1973xp, Tobochnik79, S_Ota_1992, Mattis1984} due to its simplicity and remarkable properties. 
Surprisingly, it exhibits an infinite-order topological phase transition, known as the Berezinskii-Kosterlitz-Thouless (BKT) phase transition. 
This transition is peculiar because it does not follow the usual Ehrenfest classification of phase transitions, which describes the order of a phase transition in terms of the lowest discontinuous derivative of the partition function. 
This transition does not violate the HMWC theorem since the transition is not due to breaking of any symmetry but due to the unbinding of vortices and antivortices (topological defects) at some finite temperature. 
Across the phase transition in the XY model, all the derivatives of the free energy remain continuous. 
The transition is associated with the dissociation of the integer vortex pairs at the critical temperature. 
Due to the wide-ranging applications of this model in explaining phenomena related to superfluid Helium, thin films, superconductivity, liquid crystals, and the melting of two-dimensional crystals, several extensions of this model have been considered. 

One such extension, which we study here, was first proposed by Korshunov, Lee, and Grinstein (KLG) in Refs.~\cite{Korshunov85, Lee85}. 
This model has several interesting features, some of which include the possibility of fractional vortices and signs of passing directly from the disordered (high temperature) phase to the single particle (quasi) condensate phase via an Ising transition, a situation reminiscent of the `deconfined criticality' scenario as studied in Ref.~\cite{Shi2011}. 
The competition between the different terms in the Hamiltonian also leads to a richer phase structure. 
This model, discussed in Section ~\ref{sec:gXY}, also provides a good example to understand the interplay between the discrete $\mathbb{Z}_{2}$ symmetry and U(1) symmetry. 

\section{The Generalized XY Model}
\label{sec:gXY}

In real materials and theoretical studies, the pure XY model is often insufficient. 
Crystalline anisotropy, substrate pinning, interlayer couplings, or multi-component order can break the continuous symmetry or add new interactions. 
To capture these effects, physicists have introduced generalized XY models with extra terms or constraints in the Hamiltonian. 
The goals of these generalizations are to interpolate between XY and other known models (e.g., the $q$-state clock model or the Ising model), to explore new fixed points, or to model experimental systems more faithfully. 

For example: 

\textbf{Discrete (clock) symmetry:} One can restrict the phase to $q$ equally spaced values (the $q$-state clock model), or equivalently add an anisotropy term $\cos(q \theta)$. 
Physically, this might represent, say, a crystal field favoring spin orientations along $q$ directions. 
Even a small $\cos(q \theta)$ term breaks the U(1) symmetry down to $\mathbb{Z}_q$. 
Notably, for $q \ge 5$ the model exhibits two transitions, an Ising-like ordering of the discrete symmetry and a BKT transition, whereas for $q \le 4$ it tends to one transition or a first-order jump. 
This was analysed in detail by Jose \emph{et al.} (1977) \cite{Jose1977} and many others. 

\textbf{Nematic interactions:} A special case of the above is $q = 2$, the so-called nematic XY model. 
Here, one adds terms favoring headless ($180^\circ$-periodic) alignment. 
Such a model can have a distinct nematic phase (director order) between the disordered and full XY phases. 
Numerical studies (e.g., Canova \emph{et al.}, 2016 \cite{Canova2016}) show that with strong nematic coupling the system can exhibit multiple transitions or new critical phases not found in the pure XY case. 

\textbf{Coupled layers (3D crossover):} Realistic 2D systems (e.g., layered superconductors or quasi-2D magnets) have a small interlayer coupling. 
One models this by stacking XY layers with an interplane coupling $J_\perp$. 
The 2D--3D crossover is interesting: for very weak $J_\perp$, each layer might undergo a BKT transition almost independently, while for stronger coupling a single 3D XY ordering at a nonzero temperature occurs. 
Understanding this required extending RG ideas to anisotropic 3D (see Nelson and Pelcovits, 1977 \cite{NelsonPelcovits1977}), and experiments on layered materials often test these ideas. 

\textbf{Anisotropic XY:} Even within a single layer, one can consider different couplings along $x$ and $y$ directions (or hexagonal anisotropy). 
While such anisotropy does not change the universality class (the model still has a U(1) symmetry globally), it can modify spin-wave spectra and vortex core energies, slightly shifting $T_c$ and crossovers. 
It also serves as a step towards coupled systems (since anisotropy can be thought of as infinitely large interlayer separation in one direction). 

More exotic extensions, like including long-range interactions, external fields, or higher-harmonic terms, have also been studied. 
The unifying theme is that each generalisation breaks or modifies the symmetry and thus can produce richer phase diagrams: e.g., additional transitions (two-step or multi-critical points), phases with partial order (nematic vs ferromagnetic), or even weakly first-order behaviour. 

\begin{figure}[h]
\centering
\includegraphics[width=8.5cm]{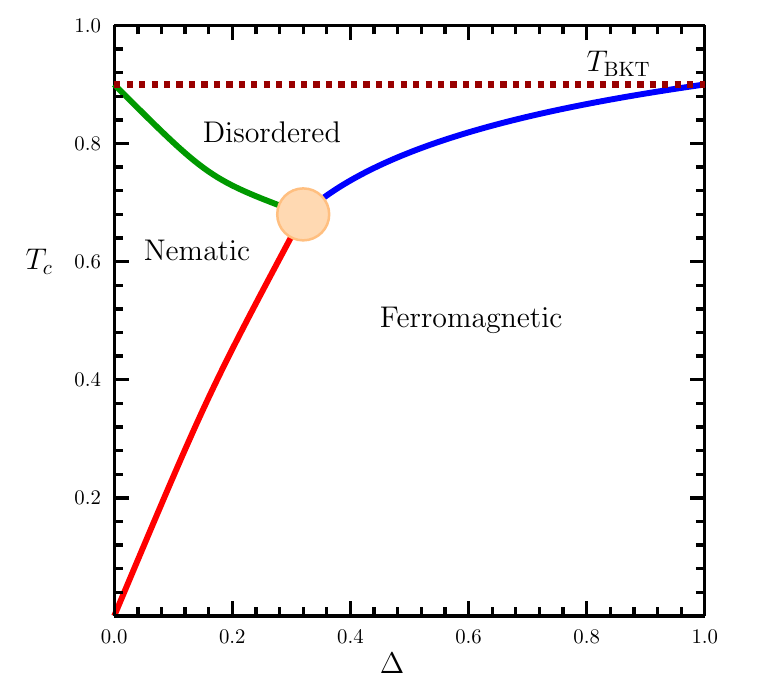}
\caption{The conjectured phase diagram of the gXY model. The BKT line (blue) separates the ferromagnetic and disordered phases; the half-BKT line (green) separates the nematic and disordered phases; whereas the Ising line (red) separates the nematic and ferromagnetic phases. The blob represents the region where the transition lines meet.}
\label{fig:cartoon} 
\end{figure}

The modification of the XY model, as proposed by KLG, is given by the Hamiltonian:
\begin{equation}
\label{eq:gXY_ham} 
\mathcal{H} = - J \sum_{\langle jk \rangle} \cos(\theta_j - \theta_k) - J_1  \sum_{\langle jk \rangle} \cos(q (\theta_j - \theta_k)), 
\end{equation}
where we use the standard notation $\langle jk \rangle$ to denote the nearest-neighbor and $\theta_j \in [0, 2 \pi)$ with $J, J_1 > 0$. 
We can consider these couplings to depend on a parameter $\Delta$ such that $J = \Delta$ and $J_1 = 1 - \Delta$ with $0 \le \Delta \le 1$. 
The limit $\Delta = 0$ corresponds to a pure spin-nematic model, while $\Delta = 1$ is the usual XY model. 
The presence of the nematic term $\cos(q \theta_j - q \theta_k)$ gives rise to fractional excitations such as half-integer vortices (for $q = 2$), and these exhibit invariance under $\theta \to \theta + (2\pi /q)$. 
In this work, we will only consider the case $q = 2$. 
The model defined by Eq. \eqref{eq:gXY_ham} is known as the generalized XY (gXY) model, and the phase diagram of this model has been subject to a lot of investigations. 
The basic conclusion is that there are three phases denoted by BKT, half-BKT, and Ising-like transitions \cite{DB_Carpenter_1989, Lee85, Hubscher2012, Canova2016, Serna_2017} and they connect~\cite{Nui2018} around a special point in the $T - \Delta$ phase space. 
We show the expected phase structure in Fig.~\ref{fig:cartoon}. 
In this work, we perform numerical computations to precisely sketch this diagram, using the numerical tensor network method, finding evidence that the Ising transition line might preempt the BKT transition for a finite range of $\Delta$. 

When $\Delta = 0$, there is a topological transition corresponding to the dissociation (unbinding) of half-integer vortex and anti-vortex, while for $\Delta = 1$, there is a dissociation of integer vortex-pairs. 
Both these transitions happen around the same temperature~\cite{DB_Carpenter_1989}. 
Though the qualitative behavior of the phase diagram of the gXY model is known, the location of the multicritical point and how the transition lines connect are mostly determined from Monte Carlo numerical work. 
The goal of this work is to study the phase diagram by exploring several $\Delta$ values for $q = 2$, and to provide an alternate method using real-space tensor renormalization group techniques to approximate the partition function of the gXY Hamiltonian. 
There has also been a lot of work for $q > 2$ finding interesting additional phases. 
We refer the reader to Ref.~\cite{Poderoso2011} to start the reference trail. 

The study of phase transitions and symmetry breaking in statistical models with short-range interactions in two dimensions is special because of the famous HMWC theorem, which states that continuous symmetries cannot break spontaneously. 
An equivalent statement is that there are no Goldstone bosons that accompany the symmetry breaking in two Euclidean dimensions. 
This makes the XY model interesting in itself, and adding the additional term makes it even more interesting. 
The richness of this model can be understood as follows: there are three primary phase regions - a disordered region, a region with integer vortices (IV), and a region with half-integer vortices (HIV). 
The integer-vortex pair phase is also referred to as the `ferromagnetic,' whereas the `nematic' represents the half-integer vortex pair phase. 
The transition line, corresponding to a continuous phase transition, also referred to as the `Ising-line,' can be captured by the logarithmic divergence of the specific heat near the critical temperature. In contrast, this method is not very useful for BKT-like transitions. 
To locate the transition between the integer-vortex pair phase and the disordered phase (where the vortices dissociate), we compute the magnetization for a small symmetry-breaking external field and then take the vanishing field limit to extract the critical temperature. 
We cannot compute the spontaneous magnetization in the thermodynamic limit without an external field, since, by the HMWC theorem, it is zero. 
This method of determining the infinite-order transition was pursued in Ref.~\cite{Jha:2020oik}. 
For the IV to HIV, the transition can simply be tracked by locating the divergence in specific heat since it is of Ising-type. 

To carry out a precise study of the phase transitions in this model, we apply the tensor network methods based on the higher-order tensor renormalization group (HOTRG) as introduced in \cite{Xie2012}. 
Some preliminary investigation of this model was carried out in \cite{Longia:2023dus}. 
This model has been studied using the variational uniform matrix product state (VUMPS) algorithm \cite{Zauner-Stauber:2017eqw} in Ref.~\cite{Song_2021}. 
Here, we use an alternative approach using the numerical real-space tensor renormalization group methods. 
Since the proposal by Levin and Nave \cite{Levin:2006jai}, the scheme of performing real-space coarse-graining using tensors has seen a lot of progress with extension to higher dimensions and innovative procedures to carry out the tensor renormalization ~\cite{Kadoh:2019kqk, Adachi2020, Az-zahra:2024gqr}. 
We refer the interested reader to Ref.~\cite{Ran2020} for a summary of tensor networks based on both real-space (renormalization group) methods and approximation of the ground state of slightly entangled many-body systems. 
The tensor renormalization group (TRG) approach has been successfully applied to the two-dimensional XY model and related models \cite{Yu:2013sbi, Akiyama:2019xzy, Jha:2020oik, Hostetler:2021uml, Jha:2022pgy, Butt:2022qqx} and also to the three-dimensional XY model with finite chemical potential \cite{Bloch:2021mjw}. 
Using the dual variable approach (from the character expansion), several gauge theories have also been studied \cite{Bazavov:2019qih, Akiyama:2022eip, Kuwahara:2022ubg} with tensors and have found agreement with results obtained using other numerical methods. 
In addition, recently, some progress was also made to extract the critical behavior by studying the renormalization group collapse of magnetization for the three-dimensional O(4) model \cite{Akiyama:2024qgv}. 
It seems likely that one can compute critical exponents accurately using TRG methods in the coming years, providing an alternative to the existing methods, such as Monte Carlo and conformal bootstrap. 
The major part of these computations is the contraction of the network during coarse-graining and the singular value decomposition (SVD) of the growing size of tensors to restrict to a fixed size after each step. 
This amounts to truncating the singular values to some threshold $D$, which is known as the bond dimension. 
(See the previous chapter for the definition and more details.) 
To improve the performance of the numerical computations in this work and get to $D = 91$ in a reasonable amount of computational time, we use the graphical processing unit (GPU) accelerated code described in Ref.~\cite{Jha:2023bpn}. 

\section{Tensor Formulation}
\label{sec:TF}

The starting point of the real-space tensor renormalization approach (on a square lattice) is to decompose the Boltzmann weight in terms of a tensor with $2d$ indices, where $d$ is the number of (Euclidean) dimensions. 
This can either be done exactly or approximately based on the symmetries of the action. 
One of the standard methods to do this decomposition is to use the character expansion \cite{Liu:2013nsa}, but different truncation schemes also exist \cite{Kadoh:2018tis}. 
Using this initial tensor, one can create a network of these tensors such that when the network is contracted, it provides a good approximation to the partition function $Z$. 

The partition function of the classical spin model with external field $h$ is:  
\begin{equation}
\begin{split}
Z = \prod_j \int \frac{d \theta_j}{2 \pi} \prod_{\langle jk \rangle} e^{\beta [\Delta \cos(\theta_j - \theta_k) + (1 - \Delta) \cos(2 (\theta_j - \theta_k))]} \times \prod_j e^{\beta [h \cos(\theta_j)]},
\end{split}
\end{equation}
where $\langle jk \rangle$ denotes neighboring lattice sites and $\beta$ is the inverse temperature. 
We expand the argument of the exponential (Euclidean action) using the Jacobi-Anger expansion and obtain: 
\begin{equation}
\begin{split} 
Z = \prod_j \int \frac{d \theta_j}{2 \pi} \prod_{l \in L} \sum_{n_l} a_{n_l}(\beta, \Delta)e^{i n_l(\theta_{j} - \theta_{k})} \times \sum_{p_l} I_{p_{l}}(\beta h) e^{i p_l \theta_{j}}, 
\end{split}
\end{equation}
where
\begin{equation}
\label{eq:a_n} 
a_n (\beta, \Delta) = \sum_{m = - \infty}^\infty I_{n - 2m}(\beta \Delta) I_m(\beta(1 - \Delta)),
\end{equation}
and $I_n$ is the modified Bessel function of the first kind. 

On integrating over the $\theta_j$ variables, we obtain the partition function:  
\begin{equation}
\label{eq:iniT} 
Z \approx {\rm tTr} \left( \prod_s T_{n_1, n_2, n_3, n_4}(s) \right),
\end{equation}
where
\begin{equation}
\label{eq:Tijkl} 
\begin{split}
T_{n_1, n_2, n_3, n_4}(s) = \sqrt{\prod_{k = 1}^4a_{n_k}(\beta, \Delta)} \times I_{n_1 + n_2 - n_3 - n_4}(\beta h).
\end{split} 
\end{equation}
This tensor represents the four-legged object at site $s$ where $n_1$, $n_2$, $n_3$, and $n_4$ denote the top, right, bottom, and left legs, respectively. 
In the case of zero field, i.e., $h = 0$, we have the conservation of the U(1) charges due to the $\delta$-function in Eq. \eqref{eq:Tijkl}. 
To write the initial tensor for numerical computations, we have to choose a suitable range of values for $m$ so that the infinite sum in Eq. \eqref{eq:a_n} can be truncated to a finite interval. 
We have fixed this to be an integer-valued range, $m \in [-50, 50]$. 
The second truncation we have to do is over the indices in Eq. \eqref{eq:Tijkl}, which also runs from $-\infty$ to $\infty$. 
For this truncation, we choose $n_k \in [-45, 45], \; n_k \in \mathbb{Z}$, i.e., $D = 91$. 
Using these truncation procedures, we use the initial tensor to perform a fixed number of coarse-graining steps, $N = 30$ for most cases, using the higher-order singular value decomposition of the tensors (HOTRG algorithm), as described in Refs. ~\cite{Yu:2013sbi, Jha:2020oik}, to obtain the partition function in the thermodynamic limit.

\section{Derivation of Tensors for the Generalized XY Model}
\label{sec:derivation}

For the tensor formulation of the generalized XY model, and the impure tensor for magnetization, we start with the following Hamiltonian
\begin{equation}
\begin{split}
H = & - \Delta \sum_{\langle j k \rangle} \cos{(\theta_j - \theta_k)} - (1 - \Delta) \sum_{\langle j k \rangle} \cos{(2(\theta_j - \theta_k))} 
- h \sum_j \cos{(\theta_j)} - h_1\sum_j \cos{(2\theta_j)},
\label{eq:SectionH} 
\end{split}
\end{equation}
which represents the generalized XY model with the symmetry-breaking fields corresponding to the integer and half-integer terms. 

The partition function then reads 
\begin{equation}
\begin{split} 
Z = & \int \Bigg[\prod_j \frac{d \theta_j}{2 \pi}\Bigg] \Bigg(e^{\beta \Delta \sum_{\langle j k \rangle} \cos{(\theta_j - \theta_k)} + \beta h \sum_j \cos{(\theta_j)}} \times 
e^{\beta(1 - \Delta) \sum_{\langle j k \rangle} \cos{(2(\theta_j - \theta_k))} + \beta h_1 \sum_j \cos{(2 \theta_j)}}\Bigg).
\end{split} 
\end{equation}
Using the expansion in terms of the dual variables, we get 
\begin{equation}
e^{\beta \Delta \cos{(\theta_j - \theta_k)}} = \sum_{a = -\infty}^\infty I_a(\beta \Delta)e^{i a (\theta_j - \theta_k)}.
\end{equation} 
Here, $I_a$ is the modified Bessel function of the first kind with integral order $a$. 
If we write a summation for each term in the Hamiltonian and then collect all the terms corresponding to $\theta_j$, we can write the partition function as a tensor network contraction of four-ranked tensors living on each lattice site. 
Such a tensor at lattice site $s$ is given by
\begin{equation}
\begin{split}
T^s_{abcd} = & \sqrt{ \sum_{m, n, o, p} I_{a}(\beta \Delta) I_{b}(\beta \Delta) I_{c}(\beta \Delta)I_{d}(\beta \Delta)} \; \\ 
& \times \sqrt{I_m(\beta(1 - \Delta)) I_n(\beta(1 - \Delta)) I_o(\beta(1 - \Delta)) I_p(\beta(1 - \Delta))} \; \\ 
& \times \sum_{r} I_r(\beta h_1) \sum_{l} I_l(\beta h) \int \frac{d \theta_s}{2 \pi} \; \times e^{i \theta_s((a - b + c - d) + 2(m - n + o - p) + l + 2r)}. \\
\end{split} 
\end{equation} 

To evaluate this integral, we can relabel the indices as follows:
\begin{equation} 
\begin{split} 
a ~\to~ & a - 2m, \\ 
b ~\to~ & b - 2n, \\ 
c ~\to~ & c - 2o, \\ 
d ~\to~ & d - 2p. \\
\end{split} 
\end{equation}
Such a relabeling does not affect the integral since the indices of the Bessel functions run from $-\infty$ to $+\infty$. 
The tensor $T^s_{abcd}$ becomes: 
\begin{equation}
\begin{split}
T^s_{abcd} = 
& \sqrt{ \sum_{m, n, o, p} I_{a - 2 m}(\beta \Delta)I_{b - 2 n}(\beta \Delta)I_{c - 2 o}(\beta \Delta)I_{d - 2 p}(\beta \Delta)} \; \\ 
& \times \sqrt{I_m(\beta(1 - \Delta)) I_n(\beta(1 - \Delta)) I_o(\beta(1 - \Delta))} \; \\ 
& \times \sqrt{I_p(\beta(1 - \Delta))} \; \sum_{l} I_l(\beta h) \sum_{r} I_r(\beta h_1) \int \frac{d \theta_s}{2 \pi} e^{i \theta_s((a - b + c - d) + l + 2r)}.
\end{split}
\end{equation}

\begin{figure}[H]
\centering

\begin{subfigure}{0.35\textwidth}
    \centering
    \includegraphics[width=\textwidth]{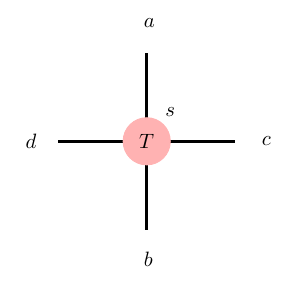}
    \caption*{(a) The rank four site tensor. We set $a(n_1)$ and $c(n_2)$ indices positive and $b(n_3)$ and $d(n_4)$ indices negative as mentioned in Eq. \eqref{eq:App_T_tensor}.}
    \label{fig:T_abcd}
\end{subfigure}
\hspace{0.05\textwidth}
\begin{subfigure}{0.35\textwidth}
    \centering
    \includegraphics[width=\textwidth]{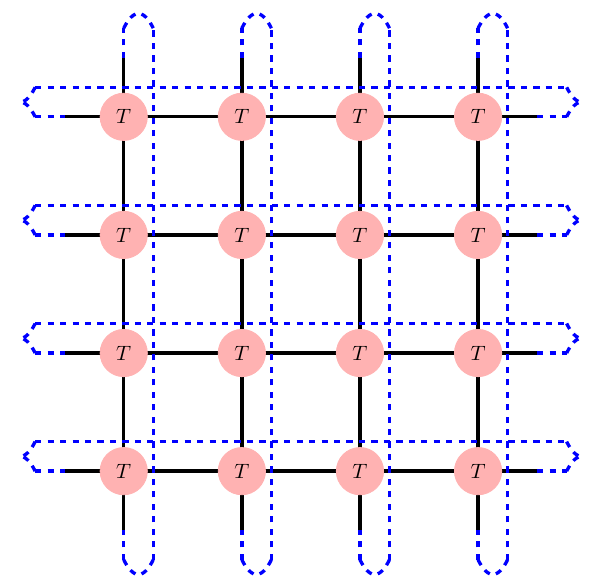}
    \caption*{(b) A fully contracted tensor network $T$ with periodic boundary condition that gives the partition function $Z$ on a $4 \times 4$ lattice. Note that all the $T$ are the same due to translational symmetry.}
    \label{fig:tensor_network}
\end{subfigure}

\caption{Schematic representation of the tensor-network formulation: 
(a) the rank-four site tensor and its index convention, and 
(b) a fully contracted tensor network on a $4\times4$ lattice with periodic boundary conditions.}
\label{fig:tensor_network_construction}

\end{figure}

The integral is the familiar Fourier transform of the delta function; hence, contracting all the other terms with this delta function, we get the final form of the four-ranked site tensor as 
\begin{equation} 
\begin{split} 
T^s_{abcd} = 
& \sqrt{ \sum_{m, n, o, p} I_{a - 2 m}(\beta \Delta) I_{b - 2 n}(\beta \Delta) I_{c - 2 o}(\beta \Delta) I_{d - 2 p}(\beta \Delta)} \; \\ 
& \times \sqrt{I_m(\beta(1 - \Delta)) I_n(\beta(1 - \Delta)) I_o(\beta(1 - \Delta))} \; \\ 
& \times \sqrt{I_p(\beta(1 - \Delta))} \; \sum_{r} I_{a - b + c - d + 2r}(\beta h) I_r(\beta h_1). \\ 
\end{split} 
\end{equation} 
To write the site tensor in a more compact notation, we define 
\begin{equation} 
a_{n_k}(\beta, \Delta) = \sum_{\nu_k = -\infty}^\infty I_{n_k - 2 \nu_k}(\beta \Delta) I_{\nu_k}(\beta(1 - \Delta)), 
\end{equation} 
using which the tensor becomes: 
\begin{equation}
\begin{split}
T_{n_1, n_2, n_3, n_4} = 
& \sqrt{\prod_{k = 1}^4 a_{n_k}(\beta, \Delta)} \; \times \sum_{l = -\infty}^\infty I_{n_1 + n_2 - n_3 - n_4 + 2l}(\beta h) I_l(\beta h_1). 
\label{eq:App_T_tensor} 
\end{split} 
\end{equation} 
The partition function can be approximated using this tensor description as a trace of the network: 
\begin{equation}
Z \approx {\rm tTr} \Big( \prod_s T_{n_1, n_2, n_3, n_4}(s) \Big).
\label{eq:Z} 
\end{equation} 
Here, ${\rm tTr}$ implies a tensor trace of the tensor network. 
We show the site tensor and the partition function as a fully contracted tensor network consisting of a $4 \times 4$ lattice in Fig. \ref{fig:tensor_network_construction}.


\section{Numerical Results} 
\label{sec:gxy_results} 

We now discuss the results for the generalized XY model using tensor network methods. 
The main quantity of interest in the real-space tensor computation is the partition function, which has to be approximated accurately. 
Using this, we compute two main observables - the specific heat, $C_{\rm{v}}$ and the magnetic susceptibility normalized by the lattice volume $V$, $\chi$. 
They are defined as 
\begin{equation} 
\label{eq:spech}
C_{\rm{v}} = \frac{\beta^2}{V} \frac{\partial^2{\ln Z}}{\partial \beta^2}, 
\end{equation} 
and 
\begin{equation} 
\label{eq:susc} 
\chi = \frac{1}{V} \frac{\partial M}{\partial h} = \frac{1}{\beta V} \frac{\partial^2 \ln Z}{\partial h^2}, 
\end{equation} 
where $M$ is the magnetization, which is defined using the free energy
\begin{equation}
F = -T \ln(Z) = - \beta^{-1} \ln(Z) \nonumber
\end{equation}
as 
\begin{equation} 
\label{eq:mag} 
M = - \frac{\partial F}{\partial h} = \frac{1}{\beta} \frac{\partial {\rm ln} Z}{\partial h}. 
\end{equation} 
The computation of magnetization from the first derivative of free energy (or partition function) is often prone to numerical errors. 
A better approach is to compute the impure tensor corresponding to the magnetization and insert it into the network to compute the partition function~\cite{Jha:2020oik}.  
The magnetization is calculated using Eq. \eqref{eq:mag} and is $M = P / Z$, where $Z$ is the partition function given in Eq. \eqref{eq:Z} and $P$ is the modified contracted tensor network with an impure tensor inserted. 
The impure tensor corresponding to $M$ is given by 
\begin{equation}
\begin{split} 
\mathcal{I}_{n_1 n_2 n_3 n_4}^{M} = 
& \; \sqrt{\prod_{k = 1}^{4} a_{n_k}(\beta, \Delta)} \; \times \; \sum_{l = -\infty}^{+\infty} I_l(\beta h_1) \; \\ 
& \times \Bigg(\frac{I_{n_1 + n_2 - n_3 - n_4 + 2l - 1}(\beta h) + I_{n_1 + n_2 - n_3 - n_4 + 2l + 1}(\beta h)}{2} \Bigg). \\ 
\end{split} 
\end{equation} 
We can also compute the nematic magnetization $M_1 = P_1 / Z$ with respect to the external field $h_1$, and its impure tensor is given as  
\begin{equation} 
\label{eq:nematic_mag} 
\begin{split} 
\mathcal{I}_{n_1 n_2 n_3 n_4}^{M_{1}} = 
& \;  \sqrt{\prod_{k = 1}^{4} a_{n_k}(\beta, \Delta)} \; \times \sum_{l = -\infty}^{+\infty} I_{n_1 + n_2 - n_3 - n_4 + 2l}(\beta h)  \; \\ 
& \times \Bigg(\frac{I_{l+1}(\beta h_1) + I_{l-1}(\beta h_1)}{2} \Bigg). \\ 
\end{split} 
\end{equation} 
Using $M$ and $M_1$, we can compute the magnetic susceptibility for a range of external fields and locate the BKT and half-BKT transitions. 

To compute the dominant magnetization, depending on $\Delta$, we use either $h$ or $h_{1}$ corresponding to the different symmetry-breaking terms in the partition function. 
We use external magnetic field $h$ for $\Delta > 0.36$ to compute $M$ and magnetic field $h_{1}$ for $\Delta \le 0.36$ related to $\cos{(2 \theta)}$, term which we refer to as $M_{1}$. 

\begin{figure}[h] 
\centering
\includegraphics[width=0.7\linewidth]{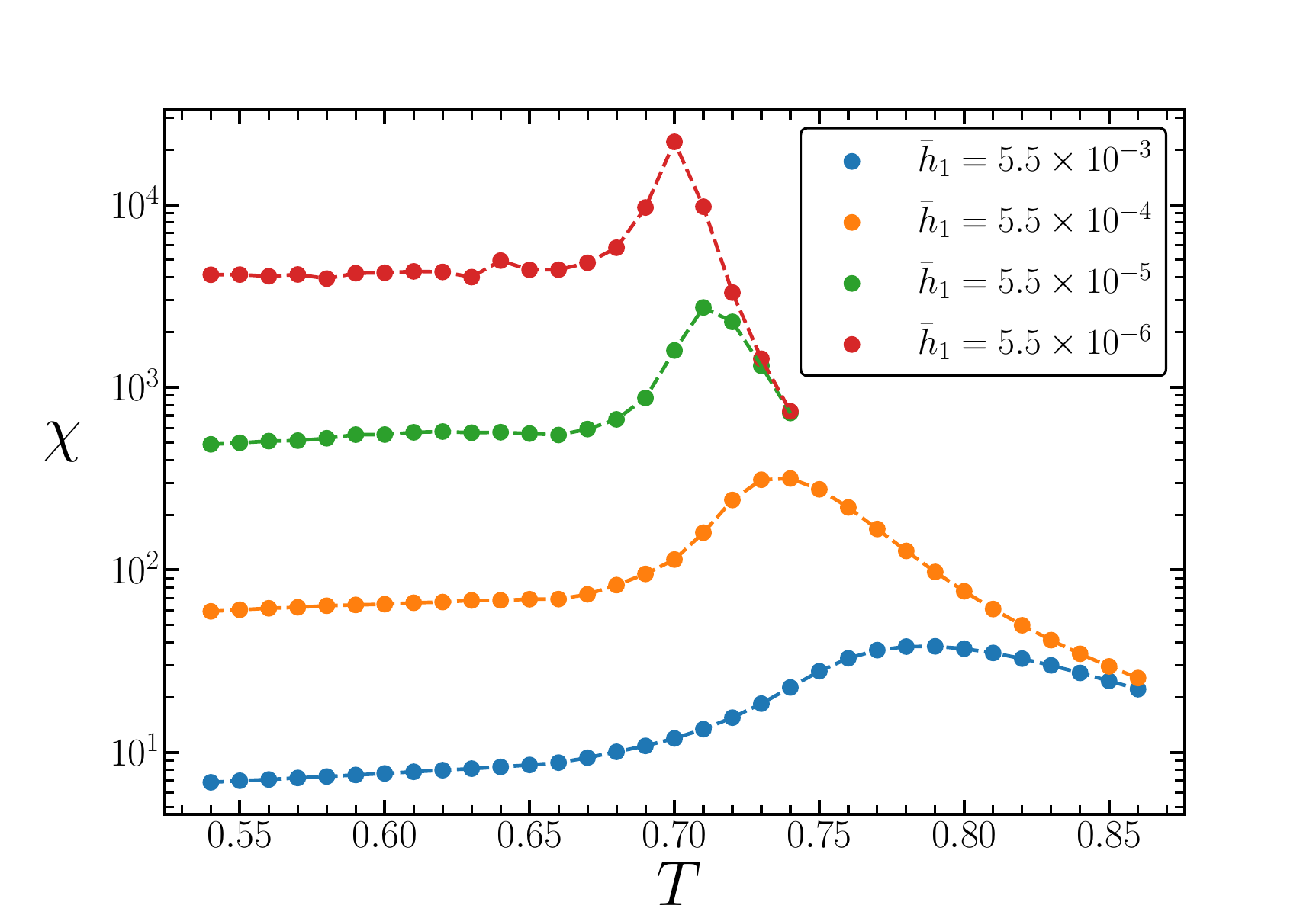} 
\caption{\label{fig:chi0p32}The magnetic susceptibility, defined in Eq. \eqref{eq:susc}, against $T$ for $\Delta = 0.32$.} 
\end{figure}

\begin{figure}[h] 
\centering
\includegraphics[width=0.7\linewidth]{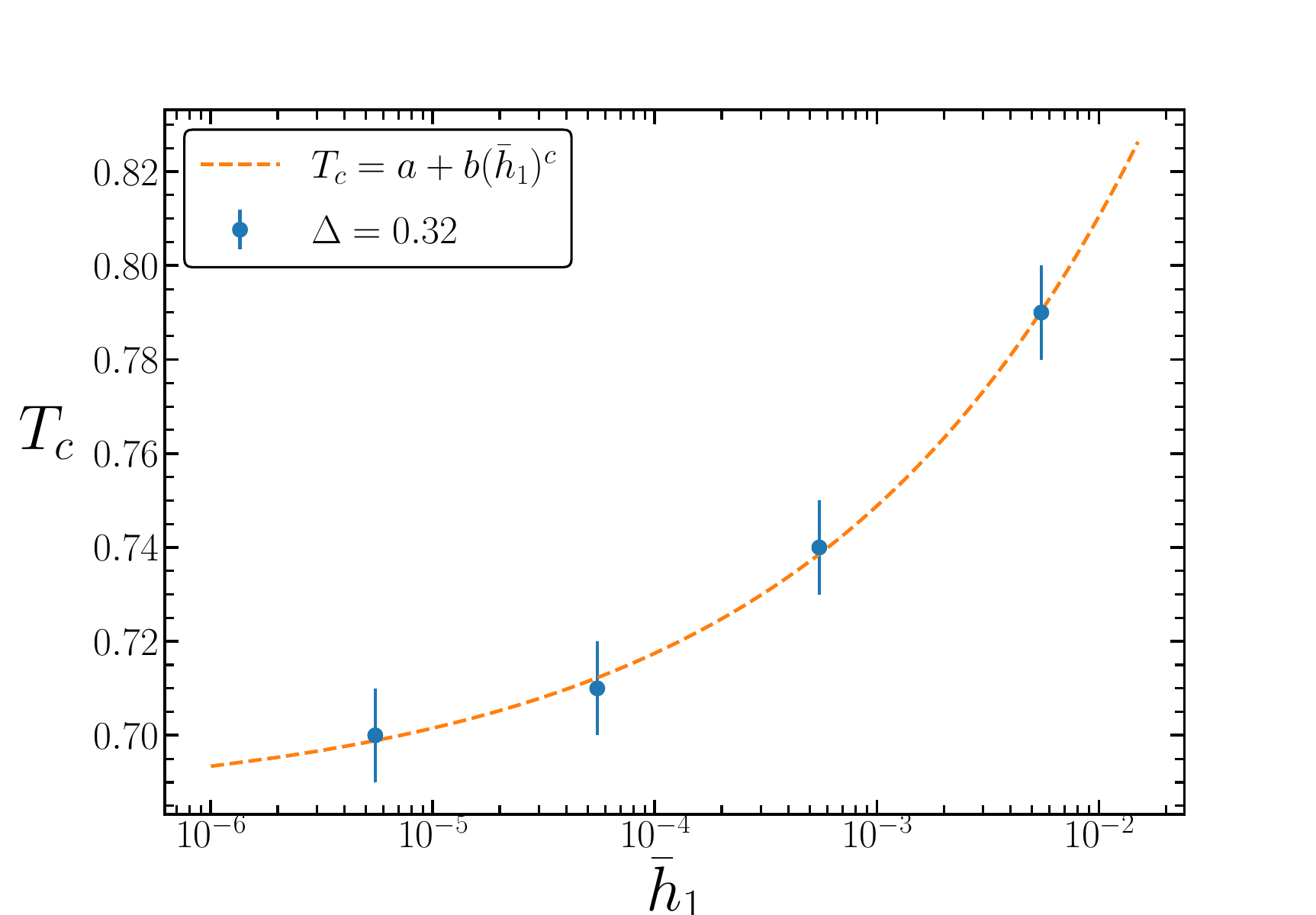} 
\caption{\label{fig:Tc0p32}The critical temperature, $T_c$ for different $\overline{h}_{1}$ and $\Delta = 0.32$. The fit parameters are $a = 0.685(3), b = 0.48(1), c = 0.29(5)$.} 
\end{figure}

Since this model has Ising and BKT-like transitions, the specific heat cannot always conclusively determine the phase transition. 
For such cases, we look at the first derivative of the magnetization computed using a simple finite difference method and compute the susceptibility. 
The peak in susceptibility signals a transition, which we then extrapolate to the zero-field limit. 
The extraction of this zero-field critical temperature is obtained by doing functional fits of the form discussed in Ref.~\cite{Jha:2020oik}. 
We show the magnetic susceptibility plot and zero-field limit extraction of the critical temperature for $\Delta = 0.32$ in Fig.~\ref{fig:chi0p32} and Fig.~\ref{fig:Tc0p32}, respectively, where $\overline{h}_1$ is the central value used for the finite-difference of the susceptibility measurement. 

We refer the reader to Section \ref{sec:plots_diff_delta} for plots corresponding to a wide range of $\Delta$ values and to Sec. \ref{app:AppD} for plots corresponding to systematic error analysis when $m$ or $D$ is varied. 
For the computation of magnetization, we use a lattice volume of $2^{35} \times 2^{35}$ while for $h = h_{1} = 0$, we use a volume of $2^{30} \times 2^{30}$. 
For all computations, we use $D = 91$ and a range of $m$ as defined in Eq. \eqref{eq:a_n} to be $[-50, 50]$. 

As discussed before, the gXY model reduces to the XY model when $\Delta = 1$. Much work has been done using Monte Carlo and tensor methods to compute the critical temperature and critical exponents. 
The critical temperature on a square lattice was computed to be $T_c = 0.89290(5)$ using tensor network methods in \cite{Jha:2020oik}, while the critical exponent $\delta \approx 15$ was computed within errors in Ref.~\cite{Jha:2023bpn}. 

In this work, we start with $\Delta = 0.8$, which is expected to have a single BKT-like transition, and move to smaller $\Delta$ values passing through the region where the transition lines meet and going all the way down to the limit of $\Delta \to 0$. 
For the BKT transition, it is well-known that the peak of specific heat is not the correct way to determine the transition temperature (it is an overestimate, as clear from the data in Table~\ref{table1}). 
From our computation, we find that for $\Delta = 0.8$, the peak of $C_{\rm{v}}$ is observed at $T = 0.95(1)$. 
This was reported to be around $T \approx 0.91$ in Ref.~\cite{Song_2021}. 
To accurately determine the transition, we compute magnetic susceptibility for a small external field $h$ and then took the zero-field limit as explained above. 
For $\Delta = 0.8$, we obtain $T \sim 0.890(4)$. 

As we decrease $\Delta$ approaching the multi-critical point (or the region where the phases meet), the difference between the transition temperatures deduced from the peak of specific heat and magnetic susceptibility, respectively, decreases for the BKT transition. 
As we cross the region where transition lines meet and move to smaller values of $\Delta$, we again see that the difference increases for the half-BKT line. 
This is evident from the data given in Table~\ref{table1}, in Sec.~\ref{sec:gxy_table}. 
Such behavior has also been noted previously in Ref.~\cite{Hubscher2012} where for the half-BKT line, the critical temperature values almost agree with each other around $\Delta = 0.35$. 
We find that the specific heat peak fails to capture the correct transition temperature for $\Delta \ge 0.50$. 

\begin{figure} 
\centering 
\includegraphics[scale=0.45]{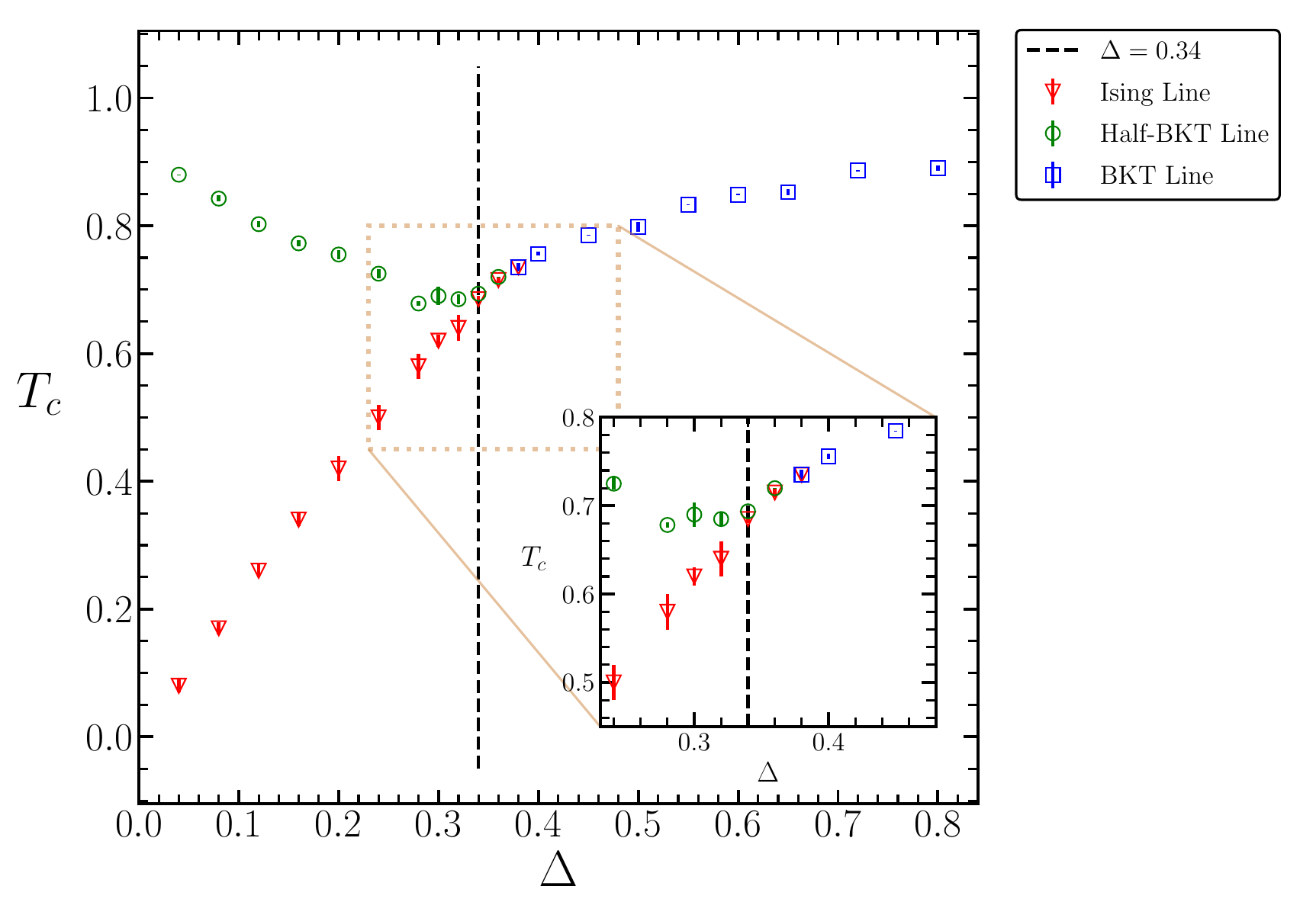} 
\caption{The critical temperatures for a range of deformation parameter $\Delta$ obtained using TRG. The red triangles represent the continuous Ising transition, while the green circles and blue squares represent the transition of the BKT class.} 
\label{fig:phase_diag} 
\end{figure}

From the results of our tensor network computations, we find the phase diagram for this model as shown in Fig.~\ref{fig:phase_diag}. 
The critical temperatures for the half-BKT and BKT lines are extracted from the zero-field limit method, as stated before, and the critical temperatures for the Ising line are inferred using the peak in the specific heat. 
We collect the numerical results in Table~\ref{table1} for critical temperature values deduced from the peak of the specific heat as well as magnetic susceptibility corresponding to the standard and nematic magnetization. 
Figures~\ref{fig:chi_T1} to \ref{fig:delta_T5} show the magnetic susceptibility v/s temperature plots as well as critical temperature scaling with finite external magnetic field for a variety of $\Delta$ values. 
In Sec.~\ref{app:AppD}, Table~\ref{table2}, we show the variation of free energy and magnetization $M_1$ as one varies $m$ parameter at fixed $\Delta = 0.34$ near the critical temperature and Figs.~\ref{fig:scale1} - \ref{fig:scale3} show the free energy scaling with increasing bond-dimension near the transition point for different deformation parameter values. 

\clearpage


\section{Numerical Data Table} 
\label{sec:gxy_table}

The following table shows the data for the critical temperature computed using a real-space tensor network method. 

\begin{table}[H] 
\renewcommand{\arraystretch}{1.25}
\setlength{\tabcolsep}{18pt}
\centering
\begin{tabular}{| c | c | c | c |} 
 \hline
 $\Delta$ & $T_{\rm{BKT}; h, h_1 \to 0}$ & $T_{\rm{BKT}, \rm{C_{v}}}$ & $T_{\rm{Ising},\rm{C_{v}}}$ \\ 
 [0.5ex] 
 \hline
0.04 & 0.880(1) & 1.00(1) & 0.08(1)\\
\hline      
0.08 & 0.843(5) & 0.95(1) & 0.17(1)\\
\hline      
0.12 & 0.803(5) & 0.90(1) & 0.26(1)\\
\hline      
0.16 & 0.773(5) & 0.86(1) & 0.34(1)\\
\hline      
0.20 & 0.755(7) & 0.82(2) & 0.42(2)\\
\hline      
0.24 & 0.725(7) & 0.78(2) & 0.50(2)\\
\hline      
0.28 & 0.678(3) & 0.74(2) & 0.58(2)\\
\hline      
0.30 & 0.690(10) & 0.72(1) & 0.62(1)\\
\hline      
0.32 & 0.685(8) & 0.705(5) & 0.64(2)\\
\hline      
0.34 & 0.694(2) & - & 0.685(5)\\
\hline      
0.36 & 0.720(1) & - & 0.715(5)\\
\hline      
0.38 & 0.735(6) & - & 0.735(5)\\
\hline      
0.40 & 0.756(3) & 0.74(1) & -\\
\hline      
0.45 & 0.785(1) & 0.78(1) & -\\
\hline      
0.50 & 0.798(8) & 0.815(3) & -\\
\hline      
0.55 & 0.833(1) & 0.84(1) & -\\
\hline      
0.60 & 0.849(1) & 0.86(1) & -\\
\hline      
0.65 & 0.853(5) & 0.88(1) & -\\
\hline      
0.72 & 0.886(1) & 0.91(1) & -\\
\hline      
0.80 & 0.890(4) & 0.95(1) & -\\
 [1ex] 
 \hline
\end{tabular}
 
\caption{The summary of the numerical results obtained for the generalized XY model. }
\label{table1}
\end{table}

$T_{\rm{Cv}}$ is the critical temperature determined from the peak of specific heat with no external magnetic field, whereas $T_{h, h_{1} \to 0}$ is the critical temperature determined from the peak of magnetic susceptibility in the limit of vanishing external magnetic field, $h$ and $h_1$ respectively. We use the symmetry breaking field $h$ for $q = 1$ ($\Delta \in [0.38, 0.80]$) and $h_1$ for $q = 2$ ($\Delta \in [0.04, 0.36]$) in Eq.~\eqref{eq:SectionH} to compute the critical temperature. For $T > 0.38$, there is a single transition of the BKT universality class. Until $\Delta = 0.32$, we can resolve the half-BKT and Ising line, but for $\Delta = 0.34, 0.36$ it is likely, based on our numerical results, that the two transition lines, i.e., Ising and half-BKT, have merged. For $\Delta \ge 0.40$, there is no ambiguity, and the transition corresponds to the BKT class. If all the transition lines meet, they do so at $\Delta = 0.36(2)$. Our results are slightly more consistent with the picture that first half-BKT and Ising lines meet around $\Delta \sim 0.34$, and then the Ising line continues to merge with the BKT line around $\Delta = 0.36(2)$. In the interval $\Delta\simeq0.34-0.36$, the numerical resolution of the two transition lines becomes increasingly difficult. Even though the available data are compatible with the half-BKT and Ising transition lines coming together and merging, the current calculations do not entirely rule out the possibility of a very narrow intermediate nematic region lying below the resolution of the present parameter scan. We thus regard the results in this interval as indicating a merging scenario rather than as definitive proof of a direct merger. The picture of sequential merging, in which the half-BKT and Ising lines first approach one another in the region $\Delta\sim0.34$ and then merge with the ordinary BKT line near $\Delta\simeq0.36(2)$, is the one most in agreement with our numerical data.

\newpage

\section{Collection of Plots for a Range of $\Delta$} 
\label{sec:plots_diff_delta} 

In this Section, we collect additional plots corresponding to the data in Table~\ref{table1}. 

\begin{minipage}{0.9\linewidth}
\centering
\begin{minipage}{0.45\linewidth}
\begin{figure}[H]
\includegraphics[width=\linewidth]{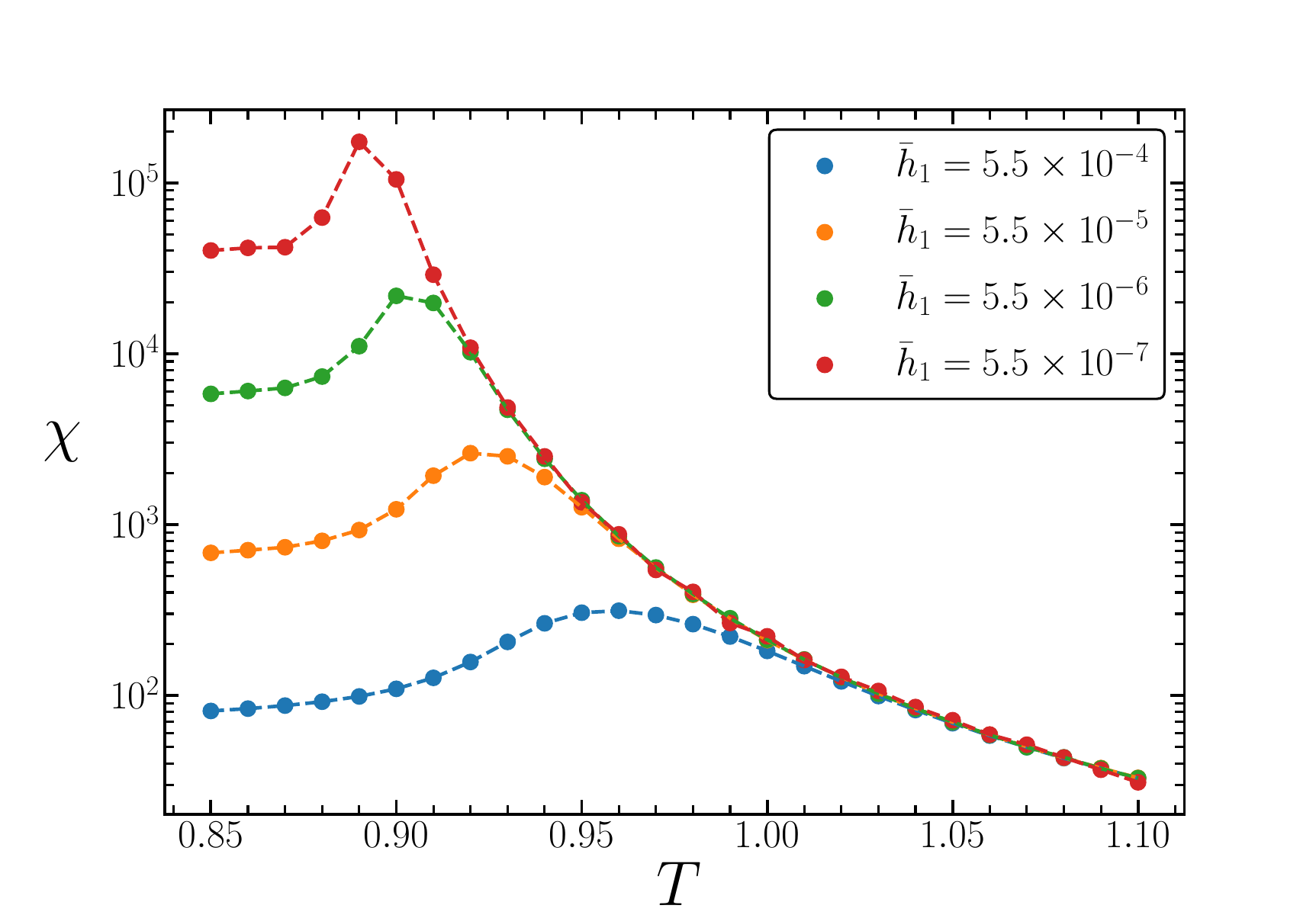}
\caption{The variation of magnetic susceptibility, $\chi$, with temperature $T$ for $\Delta = 0.04$.}
\label{fig:chi_T1}
\end{figure}
\end{minipage}
\hspace{0.05\linewidth}
\begin{minipage}{0.45\linewidth}
\begin{figure}[H]
\includegraphics[width=\linewidth]{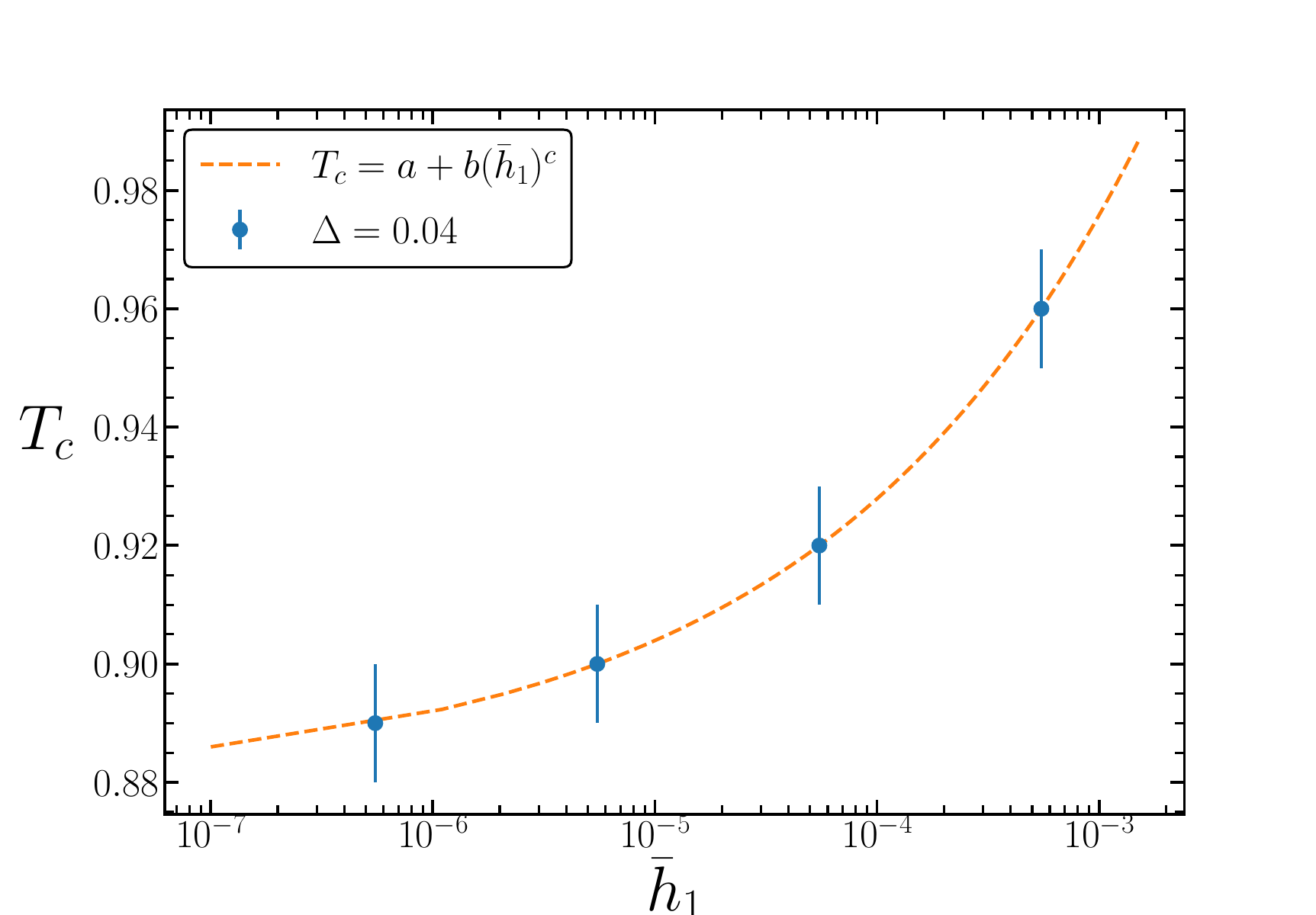}
\caption{$T_c$ v/s $\overline{h}_{1}$ for $\Delta = 0.04$. $a = 0.880(1), b = 0.766(1), c = 0.301(1)$.}
\label{fig:delta_T1}
\end{figure}
\end{minipage}
\end{minipage}

\begin{minipage}{0.9\linewidth}
\centering
\begin{minipage}{0.45\linewidth}
\begin{figure}[H]
\includegraphics[width=\linewidth]{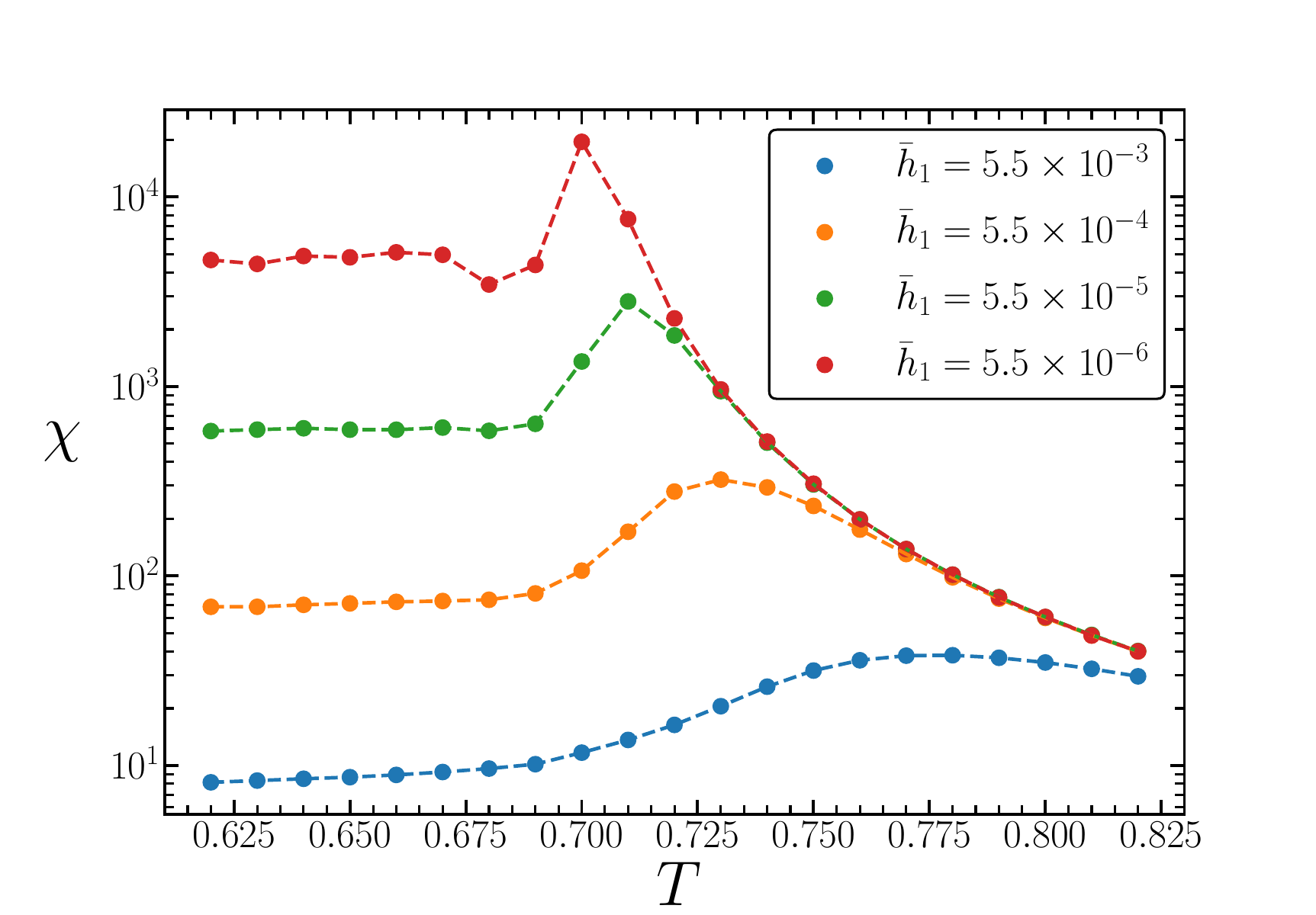}
\caption{The variation of magnetic susceptibility, $\chi$, with temperature $T$ for $\Delta = 0.34$.}
\label{fig:chi_T2}
\end{figure}
\end{minipage}
\hspace{0.05\linewidth}
\begin{minipage}{0.45\linewidth}
\begin{figure}[H]
\includegraphics[width=\linewidth]{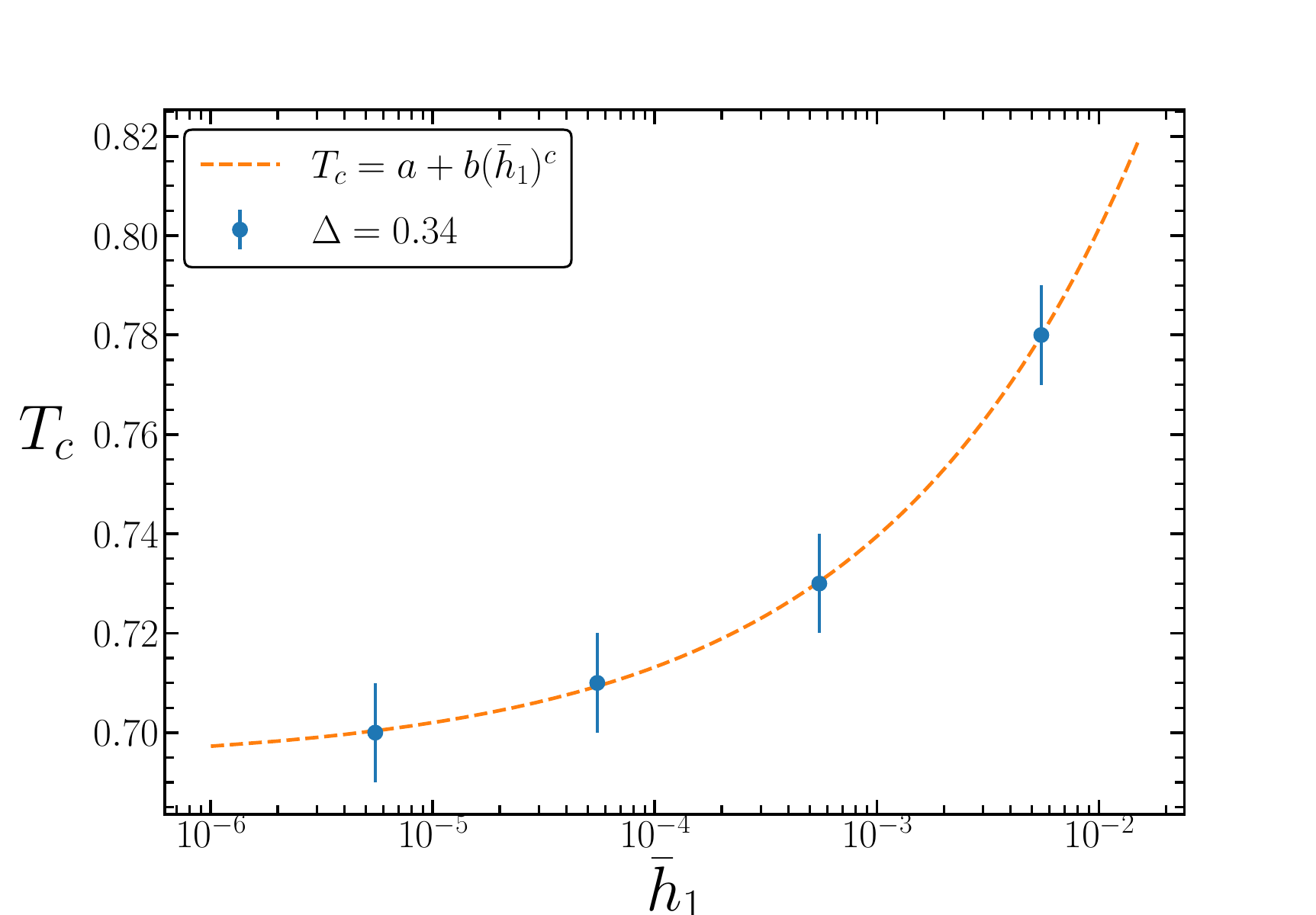}
\caption{$T_c$ v/s $\overline{h}_{1}$ for $\Delta = 0.34$. $a = 0.694(2), b = 0.60(5), c = 0.37(2)$.}
\label{fig:delta_T2}
\end{figure}
\end{minipage}
\end{minipage}

\begin{minipage}{0.9\linewidth}
\centering
\begin{minipage}{0.45\linewidth}
\begin{figure}[H]
\includegraphics[width=\linewidth]{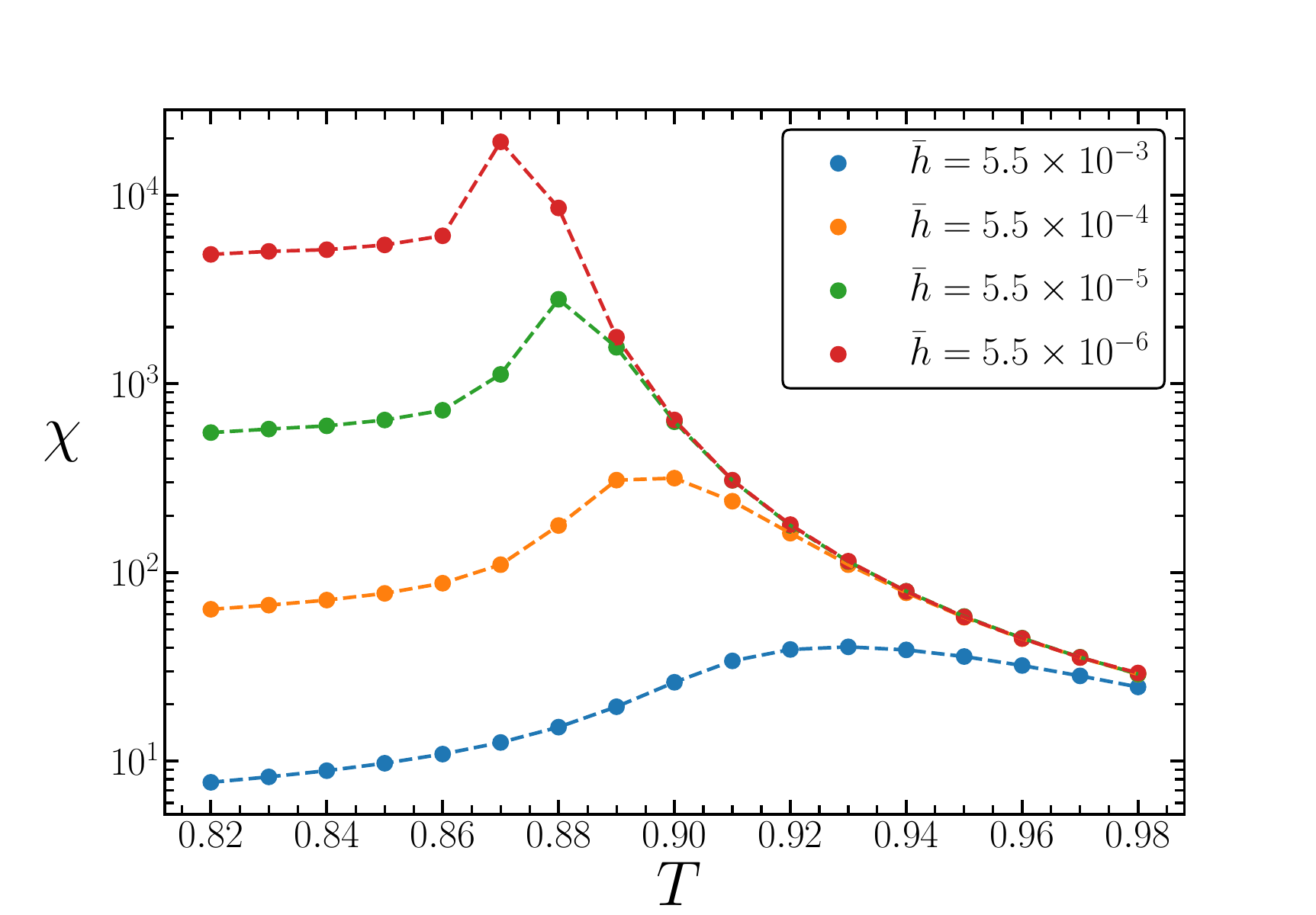}
\caption{The variation of magnetic susceptibility, $\chi$, with temperature $T$ for $\Delta = 0.65$.}
\label{fig:chi_T3}
\end{figure}
\end{minipage}
\hspace{0.05\linewidth}
\begin{minipage}{0.45\linewidth}
\begin{figure}[H]
\includegraphics[width=\linewidth]{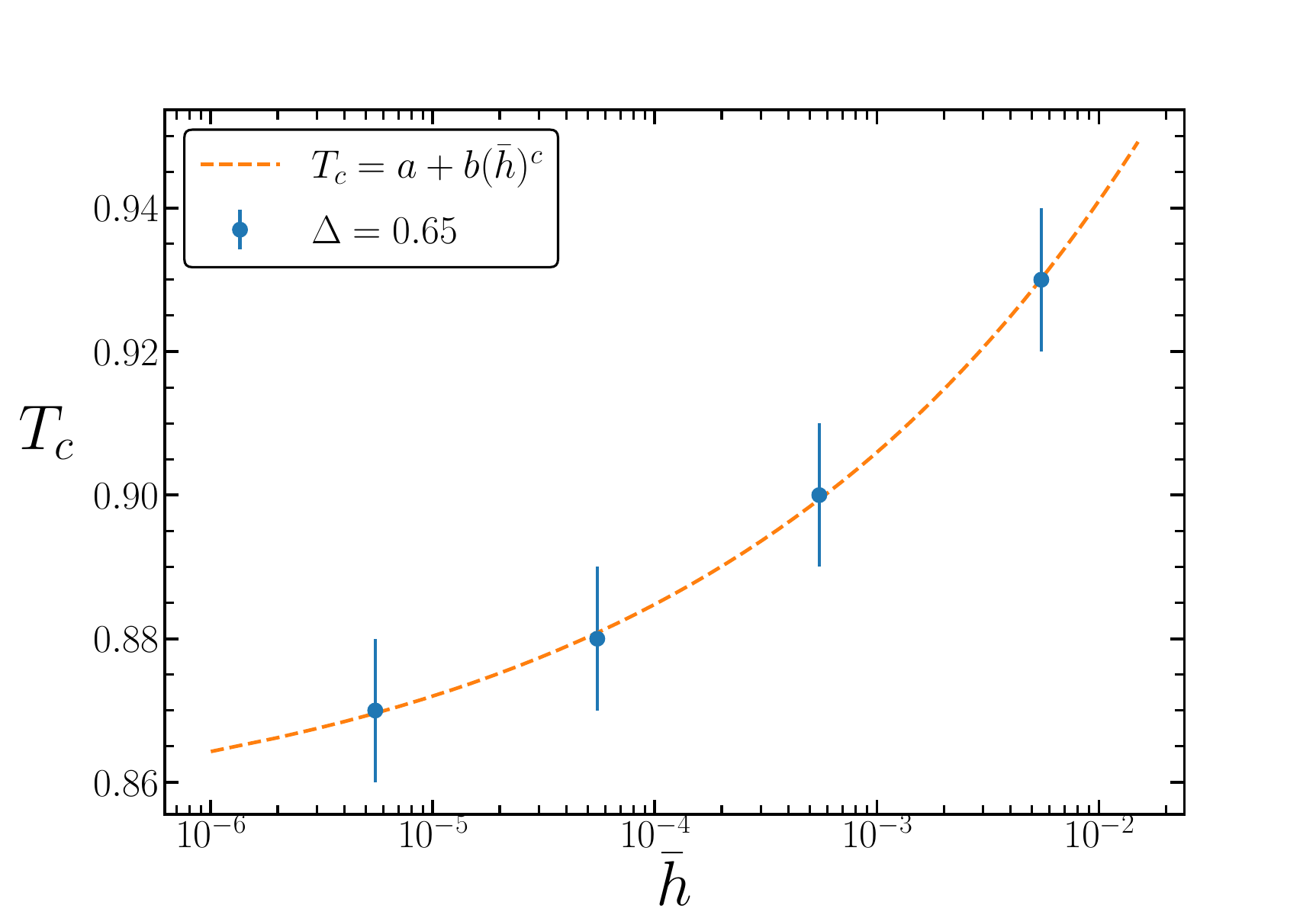}
\caption{$T_c$ v/s $\overline{h}$ for $\Delta = 0.65$. $a = 0.853(5), b = 0.24(2), c = 0.22(3)$.}
\label{fig:delta_T3}
\end{figure}
\end{minipage}
\end{minipage} 

\begin{minipage}{0.9\linewidth}
\centering
\begin{minipage}{0.45\linewidth}
\begin{figure}[H]
\includegraphics[width=\linewidth]{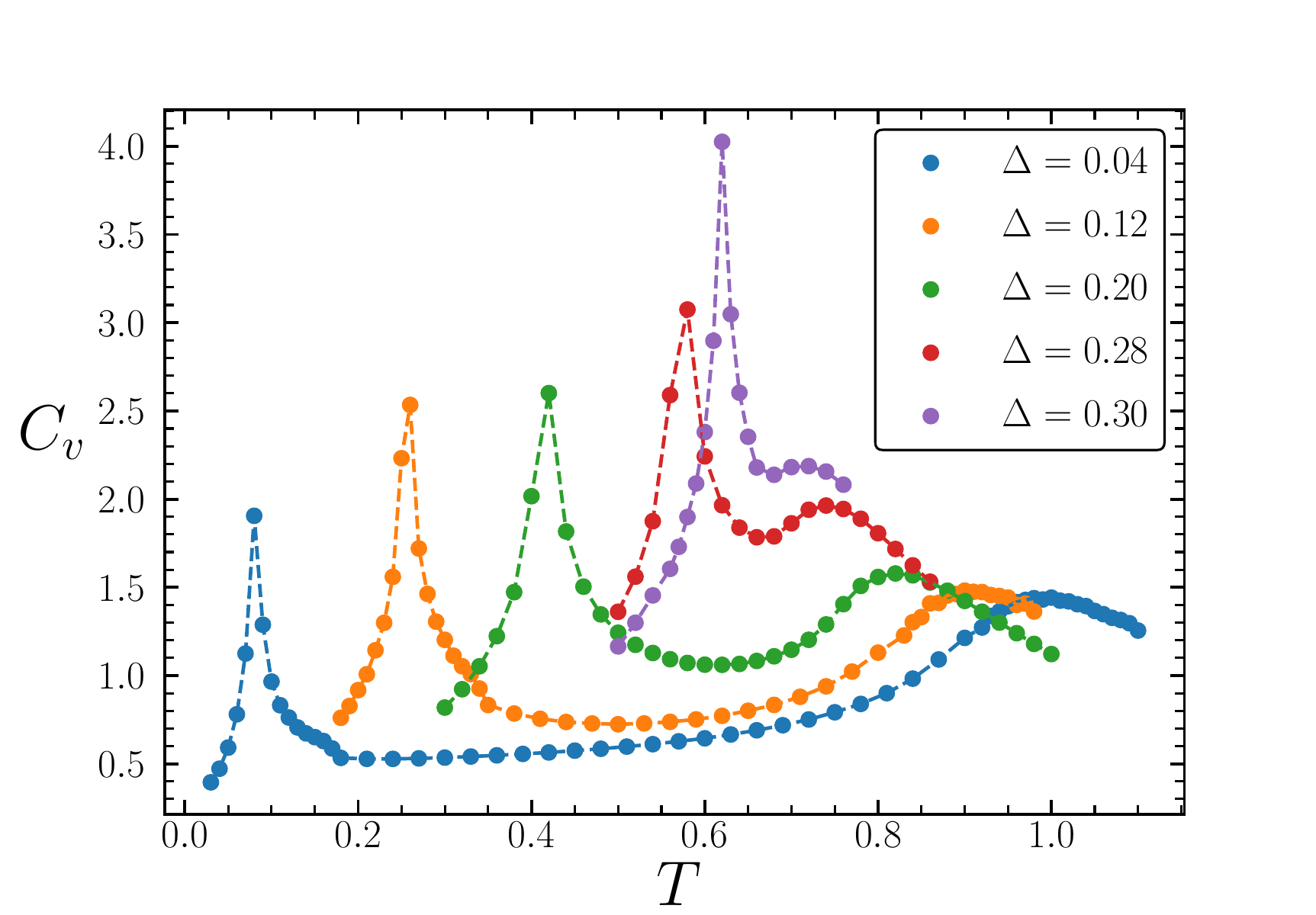}
\caption{The specific heat, $C_{\rm{v}}$ with $T$ for $\Delta < 0.32$, $h, h_1 = 0$.}
\label{fig:chi_T4}
\end{figure}
\end{minipage}
\hspace{0.05\linewidth}
\begin{minipage}{0.45\linewidth}
\begin{figure}[H]
\includegraphics[width=\linewidth]{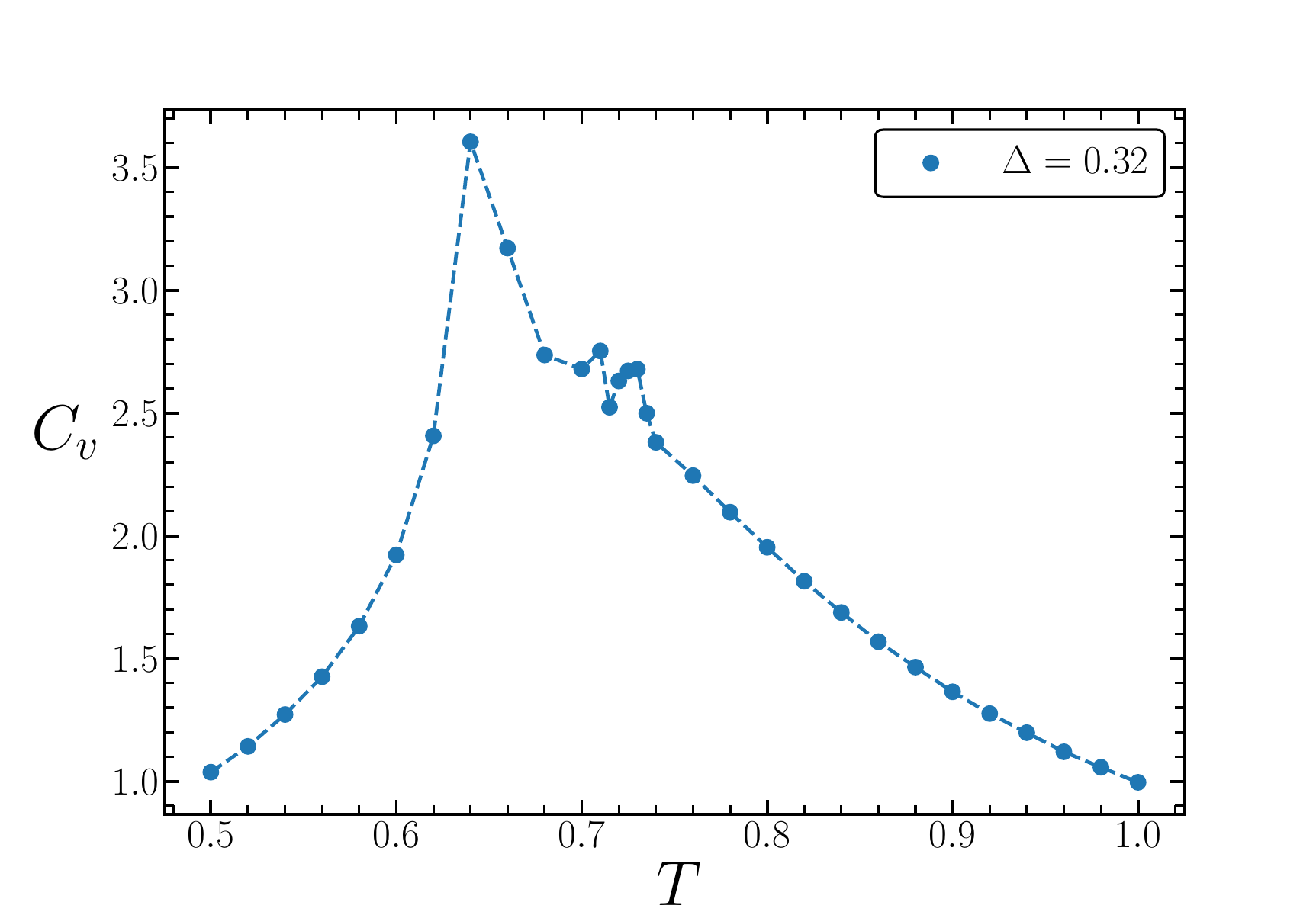}
\caption{The specific heat, $C_{\rm{v}}$ with $T$ for $\Delta = 0.32$, $h, h_1 = 0$.}
\label{fig:delta_T4}
\end{figure}
\end{minipage}
\end{minipage} 

\begin{minipage}{0.9\linewidth} 
\centering 
\begin{minipage}{0.45\linewidth} 
\begin{figure}[H] 
\includegraphics[width=\linewidth]{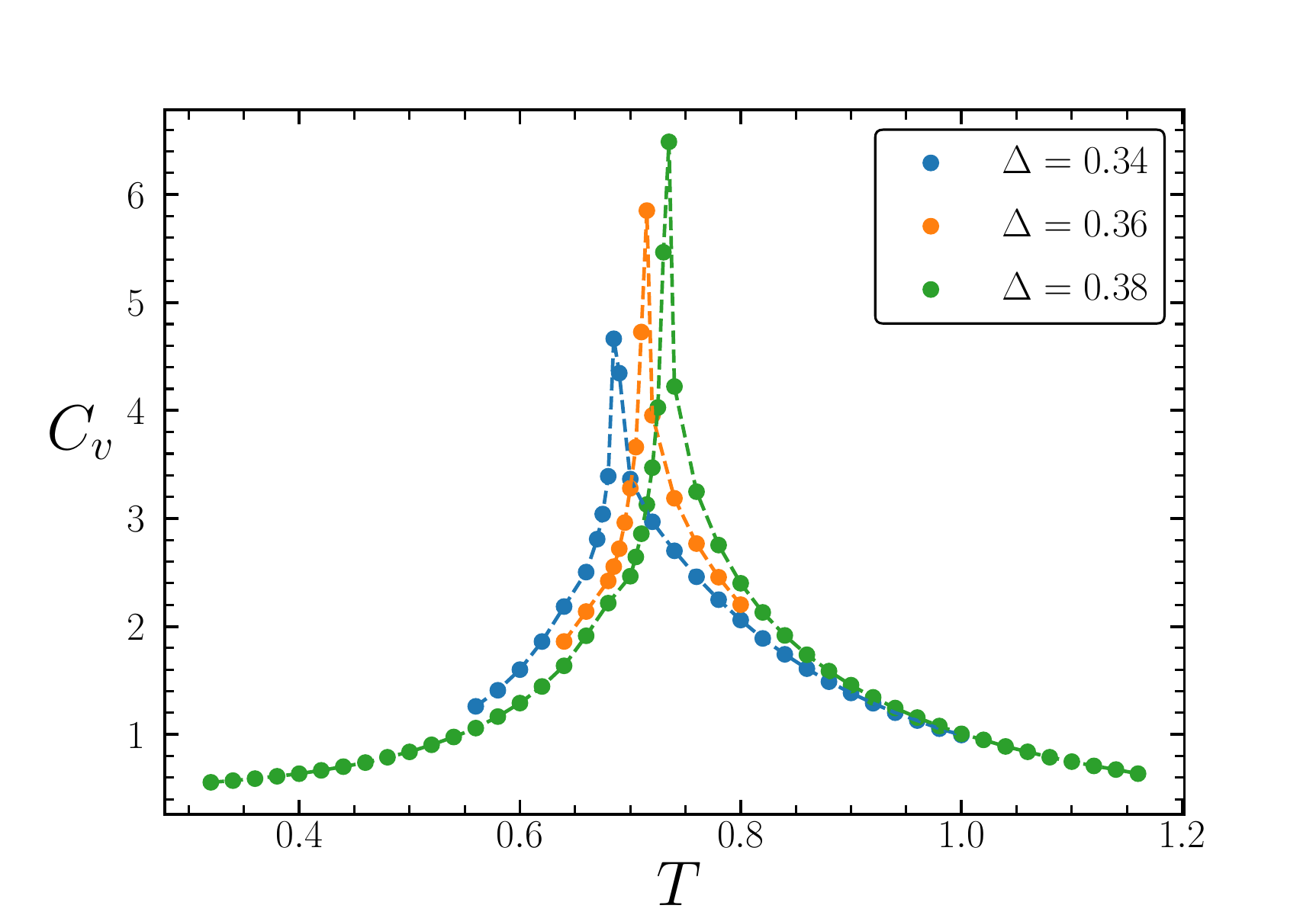} 
\caption{The specific heat, $C_{\rm{v}}$ with $T$ for $\Delta \in [0.34,0.4)$, $h, h_1 = 0$.} 
\label{fig:chi_T5} 
\end{figure} 
\end{minipage} 
\hspace{0.05\linewidth} 
\begin{minipage}{0.45\linewidth} 
\begin{figure}[H] 
\includegraphics[width=\linewidth]{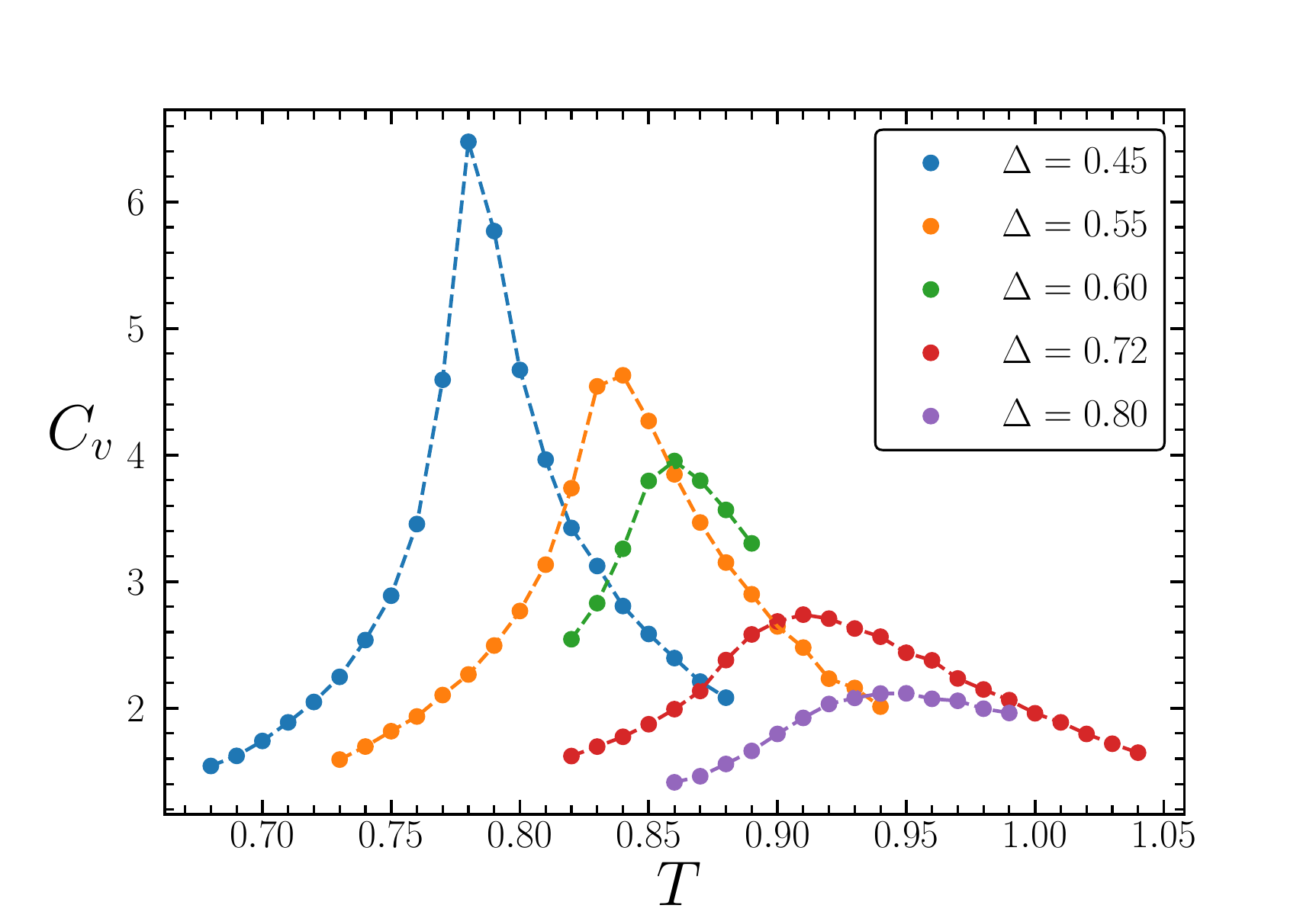} 
\caption{The specific heat, $C_{\rm{v}}$ with $T$ for $\Delta > 0.40$, $h, h_1 = 0$.} 
\label{fig:delta_T5} 
\end{figure} 
\end{minipage} 
\end{minipage} 

\section{Systematic Error Analysis for $m$ and $D$}
\label{app:AppD}

The following table shows that free energy $F$ and magnetization $M_1$ do not show much variation as the range of allowed $m$ values increase at fixed volume, $\Delta$, $T$ and $D$.

\begin{table}[H] 
\renewcommand{\arraystretch}{1.25}
\setlength{\tabcolsep}{16pt}
\centering
\begin{tabular}{|c|c|c|} 
 \hline
 $m$ & $F$ & $M_1$ \\ 
 [0.5ex] 
 \hline
    [-20, 20] & -0.5436041  & 0.486666  \\ 
    \hline
    [-30, 30] & -0.5436043  & 0.486616  \\ 
    \hline
    [-40, 40] & -0.5436043  & 0.486624  \\ 
    \hline
    [-50, 50] & -0.5436043  & 0.486863  \\ 
    \hline
    [-60, 60] & -0.5436046  & 0.486586  \\ 
    \hline
    [-70, 70] & -0.5436066  & 0.486928  \\ 
    \hline
    [-80, 80] & -0.5436041  & 0.486907  \\ 
    [1ex]
    \hline
\end{tabular}
\caption{The free energy and magnetization computed using $h_1$ field for $\Delta = 0.34$, $T \approx T_c = 0.694$, and lattice volume of $2^{30} \times 2^{30}$ for different range of values for $m$ with $D = 91$.}
\label{table2}
\end{table}

\begin{figure}[h] 
\centering 
\includegraphics[width=10cm]{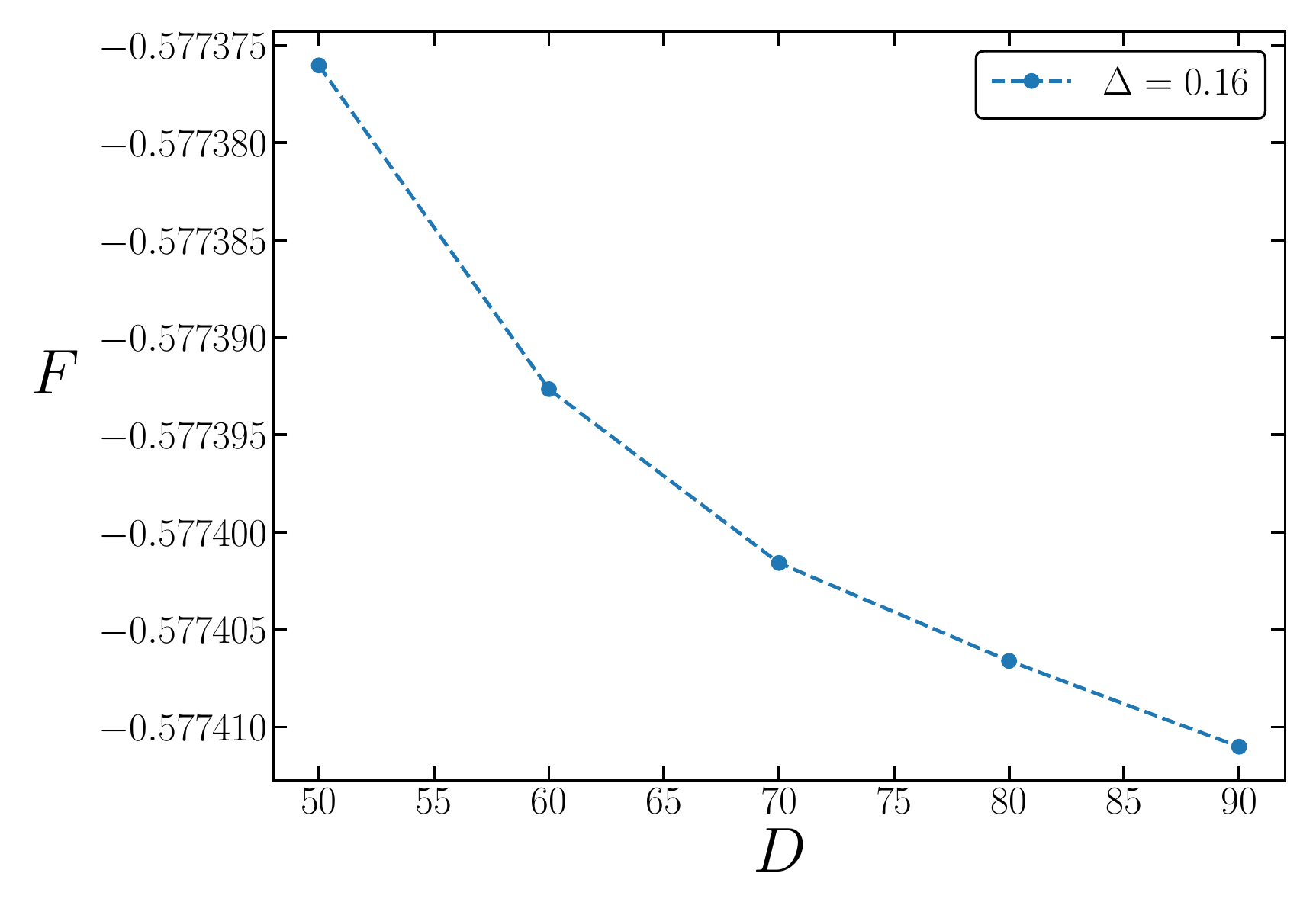} 
\caption{$D$-scaling for $\Delta = 0.16$, $m \in [-50, 50]$, $T \approx T_c = 0.773$ and volume is $2^{30} \times 2^{30}$.} 
\label{fig:scale1}
\end{figure}

\begin{figure}[h] 
\centering 
\includegraphics[width=10cm]{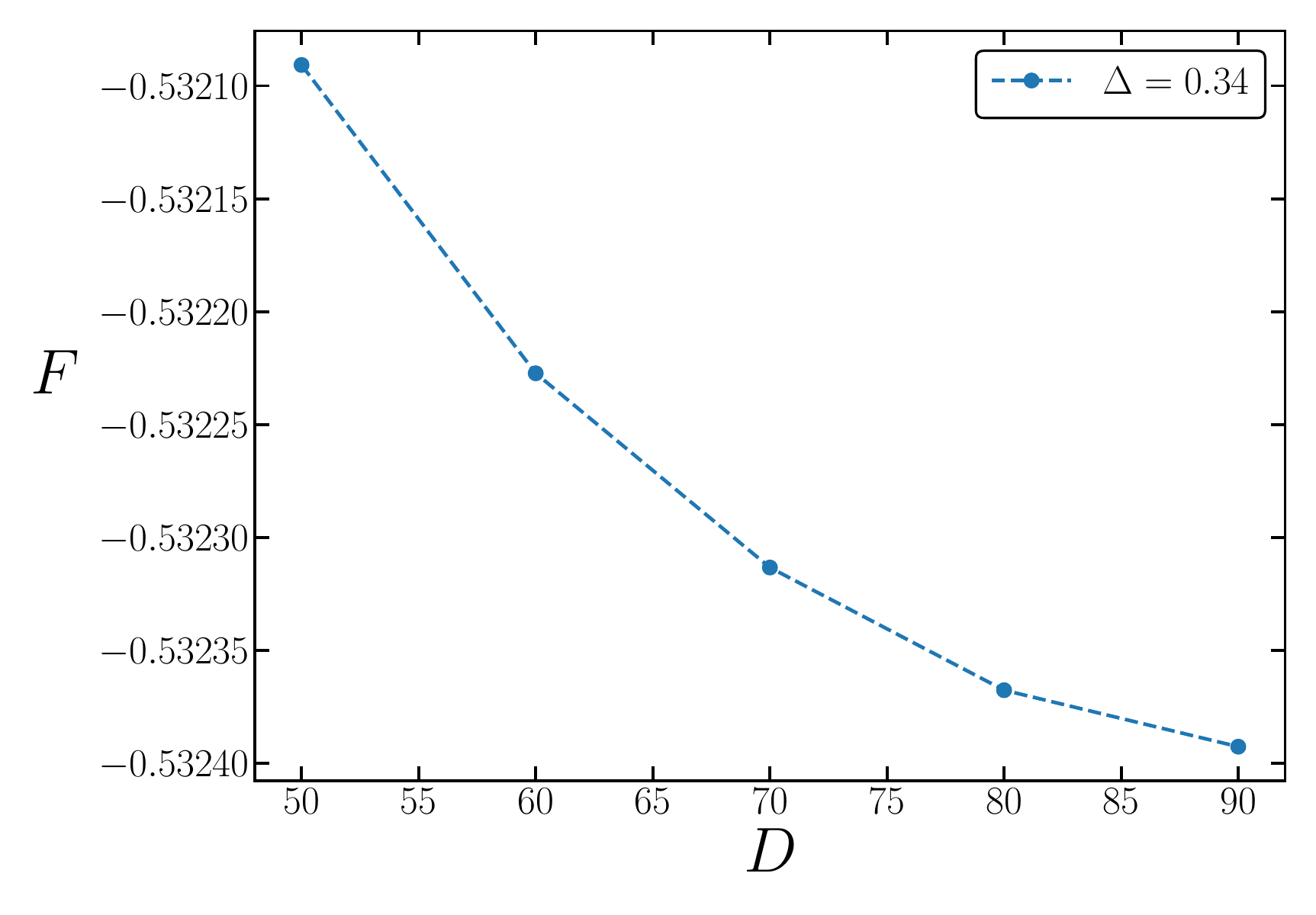} 
\caption{$D$-scaling for $\Delta = 0.34$, $m \in [-50, 50]$, $T \approx T_c = 0.694$ and volume is $2^{30} \times 2^{30}$.} 
\label{fig:scale2}
\end{figure}

\begin{figure}[h] 
\centering 
\includegraphics[width=10cm]{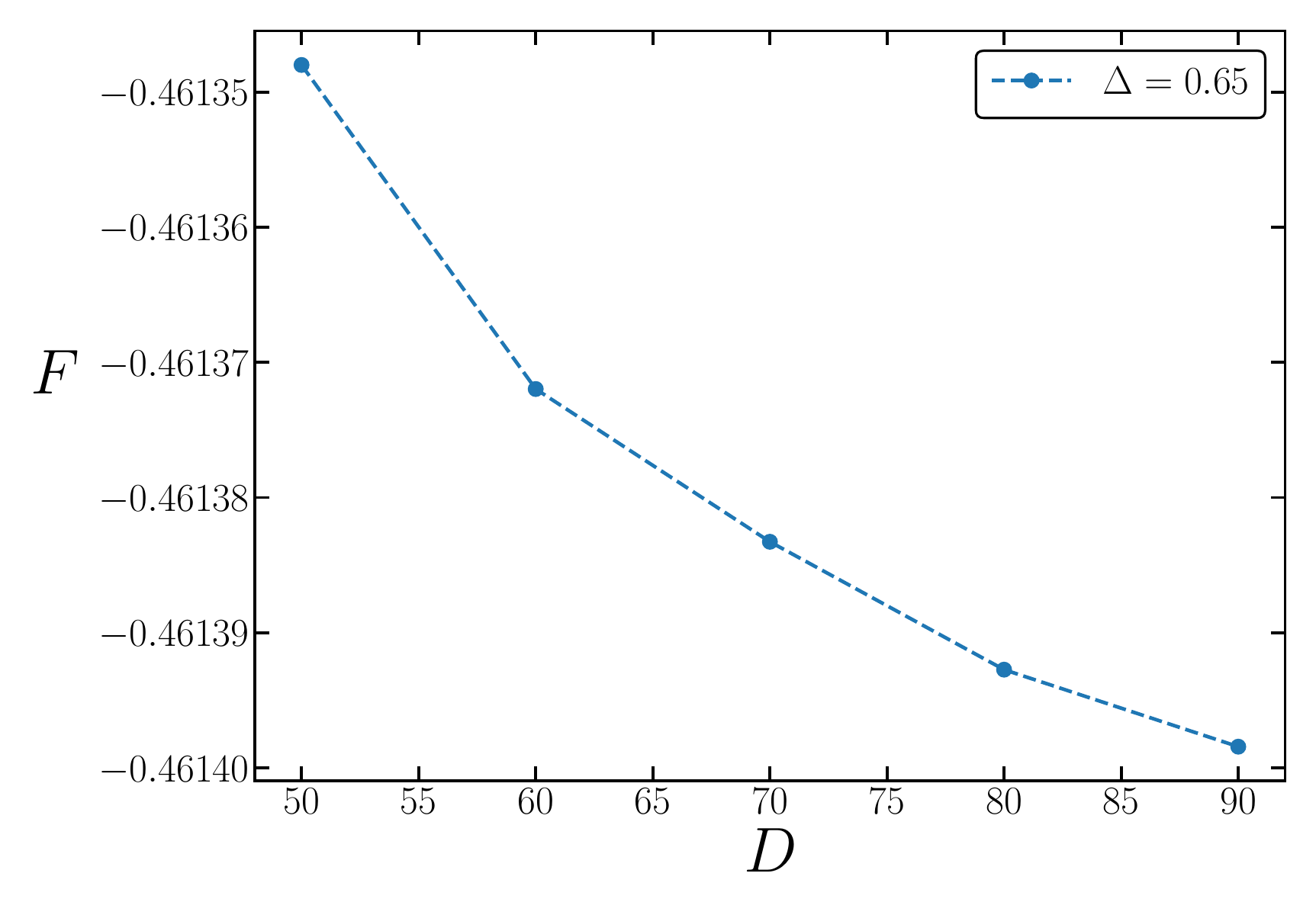} 
\caption{$D$-scaling for $\Delta = 0.65$, $m \in [-50, 50]$, $T \approx T_c = 0.853$ and volume is $2^{30} \times 2^{30}$.} 
\label{fig:scale3}
\end{figure}

Figures \ref{fig:scale1} - \ref{fig:scale3} show the scaling of free energy with bond dimension $D$ for different values of $\Delta$ and its corresponding critical temperature.

The results presented above also provide an assessment of the main sources of systematic uncertainty in the numerical determination of the transition temperatures and the resulting phase diagram. The finite bond dimension $D$ introduces a truncation error, while the Fourier cutoff $m$ leads to an additional systematic error; both are controlled by increasing $D$ and $m$ and checking the stability of the relevant observables and transition estimates. The discretization of the temperature grid and finite external fields $h$ and $h_1$ also affect the determination of the transition points. The latter effects are accounted for by extrapolating the transition temperatures to the zero-field limit, $h,h_1 \rightarrow 0$, with the stability of the extrapolation checked against reasonable variations of the fitting range. Finally, the classification of the transitions as BKT or Ising is based on a combined analysis of susceptibility, specific heat, magnetization, and other characteristic signatures.

The reliability of the final phase diagram is evaluated by examining its convergence as a function of both $D$ and $m$, by looking at the temperature resolution, by carrying out a zero-field extrapolation, by checking the stability of the fitting procedure, and by ensuring that the criteria used for classifying the transitions are consistent. Each of these individual aspects is assessed using convergence studies and variations in the parameters; the overall impact of these factors is represented by the uncertainty attached to the quoted transition temperatures.

\section{Conclusions} 
\label{sec:Conclusions} 

We studied the phase diagram of the generalized XY model in two dimensions using the GPU-improved real-space tensor network methods. 
We located the phase transitions belonging to the Ising and the BKT universality classes and identified the multicritical region around $\Delta = 0.36(2)$ and $T_c = 0.716(3)$ in the phase diagram where the multiple transition lines appear to merge. 
Our result is consistent with an interpretation where the half-BKT and Ising transition lines first meet (as $\Delta$ is increased from zero), and then the Ising line continues for a small range of $\Delta$ to merge with the BKT line, as observed in Ref.~\cite{Serna_2017} for $q = 2$ and for $q = 3$ Potts model in Ref.~\cite{Drouin-Touchette2022}. 
This work refines the previous computation of this model using matrix product state (MPS) methods, and we obtain results qualitatively similar to those from the standard Monte Carlo method in the literature. 
It remains an open problem to apply and extend the methods used in this work to models involving fractional vortices, corresponding to terms like $\cos(q \theta_i - q \theta_j)$ with $q \ge 3$ in the Hamiltonian. 
It is believed that these models have a more complicated phase diagram, as studied in Ref.~\cite{Poderoso2011} using Monte Carlo methods. 
It would be interesting to revisit this model using tensor network methods in the future.

\clearpage\mbox{}\clearpage
\chapter{Configurational Temperature Estimator} 
\label{ch:tmp_est}

\textit{This chapter is based on the paper:}
\begin{itemize} 
\item[] \textit{N. S. Dhindsa, A. Joseph and V. Longia, 
``Gradient and Hessian-Based Temperature Estimator in Lattice Gauge Theories: A Diagnostic Tool for Stability and Consistency in Numerical Simulations''} (Published)  
\item[] \href{https://doi.org/10.1007/JHEP10(2025)015}{JHEP 10, 015 (2025)}, \href{https://doi.org/10.48550/arXiv.2508.05595}{arXiv:2508.05595 [hep-lat]} 
\end{itemize} 

In molecular dynamics simulations, we can use conserved quantities such as energy and momentum as effective indicators of algorithmic correctness. 
Although the conservation of these quantities does not guarantee the validity of a simulation, violations typically signal programming or numerical errors, which are often easy to detect. 

In contrast, canonical Monte Carlo simulations lack straightforward conservation laws that can be used for validation. 
In particular, there has traditionally been no method for directly computing the thermodynamic temperature using only the field configurational data. 
As a result, algorithm validation in the canonical ensemble relies largely on comparison with known thermodynamic properties. 
This approach can become problematic when studying novel systems or exploring state points for which no reference data exist. 

A significant advance in this area was the work of Rugh, who derived an expression for temperature, based on a geometric analysis of phase space within the microcanonical ensemble \cite{PhysRevLett.78.772}. 
His formulation relates the temperature to the curvature of the constant-energy hypersurface and offers a purely dynamical definition grounded in the phase space structure. 

Building on Rugh's insight, Butler et al. proposed a configurational definition of temperature applicable to canonical ensembles \cite{10.1063/1.477301}. 
This expression requires only configurational information, namely, gradients and curvatures of the potential energy or action, and is therefore ideally suited to Monte Carlo simulations, which do not sample momenta. 
Interestingly, this type of definition, based on mean-square forces, was not entirely new: related ideas can be traced back to earlier works, including the 1952 Russian-language edition of {\it Statistical Physics} by Landau and Lifshitz \cite{Landau1952}, and even more indirectly to Tolman's 1938 textbook {\it The Principles of Statistical Mechanics}, which discusses generalized equipartition without explicitly formulating the concept \cite{Tolman1979}. 
A later historical account by Hoover also notes this lineage \cite{Hoover2007}. 

In Ref. \cite{PhysRevE.62.4757}, the authors demonstrate that arbitrary phase-space vector fields can be used to construct phase functions whose ensemble averages yield the thermodynamic temperature. 
They establish conditions under which these functions are valid in systems with periodic boundaries and within the molecular dynamics (MD) ensemble, and support their results with simulations using short-ranged potentials. 
In Ref. \cite{10.1063/1.477301}, the configurational temperature is shown, via Lennard-Jones simulations, to respond rapidly and accurately to changes in the input temperature, even when the system deviates from global thermodynamic equilibrium. 
A numerical test of the temperature estimator in the canonical ensemble is presented in Ref. \cite{PhysRevE.94.062113}, where the authors simulate the two-dimensional XY model using a generalized Wolff cluster algorithm. 
Broader discussions and generalizations of the underlying theoretical framework can be found in Refs. \cite{1998JPhA...31.7761R, PhysRevE.62.4757, 10.1063/1.1348024}, while further applications and validation in MD simulations are provided in Refs. \cite{10.1063/1.477301, 10.1063/1.480995}. 

Although the configurational temperature was initially proposed as a tool for thermostat design and system control in numerical simulations, Butler et al. emphasized its practical value as a diagnostic. 
They demonstrated its utility in detecting sampling inconsistencies and numerical errors in Monte Carlo simulations. 
In modern simulations, the configurational temperature can serve as a means of assessing thermodynamic consistency, independently of the microscopic momentum variables used in kinetic definitions of temperature. 

In lattice field theory simulations, the physical temperature is typically introduced through the temporal lattice extent or inverse coupling. 
However, lattice discretization artifacts can lead to significant deviations in physical observables, such as a shift in the transition temperature at finite lattice spacing. 
These effects vanish only in the continuum limit.
Thus, verifying that the configurational temperature matches the intended value is a nontrivial task. 
The configurational temperature estimator provides a nontrivial, gauge-invariant way to verify whether a system is truly sampling from the expected thermal distribution. 
It is especially useful in lattice gauge theories where discretization and algorithmic artifacts can shift critical temperatures or other thermodynamic properties. 

\section{Derivation of the Configurational Temperature Estimator} 
\label{sec:derv_temp_est} 

This section outlines the construction of the configurational temperature estimator entirely from the sampled field configurations. 
In Hybrid Monte Carlo (HMC) simulations, where the momenta are auxiliary, this estimator can serve as a nontrivial check of thermodynamic consistency. 
From elementary thermodynamics, we have the first law expressed as 
\begin{equation} 
dE = T dS - p dV. 
\end{equation} 
For an isochoric transformation, the rate of change of entropy with an increase in energy determines the inverse temperature: 
\begin{equation} 
\frac{1}{T} = \left( \frac{\partial S}{\partial E} \right)_V. 
\label{eq:dE} 
\end{equation} 

Let $\vec{\Gamma}$ denote a point in phase space, comprising all position and momentum coordinates of the system 
\begin{equation} 
\vec{\Gamma} \equiv \left( q_1, q_2, \ldots, q_N; p_1, p_2, \ldots, p_N \right).
\end{equation} 
Here, $q_i$ and $p_i$ represent the position and momentum of the $i$-th particle, respectively. 

We define a microcanonical ensemble corresponding to a Hamiltonian $H(\vec{\Gamma})$ as the set of phase space points for which the total energy lies within a narrow interval $[E, E + \Delta E]$, with $\Delta E \ll E$. 
Under the assumption of equal {\it a priori} probabilities, the entropy associated with this ensemble is proportional to the volume of this energy shell in phase space: 
\begin{align} 
S(E) &= k_B \ln \Omega_\Gamma(E, N, V) \nonumber \\ 
&= k_B \ln \left( \int_{\mu C(E)} d\vec{\Gamma} \right), 
\end{align} 
where $\mu C(E)$ denotes the microcanonical shell, defined as the set of all $\vec{\Gamma}$ satisfying $E \leq H(\vec{\Gamma}) \leq E + \Delta E$. 
To compute the derivative $\partial S / \partial E$, we examine how this phase-space volume changes under an infinitesimal shift in energy. 
Specifically, we introduce a displacement of each phase space point along a vector field $\vec{n}(\vec{\Gamma})$ that is constructed to increase the energy uniformly across the ensemble. 
This procedure amounts to mapping the set $\mu C(E)$ to a nearby energy shell $\mu C(E + \Delta E)$ so that the displacement induces a constant energy increase $\Delta E$ to leading order. 

This construction allows us to evaluate the change in entropy purely from the geometry of the phase space, ultimately leading to an expression for the temperature in terms of configurational derivatives of the Hamiltonian with respect to the underlying coordinates. 

To evaluate the change in entropy with respect to energy, let us consider an infinitesimal transformation that shifts the energy of the system from $E$ to $E + \Delta E$ by displacing phase space points along a specific direction in configuration space. 
The displaced point is defined by 
\begin{equation} 
\vec{\Gamma}'(\vec{\Gamma}) = \vec{\Gamma} + \Delta E ~ \frac{\vec{\nabla}_{\vec{q}} H(\vec{\Gamma})}{\vec{\nabla}_{\vec{q}} H(\vec{\Gamma}) \cdot \vec{\nabla}_{\vec{q}} H(\vec{\Gamma})}, 
\end{equation} 
where 
\begin{equation} 
\vec{\nabla}_{\vec{q}} = \left( \frac{\partial}{\partial q_1}, \dots, \frac{\partial}{\partial q_N} \right), 
\end{equation} 
is the configurational gradient operator. 
This transformation moves each point in configuration space along the direction of increasing energy, ensuring a uniform increase in energy by $\Delta E$. 
Only the configurational components of the Hamiltonian gradient are considered, in keeping with the goal of expressing thermodynamic quantities in terms of positional degrees of freedom alone. 

The vector field along which we displace the configurations is given by 
\begin{equation} 
\vec{n}(\vec{\Gamma}) \equiv \frac{\vec{\nabla}_{\vec{q}} H(\vec{\Gamma})}{\vec{\nabla}_{\vec{q}} H(\vec{\Gamma}) \cdot \vec{\nabla}_{\vec{q}} H(\vec{\Gamma})}. 
\label{eq:nvector_Gamma} 
\end{equation} 
We choose this vector field such that $0 < | \langle \vec{\nabla} H \cdot \vec{n}(\vec{\Gamma}) \rangle| < \infty$, $0 < | \langle\vec{\nabla} \cdot \vec{n}(\vec{\Gamma})\rangle | < \infty$ (where $\langle \dots \rangle$ represents an ensemble average) and $\langle \vec{\nabla} H \cdot \vec{n}(\vec{\Gamma})\rangle$ grows more slowly than $e^N$ in the thermodynamic limit \cite{PhysRevE.62.4757}. 
This choice ensures that the displacement gives a change in the Hamiltonian of exactly $\Delta E$, to leading order. 

To see this, consider the analogy of a topographic landscape where the height corresponds to the energy $H(\vec{q})$ of a configuration $\vec{q}$. 
Contour lines of constant energy represent level surfaces in configuration space. 
Moving tangentially to the contour leaves the energy unchanged, while motion in the direction of the gradient increases the energy most rapidly. 
The gradient $\vec{\nabla}_{\vec{q}} H$ therefore points normal to the contour and indicates both the direction and rate of steepest energy ascent. 

We now consider a small displacement of the configuration vector: 
\begin{equation} 
\vec{q}~' = \vec{q} + \epsilon ~ \vec{n}(\vec{\Gamma}), 
\end{equation} 
with $\epsilon = \Delta E$ and $\vec{n}(\vec{\Gamma})$ as defined in Eq.~\eqref{eq:nvector_Gamma}. 
The normalization by $\vec{\nabla}_{\vec{q}} H \cdot \vec{\nabla}_{\vec{q}} H$ ensures that the change in energy is precisely $\Delta E$.  
Let us expand the Hamiltonian to leading order, using a Taylor series.
Then, we have 
\begin{equation} 
H(\vec{q}~') = H(\vec{q}) + \vec{\nabla}_{\vec{q}} H \cdot (\vec{q}~' - \vec{q}) + \mathcal{O}((\vec{q}~' - \vec{q})^2). 
\end{equation} 
Substituting the displacement vector 
\begin{equation} 
\vec{q}~' - \vec{q} = \Delta E ~ \vec{n}(\vec{\Gamma}), 
\end{equation} 
we obtain 
\begin{align} 
H(\vec{q}~') &= H(\vec{q}) + \Delta E ~ \vec{\nabla}_{\vec{q}} H \cdot \vec{n}(\vec{\Gamma}) + \mathcal{O}((\Delta E)^2) \nonumber \\ 
&= H(\vec{q}) + \Delta E  + \mathcal{O}((\Delta E)^2), 
\end{align} 
which confirms that the Hamiltonian increases by exactly $\Delta E$ to linear order in the displacement.  
This controlled shift in energy forms the basis for computing thermodynamic derivatives such as $\partial S / \partial E$ in a purely configurational framework. 

To leading order in $\Delta E$, the displacement of phase space points described above gives rise to a uniform change in energy, independent of the initial phase space vector $\vec{\Gamma} \in \mu C(E)$. 
All points on the constant-energy hypersurface are mapped to a nearby surface with energy $E + \Delta E + \mathcal{O}((\Delta E)^2)$. 

Let us compute the corresponding change in entropy by evaluating the transformation of the phase space volume under this displacement. 
Given the map $\vec{q} \mapsto \vec{q}~'$, we have the entropy of the displaced ensemble 
\begin{align} 
S(E + \Delta E) &= k_B \ln \int_{\mu C(E + \Delta E)} d\vec{\Gamma} \nonumber \\ 
&= k_B \ln \int_{\mu C(E)} \left| \frac{\partial \vec{\Gamma}'}{\partial \vec{\Gamma}} \right| d\vec{\Gamma}. 
\end{align} 
The integrand includes the Jacobian determinant associated with the phase space transformation. 

Since only the positions $\vec{q}$ are displaced while momenta remain fixed, the Jacobian is determined by the following configuration space transformation
\begin{equation} 
J(\vec{q}) = \left| \frac{\partial \vec{q}~'}{\partial \vec{q}} \right| = 1 + \Delta E ~ \vec{\nabla}_{\vec{q}} \cdot \vec{n}(\vec{q}), 
\end{equation} 
where $\vec{n}(\vec{q})$ is the normalized displacement vector defined earlier. 
For completeness, we may write the full phase space Jacobian as 
\begin{equation} 
J(\vec{\Gamma}) = \left| \frac{\partial \vec{\Gamma}'(\vec{\Gamma})}{\partial \vec{\Gamma}} \right| = 1 + \Delta E ~ \vec{\nabla}_{\vec{q}} \cdot \vec{n}(\vec{\Gamma}). 
\label{eq:J_Gamma} 
\end{equation} 

This Jacobian captures the local volume change in phase space, and its logarithm contributes to the change in entropy under the transformation.
Then, we have
\begin{align} 
\Delta S &= S(E + \Delta E) - S(E) \nonumber \\ 
&= k_B \ln \left( 1 + \Delta E ~ \vec{\nabla}_{\vec{q}} \cdot \vec{n}(\vec{q}) \right) \nonumber \\ 
&\approx k_B \Delta E ~ \vec{\nabla}_{\vec{q}} \cdot \vec{n}(\vec{q}).
\end{align} 
In the above, we used the approximation $\ln(1 + \delta) \approx \delta$, valid for small $\delta$. 

Taking the limit as $\Delta E \to 0$, we obtain an expression for the derivative of entropy with respect to energy: 
\begin{eqnarray} 
\frac{\partial S}{\partial E} &=& \lim_{\Delta E \to 0} \frac{\Delta S}{\Delta E} \nonumber \\
&=& k_B ~ \left\langle \vec{\nabla}_{\vec{q}} \cdot \vec{n}(\vec{q}) \right\rangle, 
\end{eqnarray} 
where the average is taken over the ensemble. 
After substituting the definition of $\vec{n}(\vec{q})$, we get 
\begin{eqnarray} 
\frac{1}{k_B T} &=& \frac{1}{k_B} \frac{\partial S}{\partial E} \nonumber \\
&=& \left\langle \vec{\nabla}_{\vec{q}} \cdot \left( \frac{\vec{\nabla}_{\vec{q}} H}{\vec{\nabla}_{\vec{q}} H \cdot \vec{\nabla}_{\vec{q}} H} \right) \right\rangle. 
\end{eqnarray} 

We note that only the configurational contribution to the Hamiltonian is relevant in many canonical ensemble simulations, particularly Monte Carlo algorithms that omit momentum variables. 
Assuming $H(\vec{q}) = \Phi(\vec{q})$, where $\Phi$ is the potential energy (or the action), the above expression reduces to 
\begin{equation} 
\frac{1}{k_B T} = \left\langle \vec{\nabla}_{\vec{q}} \cdot \left( \frac{\vec{\nabla}_{\vec{q}} \Phi}{\vec{\nabla}_{\vec{q}} \Phi \cdot \vec{\nabla}_{\vec{q}} \Phi} \right) \right\rangle + \mathcal{O}\left( \frac{1}{N} \right). 
\label{eq:temp_eq} 
\end{equation} 

By the equivalence of ensembles, Eq.~\eqref{eq:temp_eq} remains valid in the canonical ensemble in the thermodynamic limit $N \to \infty$. 
This configurational definition of temperature thus provides a powerful diagnostic for validating sampling consistency in simulations that operate without explicit momentum variables\footnote{This formalism is not suited for systems where the momentum variables play a crucial role. For example, in Hamiltonian lattice gauge theory formalism, as opposed to Euclidean path integral formalism, the gauge field is the coordinate, the electric field is the canonical momentum, and both are physical observables.}. 

We can rewrite the expression in Eq. \eqref{eq:temp_eq} in terms of the Hessian matrix and the gradient vector of $\Phi(\vec{q})$. 
It helps express the configurational temperature in a more compact and numerically useful way. 

Let us denote the gradient and Hessian in the following way: 
\begin{equation} 
\vec{g} = \vec{\nabla}_{\vec{q}} \Phi(\vec{q}), ~~ g \in {\mathbb R}^N 
\end{equation} 
and 
\begin{equation} 
{\mathbb H} = \vec{\nabla}_{\vec{q}} \vec{\nabla}_{\vec{q}}^T \Phi(\vec{q}), ~~ {\mathbb H} \in {\mathbb R}^{N \times N}. 
\end{equation} 

Then, the quantity 
\begin{equation} 
\vec{\nabla}_{\vec{q}} \cdot \left( \frac{\vec{\nabla}_{\vec{q}} \Phi}{\vec{\nabla}_{\vec{q}} \Phi \cdot \vec{\nabla}_{\vec{q}} \Phi} \right) 
\end{equation} 
can be written using the product rule (divergence of a vector field), as 
\begin{equation} 
\vec{\nabla}_{\vec{q}} \cdot \left( \frac{\vec{g}}{\vec{g} \cdot \vec{g}} \right) = \frac{1}{\vec{g} \cdot \vec{g}} \vec{\nabla}_{\vec{q}} \cdot \vec{g} - \frac{2}{\left( \vec{g} \cdot \vec{g} \right)^2} \sum_{i, j} g_i H_{ij} g_j. 
\label{eq:another-form} 
\end{equation} 

This gives us 
\begin{equation} 
\vec{\nabla}_{\vec{q}} \cdot \left( \frac{\vec{g}}{\vec{g} \cdot \vec{g}} \right) = \frac{{\rm Tr} ({\mathbb H})}{| \vec{g} |^2} - 2 \frac{\vec{g}^T {\mathbb H} \vec{g}}{|\vec{g}|^4}. 
\end{equation} 

In higher dimensions, configurations sampled via $e^{-\Phi}$ tend to be smoother due to averaging over more directions, and the number of degrees of freedom increases. 
As a result, the gradient vector accumulates contributions from many local directions, leading to weaker directional alignment with the curvature. 
Consequently, the directional term $\vec{g}^{T} \mathbb{H} \vec{g} / |\vec{g}|^4$ becomes increasingly subdominant compared to the trace term $\mathrm{Tr}(\mathbb{H}) / |\vec{g}|^2$ in the temperature estimator.

\section{Application to Euclidean Lattice Field Theory}
\label{sec:App_Eucl_QFT}

Although the configurational temperature was originally introduced in the context of classical statistical mechanics, the Euclidean path integral formulation of quantum field theory provides a natural mathematical bridge that allows the same formalism to be applied in lattice field theory. 

In the Euclidean path integral approach, expectation values of observables are computed as
\begin{equation}
\langle O \rangle = \frac{\int \mathcal{D}\phi \; O[\phi] \, e^{- S[\phi]}}{\int \mathcal{D}\phi \; e^{- S[\phi]}} ,
\label{eq:path_integral}
\end{equation}
where $S[\phi]$ is the Euclidean action and $\mathcal{D}\phi$ denotes integration over all field configurations. 

This expression has the same mathematical structure as the canonical ensemble of classical statistical mechanics,
\begin{equation}
P(\vec{q}) \propto e^{-\Phi(\vec{q})/T},
\end{equation}
where $\Phi(\vec{q})$ is the potential energy and $T$ is the temperature. 
In the Euclidean path integral formulation the corresponding distribution is
\begin{equation}
P[\phi] \propto e^{- S[\phi]} .
\label{eq:analogy}
\end{equation}

Thus, the Euclidean action $S[\phi]$ plays a role analogous to the classical potential energy, while the effective inverse temperature is unity (in natural units). 
This formal correspondence underlies the widespread use of importance sampling Monte Carlo methods in lattice field theory.

Although the physical interpretation differs between classical statistical systems and quantum field theories, the structural equivalence of the probability measures implies that statistical identities derived for canonical ensembles can often be transferred directly to the path integral framework. 
In particular, the configurational temperature estimator (originally formulated in terms of derivatives of the potential energy) can be reformulated in terms of derivatives of the Euclidean action. 

\section{Sampling Diagnostic vs.\ Thermodynamic Temperature}

It is important to clarify the interpretation of the configurational temperature in the context of Euclidean path integrals. 
In contrast to classical statistical mechanics, the configurational temperature in lattice field theory does \emph{not} correspond to a physical thermodynamic temperature. 
Instead, it should be understood as a \emph{diagnostic of sampling correctness}. 
Specifically, it tests whether the ensemble of field configurations generated by a numerical algorithm follows the intended probability distribution determined by the Euclidean action,
\begin{equation}
P[\phi] \propto e^{- S[\phi]}.
\end{equation}

To make this interpretation explicit, it is convenient to introduce an auxiliary parameter $\beta_{\rm config}$ in the statistical weight,
\begin{equation}
P[\phi] \propto e^{-\beta_{\rm config} \, S[\phi]} ,
\end{equation}
whose target value is $\beta_{\rm config} = 1$. 
Within this formulation the configurational temperature estimator measures the \emph{effective inverse temperature} associated with the sampled distribution.

If the sampling algorithm correctly reproduces the desired path integral measure, the estimator yields the expected value $\beta_{\rm config} = 1$. 
However, if the algorithm converges to an incorrect stationary distribution, due, for example, to discretization errors, numerical instabilities, insufficient equilibration, then one generically finds \cite{Dhindsa:2025xfv, Longia:2026doi}
\begin{equation}
\beta_{\rm config} \neq 1 .
\label{eq:effective_coupling}
\end{equation}
In this sense, the configurational temperature provides a direct and quantitative probe of the correctness of the sampling procedure.

The estimator can be expressed directly in terms of the lattice action. 
Denoting the measured configurational inverse temperature in a lattice simulation by $\beta^L_{\rm config}$, we define
\begin{equation}
\beta^L_{\rm config}  \equiv \left\langle \nabla_{\phi} \cdot \frac{\nabla_{\phi} S[\phi]}{\left|\nabla_{\phi} S[\phi]\right|^2} \right\rangle,
\label{eq:lattice_estimator}
\end{equation}
where $\nabla_{\phi}$ denotes the gradient with respect to all lattice field variables $\phi$, and the expectation value is taken over the ensemble of configurations generated in the simulation.

At first sight the factor $\left|\nabla_{\phi}S[\phi]\right|^{-2}$ appearing in Eq.~\eqref{eq:lattice_estimator} may appear problematic, since it seems to become singular at stationary configurations where $\nabla_{\phi} S[\phi] = 0$. 
However, the apparent singularity is removable. 
To see this, consider an expansion of the action around a stationary configuration $\phi_0$,
\begin{equation}
S(\phi) = S(\phi_0) + \frac{1}{2}(\phi - \phi_0)^T H (\phi - \phi_0) + \cdots,
\end{equation}
where $H$ is the Hessian matrix evaluated at $\phi_0$. 
Near such a point the gradient scales linearly with the displacement $(\phi - \phi_0)$, while both the numerator and denominator of the estimator in Eq.~\eqref{eq:lattice_estimator} scale quadratically with the same power of the displacement. 
Consequently, their ratio remains finite in the vicinity of stationary points.

In practical lattice simulations we do not observe numerical instabilities associated with configurations close to stationary points of the action, confirming that the estimator is well behaved in typical Monte Carlo ensembles.

\section{Applicability of Configurational Temperature to HMC} 
\label{sec:hmc} 

We can apply the configurational temperature estimator to HMC simulations, provided some important clarifications are kept in mind. 
Consider the situation in which we are simulating a scalar field theory using HMC. 
The scalar field action $S[\phi]$ serves the role of a potential energy in an extended phase space. At the same time, the conjugate momenta are auxiliary variables introduced solely to facilitate efficient proposal generation. 

Despite the use of a fictitious Hamiltonian dynamics, the HMC algorithm samples from the same canonical ensemble as conventional Metropolis Monte Carlo:
\begin{equation}
p(\phi) \propto e^{-S[\phi]}.
\end{equation}
At the beginning of each trajectory, the momenta are drawn from a Gaussian distribution.
Then they are evolved deterministically during the Hamiltonian flow, and are discarded at the end. 
Their introduction facilitates global updates and improves sampling efficiency by reducing autocorrelations compared to local-update methods, even though they do not contribute to the statistical ensemble used for observable estimation. 

The artificial Hamiltonian guiding the dynamics takes the form
\begin{equation}
H[\phi, \pi] = \frac{1}{2} \sum_x \pi(x)^2 + S[\phi],
\end{equation}
where $\pi(x) \sim \mathcal{N}(0, 1)$ are auxiliary Gaussian momenta. 
After integrating out the momenta, the marginal distribution over field configurations remains canonical: 
\begin{align} 
p(\phi) &\propto \int \mathcal{D}\pi ~ e^{-H[\phi, \pi]} \\ 
&= e^{-S[\phi]} \int \mathcal{D}\pi ~ e^{- \frac{1}{2} \sum_x \pi(x)^2} \propto e^{-S[\phi]}. 
\end{align} 
Thus, HMC produces samples from the Boltzmann distribution $e^{-S[\phi]}$, and the configurational temperature formalism, which relies only on the distribution of configurations $\phi$, remains valid. 

The configurational temperature estimator, therefore, provides a thermodynamic consistency check for HMC simulations. 
No information from the momenta is required to compute $\beta_{\rm config}$; the estimator depends only on the sampled field configurations.

\section{Configurational Temperature Estimator as a Diagnostic in HMC Simulations}

Measuring the configurational temperature in lattice simulations offers an independent means to verify the correctness of the simulations. 
Below, we list several aspects that underscore the utility of the configurational temperature in HMC simulations.

\begin{itemize}
\item[(i.)] {\it Validation of sampling consistency}

HMC aims to generate field configurations $\phi$ according to the canonical probability distribution $p(\phi) \propto e^{-S[\phi]}$. 
However, various algorithmic issues, such as inaccurate gradient computations, incorrect acceptance criteria, integration errors, or violation of detailed balance, can lead to subtle biases in the sampled ensemble. 
Since the configurational temperature is a property of the ensemble distribution, it provides a sensitive probe for detecting such inconsistencies. 
If the simulation is thermodynamically accurate, we expect
\begin{equation}
\beta_{\text{config}} \approx 1.
\end{equation}

\item[(ii.)] {\it Momentum-free estimator}

We can compute a {\it kinetic temperature} using the auxiliary momenta
\begin{equation}
T_{\text{kin}} = \frac{1}{N} \sum_x \left\langle \pi(x)^2 \right\rangle.
\end{equation} 
However, this diagnostic merely verifies the correctness of the momentum sampling from a Gaussian distribution. 
It does not provide information about whether the field configurations are sampled correctly. 
By contrast, the configurational temperature is purely configurational and directly tests the consistency of the $\phi$-ensemble with the underlying Boltzmann distribution.

\item[(iii.)] {\it Monitoring thermalization and autocorrelations}

The configurational temperature can deviate significantly from its expected value in simulations where the system thermalizes slowly or becomes trapped in metastable states. 
Monitoring $\beta_{\text{config}}$ during early trajectories provides a quantitative check on equilibration. 
Persistent deviations may signal slow phase separation, poor mixing, or errors in the update dynamics or acceptance mechanism. 
Thus, the configurational temperature is a helpful indicator for diagnosing thermalization and autocorrelation issues.
\end{itemize} 

In summary, the configurational temperature estimator does not replace standard validation tools such as observable comparisons or autocorrelation analyses, but it constitutes a powerful and underutilized consistency check. 
Its applicability is particularly pronounced when dealing with novel actions, large-scale simulations, or custom implementations where traditional diagnostics may be insufficient. 

In the next chapter, we apply the configurational temperature estimator in compact U(1) lattice gauge theories in various dimensions.

\clearpage\mbox{}\clearpage
\chapter{Benchmarking the Estimator in U(1) Lattice Gauge Theories} 
\label{ch:u1} 

\textit{This chapter is based on the paper:} 
\begin{itemize} 
\item[] \textit{N. S. Dhindsa, A. Joseph and V. Longia, 
``Gradient and Hessian-Based Temperature Estimator in Lattice Gauge Theories: A Diagnostic Tool for Stability and Consistency in Numerical Simulations''} (Published) 
\item[] \href{https://doi.org/10.1007/JHEP10(2025)015}{JHEP 10, 015 (2025)}, \href{https://doi.org/10.48550/arXiv.2508.05595}{arXiv:2508.05595 [hep-lat]} 
\end{itemize} 

In this chapter, we apply the configurational temperature estimator to the compact U(1) lattice gauge theory. 
This is a canonical Abelian model that exhibits rich phase structure in various dimensions. 
Monte Carlo studies have shown that for spacetime dimensions $D < 4$, the compact U(1) lattice gauge theory exhibits only a confining phase. 
In contrast, for $D \geq 4$, the theory supports at least two distinct phases~\cite{Bhanot:1980pc}. 

In this theory, the link variables are elements of the group U(1). 
They are associated with directed links from a site $n$ in the direction $\mu$, and are defined as 
\begin{equation} 
U_\mu(n) = \exp(i \theta_\mu(n)), \quad \theta_\mu(n) \in (-\pi, \pi], 
\end{equation} 
with $\mu = 1, \cdots, d$ denoting the spacetime directions. 

The standard Wilson action for this theory takes the form 
\begin{equation} 
S = \beta \sum_{(\mu, \nu; n)} \left[1 - \cos\theta_{\mu\nu}(n) \right]. 
\end{equation} 
Here, $\beta$ denotes the lattice coupling.
The plaquette angle $\theta_{\mu \nu}(n)$ represents the discrete curl of the link field: 
\begin{equation} 
\theta_{\mu \nu}(n) = \theta_\mu(n) + \theta_\nu(n + \hat{\mu}) - \theta_\mu(n + \hat{\nu}) - \theta_\nu(n), 
\end{equation} 
with $\hat{\mu}$ and $\hat{\nu}$ denoting the unit vectors in $\mu$ and $\nu$ directions, respectively. 

The corresponding partition function is given by
\begin{equation} 
Z(\beta) = \sum_{\{\theta\}} \exp\left[- S(\{\theta\})\right]. 
\end{equation} 
The free energy per plaquette is defined as 
\begin{equation} 
f(\beta) = \frac{1}{N_p} \ln Z(\beta), 
\end{equation} 
with $N_p$ being the total number of plaquettes in the lattice. 

As an observable, we consider the average plaquette energy (or action density), 
\begin{equation} 
E(\beta) = - \frac{\partial f(\beta)}{\partial \beta} = \left \langle 1 - \cos(\theta_{\mu\nu}) \right \rangle. 
\end{equation} 
It serves as an effective order parameter across different coupling regimes. 
Non-analytic behavior in $E(\beta)$, such as discontinuities or rapid changes, indicates phase transitions. 
In practice, we track the first or second derivative, e.g., the plaquette susceptibility, as higher-order derivatives are numerically unstable. 
Throughout our study, we compute energy densities suited to each dimension and analyze them alongside the temperature estimator to understand the thermal behavior across different setups. 

\section{One-Dimensional Theory} 

The theory uses compact link variables $U_n$, which reside on the links between lattice sites $n$ and $n + 1$. 
They take values in the U(1) gauge group: 
\begin{equation} 
U_n = e^{i \theta_n}, \quad \theta_n \in (-\pi, \pi]. 
\end{equation} 

The Wilson action for this system reduces to the following nearest-neighbor interaction form: 
\begin{equation} 
S = \beta \sum_{n = 0}^{N_\tau - 1} \left( 1 - \cos(\theta_n - \theta_{n + 1}) \right), 
\end{equation} 
where $N_\tau$ denotes the number of lattice sites. 

This model is equivalent to the one-dimensional classical XY model, which is exactly solvable. 
Although it does not exhibit a phase transition, it is a valuable testing ground for lattice algorithms due to its analytical tractability. 

Let us denote the configurational (inverse) temperature obtained through the estimator as $\hat{\beta}$. 
The estimator, introduced in Eq.~\eqref{eq:another-form}, takes the form 
\begin{equation} 
\hat{\beta} = \frac{\sum_n h_{nn}}{\sum_n g_n^2} - \frac{2 \sum_{nm} g_n g_m h_{nm}}{\left( \sum_n g_n^2 \right)^2}, 
\label{eq:beta_lat} 
\end{equation} 
where $g_n$ and $h_{nm}$ denote the gradient and Hessian of the action, respectively: 
\begin{equation} 
g_n = \frac{1}{\beta} \frac{\partial S}{\partial \theta_n}, \quad h_{nm} =  \frac{1}{\beta} \frac{\partial^2 S}{\partial \theta_n \partial \theta_m}. 
\label{eq:grad_hess} 
\end{equation} 

In the above, we chose to scale $\beta^L_{\rm config}$ by the coupling $\beta$.
That is
\begin{equation}
\hat{\beta} = \beta^L_{\rm config} \beta. 
\end{equation}
Note that this is just a convention, we could also work directly with $\beta^L_{\rm config}$.

The explicit expressions for the gradient and Hessian terms are: 
\begin{align} 
g_n &= \sin(\theta_{n - 1} - \theta_n) - \sin(\theta_n - \theta_{n + 1}), \label{eq:grad} \\ 
h_{nn} &= \cos(\theta_{n - 1} - \theta_n) + \cos(\theta_n - \theta_{n + 1}), \label{eq:hess_diag} \\ 
h_{nm} &= -\cos(\theta_n - \theta_m) \delta_{m, n + 1} \nonumber \\ 
& ~~~ ~~~ - \cos(\theta_m - \theta_n) \delta_{m, n - 1}, \label{eq:hess_offdiag} 
\end{align} 
with periodic boundary conditions implied. 
These expressions allow for direct evaluation of the temperature estimator during the simulation. 

Using HMC, we evolve the system in fictitious Monte Carlo time and compute observables such as the internal energy, specific heat, and the measured configurational temperature $\beta_M \equiv \langle \hat{\beta} \rangle$. 
We compute the average energy 
\begin{equation} 
E(\beta) = \left\langle 1 - \cos(\theta_n - \theta_{n + 1}) \right\rangle, 
\label{eq:energy_lat_1d_u1} 
\end{equation} 
the corresponding specific heat, and the configurational temperature estimator defined in Eq.~\eqref{eq:beta_lat}, using Eqs. \eqref{eq:grad}, \eqref{eq:hess_diag}, and \eqref{eq:hess_offdiag}. 
We monitor $\beta_M$ obtained from the simulations for a given set of $\beta$ values. 
We validate our results by comparing the numerically computed energy and specific heat with the exact analytical expressions. 

For the theory with periodic boundary conditions, the analytical expressions for energy and specific heat are given by: 
\begin{align} 
E(\beta) &= 1 - \frac{I_1(\beta)}{I_0(\beta)}, \label{eq:energy_ana_1d} \\ 
C(\beta) &= \beta^2 \left[ 1- \frac{I_1(\beta)}{\beta I_0(\beta)} - \left( \frac{I_1(\beta)}{I_0(\beta)} \right)^2 \right], 
\label{eq:spec_ana_1d} 
\end{align} 
where $I_n(\beta)$ denotes the modified Bessel function of the first kind of order $n$. 

\begin{figure*}[!t]
    \centering

    \includegraphics[width=0.50\textwidth]{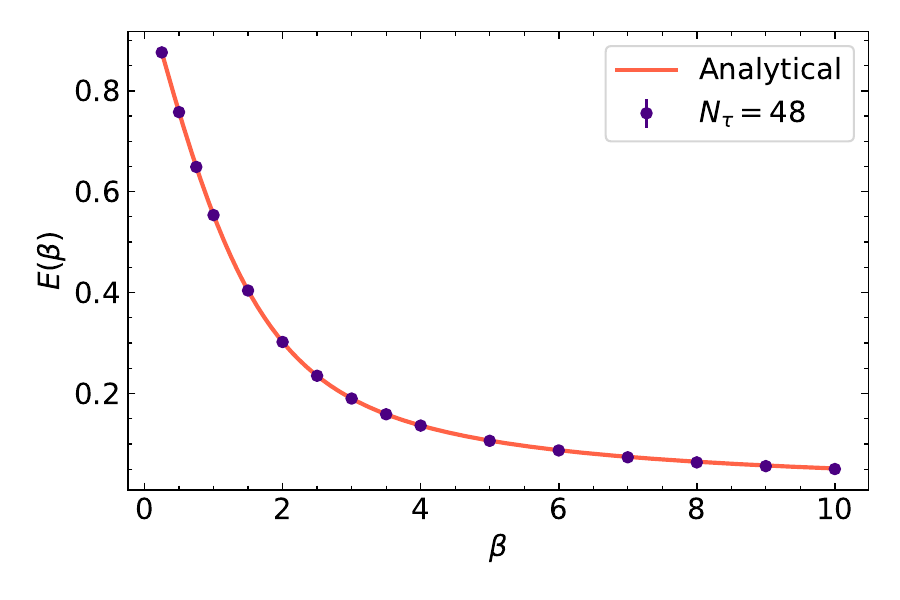}

    \vspace{0.3cm}

    \includegraphics[width=0.50\textwidth]{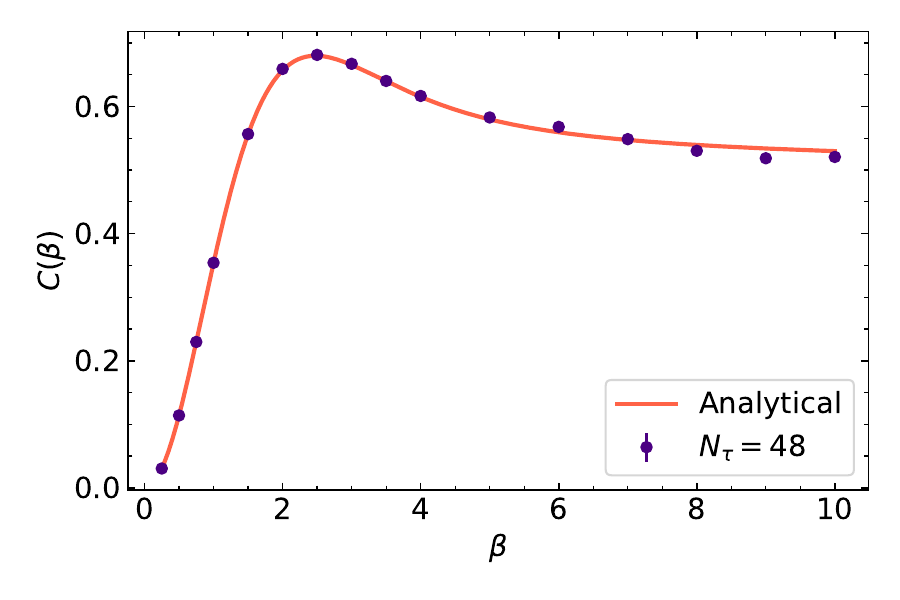}

    \vspace{0.3cm}

    \includegraphics[width=0.50\textwidth]{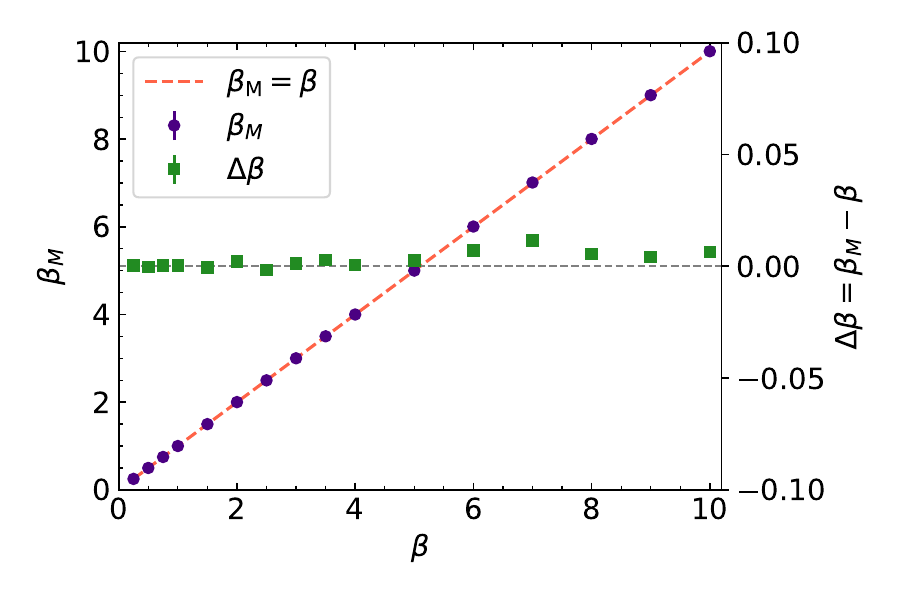}

    \caption{Thermodynamic observables for the one-dimensional U(1) lattice theory on a 48-site lattice. The energy $E(\beta)$ (top), specific heat $C(\beta)$ (middle), and the measured configurational $\beta_M$ (bottom) are plotted against $\beta$. $E(\beta)$ and $C(\beta)$ are benchmarked against the analytical results given in Eqs.~\eqref{eq:energy_ana_1d} and~\eqref{eq:spec_ana_1d}, respectively. The plot of $\beta_M$ against $\beta$ shows the expected behavior.}
    \label{fig:1d_u1_l48}
\end{figure*}

We perform simulations on a lattice with 48 sites. 
Fig.~\ref{fig:1d_u1_l48} shows the thermodynamic observables of the theory.  
The lattice energy $E(\beta)$, defined in Eq.~\eqref{eq:energy_lat_1d_u1}, and the specific heat, computed from energy fluctuations as 
\begin{equation}
C(\beta) = \beta^2 N_\tau \left( \langle E^2 \rangle - \langle E \rangle^2 \right), \nonumber
\end{equation}
are plotted as functions of $\beta$, and compared with the analytical results in Eqs.~\eqref{eq:energy_ana_1d} and~\eqref{eq:spec_ana_1d}. 
In addition, we show the configurational temperature estimator $\beta_M$, obtained using Eq.~\eqref{eq:beta_lat}, and verify its agreement with $\beta$ across a wide range. 
(When we say agreement with $\beta$, we mean $\beta_M \approx \beta$ or $\beta^L_{\rm config} \approx 1$.) 
To better visualize the proximity between $\beta$ and the estimated $\beta_M$, we also show the difference $\Delta\beta = \beta_M - \beta$ on a secondary axis in Fig.~\ref{fig:1d_u1_l48} (right). 
The values of $\Delta\beta$ are close to zero across the entire range, indicating consistency between input and output couplings. 
The consistency of the numerical results with analytic predictions provides a strong validation of both our implementation and the effectiveness of the temperature estimator. 
The values corresponding to Fig.~\ref{fig:1d_u1_l48} are given in Table~\ref{tab:1d_u1_l48} in Section \ref{sec:table_data}. 

\begin{figure*}[!t]
\centering
\includegraphics[width=0.3\textwidth]{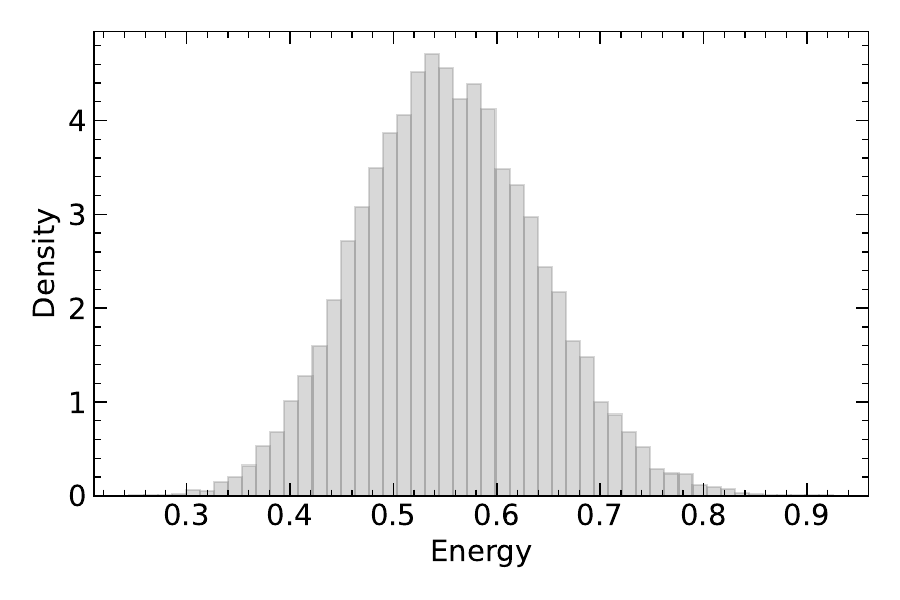}
\includegraphics[width=0.3\textwidth]{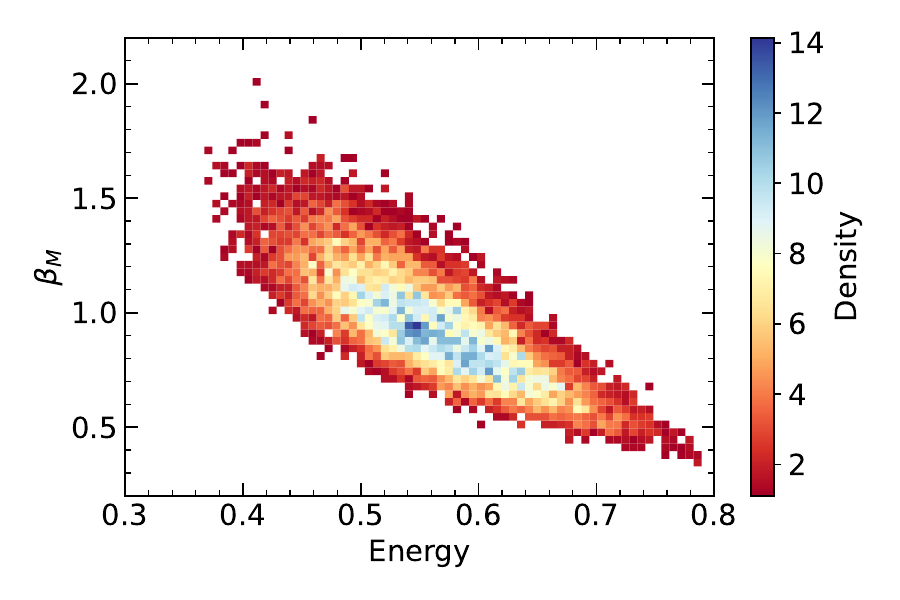}
\includegraphics[width=0.3\textwidth]{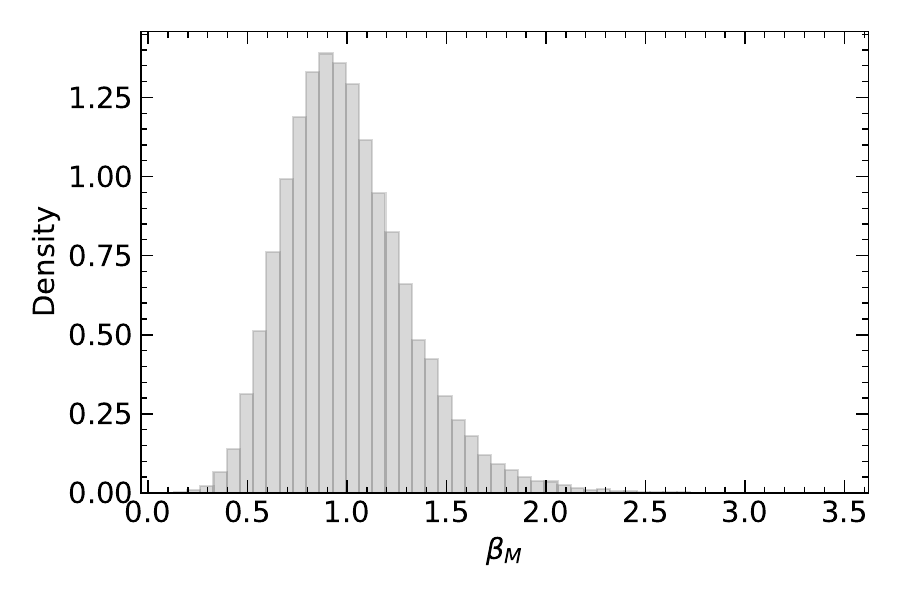}
\includegraphics[width=0.3\textwidth]{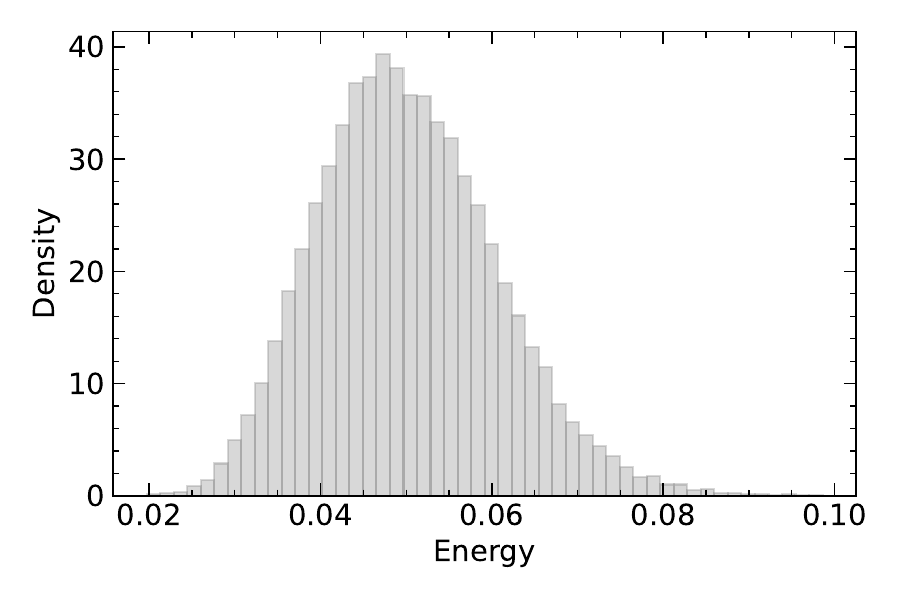}
\includegraphics[width=0.3\textwidth]{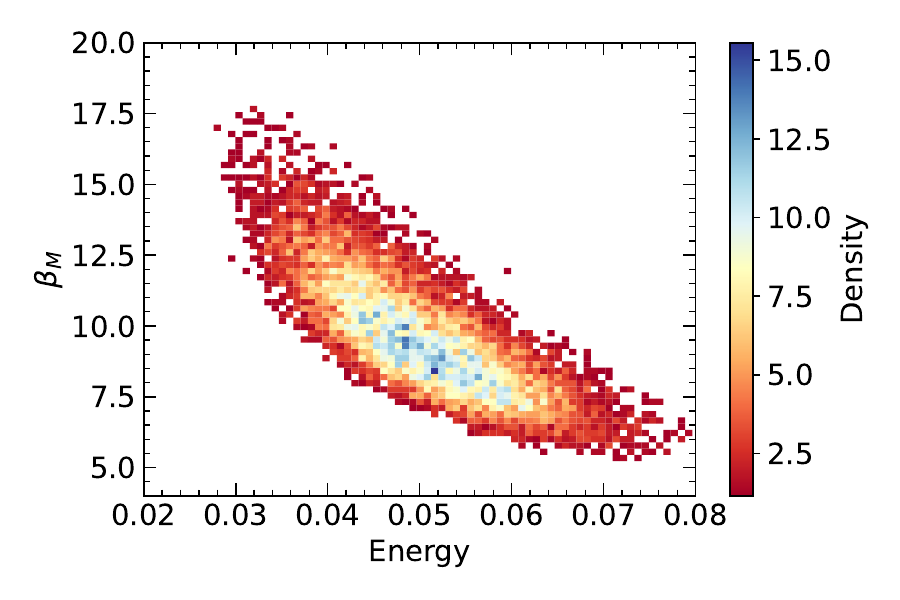}
\includegraphics[width=0.3\textwidth]{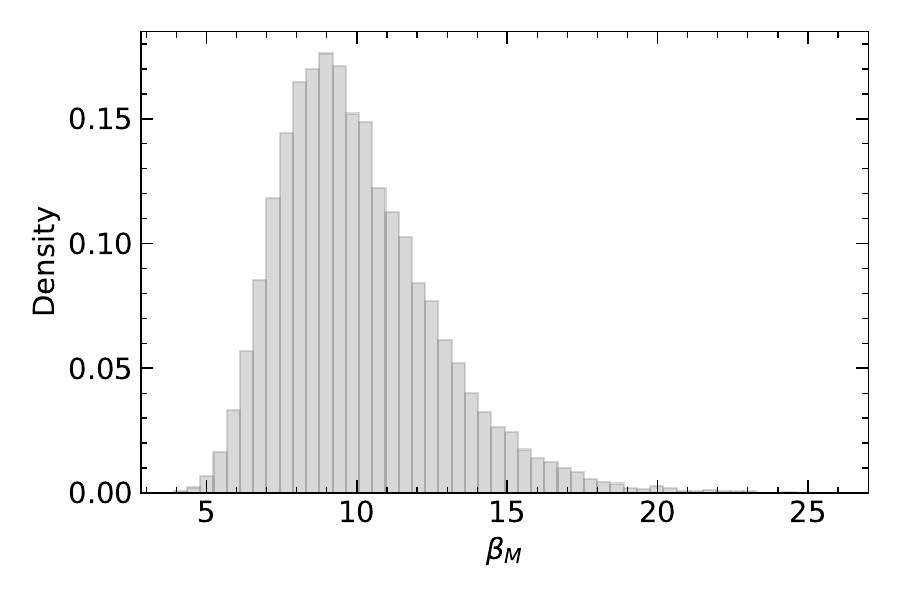}
\caption{Thermodynamic observables of the one-dimensional U(1) lattice gauge theory on a lattice with 48 sites. 
The figure presents histograms of the energy (left), $\beta_M$ (right), and scatter plots in the form of a heat-map illustrating the relationship between energy and $\beta_M$ (middle). 
The top three plots correspond to $\beta = 1$, while the bottom three are for $\beta = 10$. 
These plots highlight the fluctuations around the average values, with the heat-map scatter plots demonstrating how energy shifts as the configurational temperature deviates from the true value (which is equal to unity).
\label{fig:1d_u1_l48_combined}}
\end{figure*}

Figure \ref{fig:1d_u1_l48_combined} presents the thermodynamic observables for a fixed $\beta$. 
The histograms of energy and $\beta_M$ highlight fluctuations around their average values, reflecting statistical variations inherent in the Monte Carlo sampling. 
The heat-map scatter plots illustrate the relationship between energy and $\beta_M$, demonstrating how energy shifts when the temperature deviates from the true value. 

The system does not always maintain a constant configurational temperature in numerical simulations due to statistical fluctuations. 
It may become trapped in local vacua or metastable states distinct from the true ground state. 
As shown in the plots, the configurational temperature estimator method tracks these fluctuations, offering a more accurate estimate of the actual temperature during the simulation. 
By comparing the $\beta_M$ with $\beta$, we can assess the reliability of the simulations and ensure the system is not stuck in a false vacuum. 

\begin{figure}[h]
\centering
\includegraphics[width=0.6\textwidth]{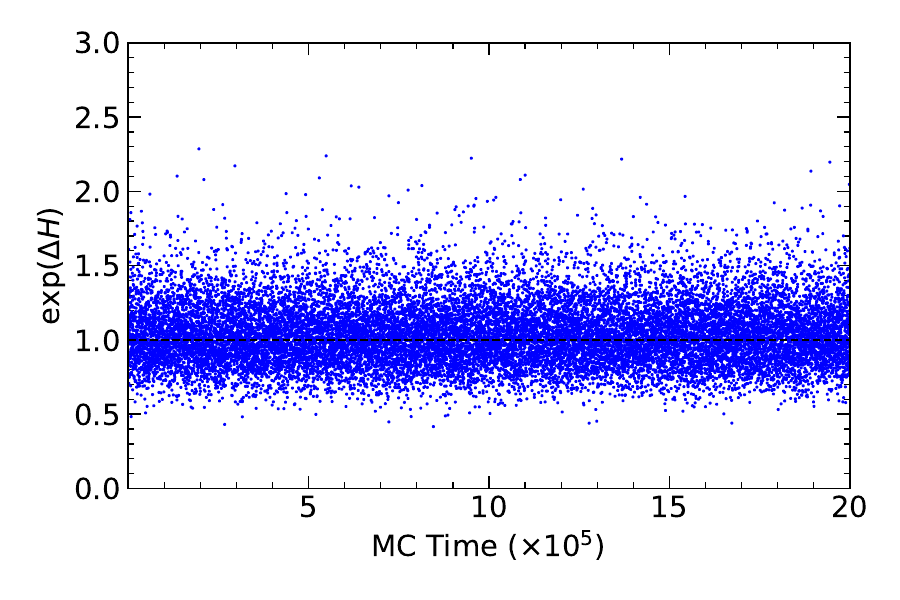}
\caption{Time series of $\exp(-\Delta H)$ for the one-dimensional U(1) lattice gauge theory on a 48-site lattice. 
We used $\beta = 3$.}
\label{fig:1d_u1_timeseries}
\end{figure}

Figure~\ref{fig:1d_u1_timeseries} shows the Monte Carlo time series of the Metropolis factor $\exp(-\Delta H)$ and $\beta_M$ of the theory at $\beta = 3$. 
The data correspond to configurations recorded after thermalization. 
The $\exp(- \Delta H)$ values fluctuate around unity, serving as a consistency check for the Monte Carlo field updates, while $\beta_M$ tracks the estimated configurational temperature across Monte Carlo time. 

\begin{figure}[h]
\centering
\includegraphics[width=0.6\textwidth]{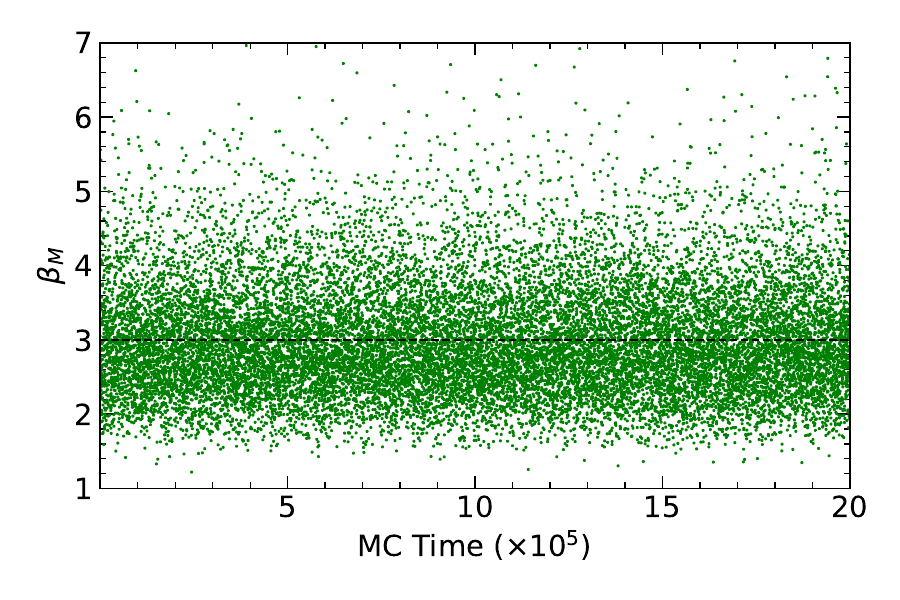}
\caption{Time series of $\beta_M$ for the one-dimensional U(1) lattice gauge theory on a 48-site lattice. 
We used $\beta = 3$.}
\label{fig:1d_u1_timeseries_beta_M}
\end{figure}

In Fig. \ref{fig:1d_u1_timeseries_beta_M} we show the time series plot of $\beta_M$. 
We see that $\beta_M$ fluctuates around $\beta = 3$, suggesting that the we are sampling from the correct distribution. 

\section{Two-Dimensional Theory}
\label{sec:2d_temp_est}

The compact U(1) lattice gauge theory in two Euclidean dimensions is defined by the Wilson action
\begin{equation}
S = \beta \sum_{(x, y)} \left[1 - \cos \theta_p(x, y)\right]. 
\label{eq:u1_2d_action}
\end{equation}
Here, $\theta_p(x, y)$ is the plaquette angle at site $(x, y)$, constructed from the oriented sum of link variables around the elementary square:
\begin{eqnarray}
\theta_p(x, y) &=& \theta_x(x, y) + \theta_y(x + 1, y) - \theta_x(x, y + 1) - \theta_y(x, y),
\end{eqnarray}
with each link variable $\theta_{x/y}(x, y) \in (-\pi, \pi]$ defined on the corresponding directed edge of the lattice. 

To extract an effective inverse temperature from field configurations, we study the local response of the action to variations in the link angles. 
The gradient of the action with respect to the $x$-oriented link $\theta_x(x, y)$ and the $y$-oriented link $\theta_y(x, y)$ is given by
\begin{align}
g_x(x, y) &= \frac{1}{\beta} \frac{\partial S}{\partial \theta_x(x, y)} \nonumber \\ 
&= \sin(\theta_p(x, y)) - \sin(\theta_p(x, y - 1)), \\
g_y(x, y) &= \frac{1}{\beta} \frac{\partial S}{\partial \theta_y(x, y)} \nonumber \\ 
&= \sin(\theta_p(x - 1, y)) - \sin(\theta_p(x, y)).
\end{align} 

The corresponding diagonal components of the Hessian, which represent the second derivatives of the action with respect to the same link variable, are
\begin{align}
h_{xx}(x, y) &= \frac{1}{\beta} \frac{\partial^2 S}{\partial \theta_x(x, y)^2} \nonumber \\ 
&= \cos(\theta_p(x, y)) + \cos(\theta_p(x, y - 1)), \\
h_{yy}(x, y) &= \frac{1}{\beta}  \frac{\partial^2 S}{\partial \theta_y(x, y)^2} \nonumber \\ 
&= \cos(\theta_p(x, y)) + \cos(\theta_p(x - 1, y)).
\end{align} 

Using the previously defined expressions for the gradients and diagonal Hessians with respect to the link angles, we have
\begin{align}
\hat{\beta} &= \frac{\sum_{x, y} \left[ h_{xx}(x, y) + h_{yy}(x, y) \right]}{\sum_{x, y} \left[ \left(g_x(x, y)\right)^2 + \left(g_y(x, y)\right)^2 \right]} \nonumber \\ 
&- \frac{2 \sum_{x, y, x', y'} \left[ g_x(x, y) \, g_y(x', y') \, h_{xy}(x, y; x', y') \right]}{\left( \sum_{x, y} \left[ \left(g_x(x, y)\right)^2 + \left(g_y(x, y)\right)^2 \right] \right)^2}.
\label{eq:beta_lat_2d}
\end{align}
In the above, the first term captures contributions from the diagonal elements of the Hessian, and the second term includes cross-derivative (off-diagonal) contributions. 
The notation is consistent with the site and direction-dependent link angles $\theta_x(x, y)$ and $\theta_y(x, y)$, and their corresponding local derivatives of the action. 

In the two-dimensional theory, where spatial directions are treated symmetrically, we can also independently analyze the contributions from $x$ and $y$-oriented links. 
In particular, evaluating only the first term in Eq.~\eqref{eq:beta_lat_2d} for either the $x$ or $y$-oriented links provides a reliable estimator that approaches the true coupling in the limit of large lattice volumes. 
To incorporate the contribution from off-diagonal Hessian components in the estimator of the effective coupling $\hat\beta$, we evaluate the crossed terms involving mixed second derivatives of the action. 
For each $x$-oriented link $\theta_x(x, y)$, the second derivative $\partial^2 S / \partial \theta_x(x, y) \partial \theta_y(x', y')$ is nonzero only when $\theta_y(x', y')$ shares a plaquette with $\theta_x(x, y)$. 
There are four such $\theta_y$ links associated with $\theta_x(x, y)$ through the two adjacent plaquettes: at $(x, y)$ and $(x, y - 1)$. 
The corresponding nonzero mixed second derivatives are:
\begin{align}
\frac{1}{\beta} \frac{\partial^2 S}{\partial \theta_x(x, y) \partial \theta_y(x, y)} &= - \cos(\theta_p(x, y)), \\
\frac{1}{\beta} \frac{\partial^2 S}{\partial \theta_x(x, y) \partial \theta_y(x + 1, y)} &= + \cos(\theta_p(x, y)), \\
\frac{1}{\beta} \frac{\partial^2 S}{\partial \theta_x(x, y) \partial \theta_y(x, y - 1)} &= + \cos(\theta_p(x, y - 1)), \\
\frac{1}{\beta} \frac{\partial^2 S}{\partial \theta_x(x, y) \partial \theta_y(x + 1,y - 1)} &= - \cos(\theta_p(x, y - 1)).
\end{align}
Each of these terms contributes to the total crossed term in the estimator as defined in Eq.~\eqref{eq:beta_lat_2d}. 
In it $h_{xy}(x, y; x', y')$ is nonzero only for the four neighboring $y$-oriented links $(x', y')$ listed above. 

In addition to local observables, we aim to extract non-local physical quantities, such as the string tension, by evaluating rectangular Wilson loops. 
In the exactly solvable two-dimensional U(1) lattice gauge theory, the expectation value of a Wilson loop $W(R, T)$ of spatial extent $R$ and temporal extent $T$ is given by \cite{Rothe:1987tc}
\begin{equation}
\langle W(R, T) \rangle = \left( \frac{I_1(\beta)}{I_0 (\beta)} \right)^{R T}.
\label{eq:wilson_loop_2d}
\end{equation}
This result demonstrates the area-law behavior explicitly, with the effective string tension defined as
\begin{equation}
\sigma = - \ln \left( \frac{I_1(\beta)}{I_0(\beta)} \right).
\end{equation}
In particular, taking $R = T = 1$ yields the plaquette expectation value,
\begin{equation}
P \equiv \langle \cos \theta_p \rangle = \langle W(1, 1) \rangle = \frac{I_1(\beta)}{I_0(\beta)}.
\label{eq:plaq_2d}
\end{equation}

\begin{figure*}[!t]
\centering
\includegraphics[width=0.60\textwidth]{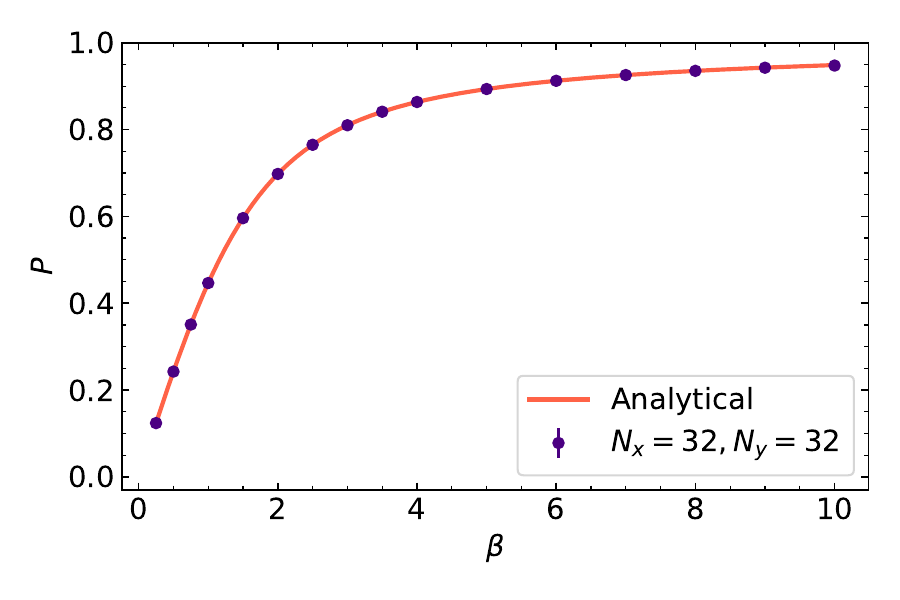}
\par\vspace{0.3cm}   
\includegraphics[width=0.60\textwidth]{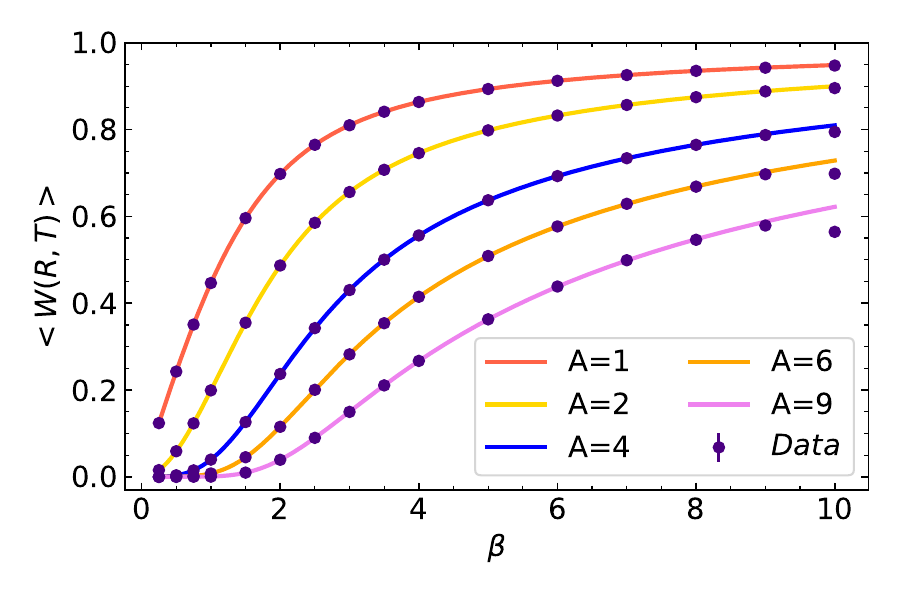}
\par\vspace{0.3cm}
\includegraphics[width=0.60\textwidth]{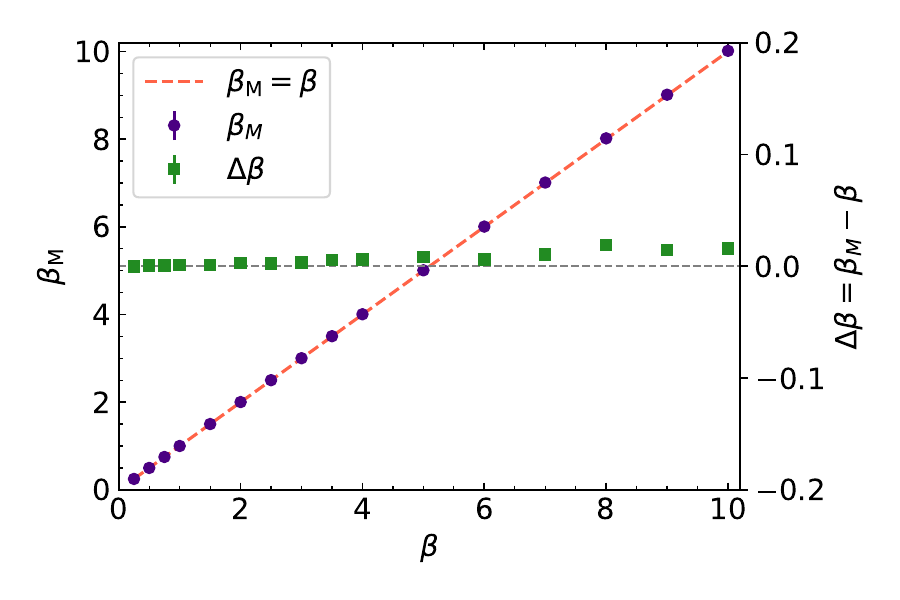}
\caption{Two-dimensional U(1) lattice gauge theory. The expectation value of plaquette (top), Wilson loops of sizes $A = R \times T$ (middle) and estimated inverse temperature $\beta_M$ as functions of input $\beta$ are shown on a $32 \times 32$ lattice (bottom).}
\label{fig:plaquette_beta_2d}
\end{figure*}

We show the plaquette expectation value $P$ and the estimated inverse temperature $\beta_M$ in Fig.~\ref{fig:plaquette_beta_2d} for a $32 \times 32$ lattice. 
We also detail the results in Table~\ref{tab:2d_u1_l32_l32} in Section \ref{sec:table_data}. 
Beyond the plaquette, we can test the consistency of the field configurations by evaluating Wilson loops of larger sizes. 
Instead of only the $1 \times 1$ plaquette, we compute expectation values of rectangular Wilson loops $W(R, T)$ with various $R$ and $T$ on the same $32 \times 32$ lattice. 
These are then compared with the corresponding analytical predictions as given in Eq. \eqref{eq:wilson_loop_2d} in Fig.~\ref{fig:plaquette_beta_2d}. 
While the analytical expression for the Wilson loop is valid for all values of $\beta$, in practice, numerical deviations can appear at large $\beta$, especially for Wilson loops with large area $R T$. 
This behavior is evident in Fig.~\ref{fig:plaquette_beta_2d}. 
For a large area $R T$, even a slight deviation from unity in the base gets exponentially amplified in the exponent. 
For instance, if we write $\left(1 - \epsilon\right)^{R T} \approx e^{-\epsilon R T}$, then even a tiny $\epsilon > 0$ leads to significant suppression when $R T$ is large. 
It explains why Wilson loop values can become extremely small and prone to numerical underestimation in practice, particularly at large $\beta$ and for large loop areas. 

The results, shown in Fig.~\ref{fig:plaquette_beta_2d}, demonstrate that the estimated configurational temperature values are consistent with the expected value, confirming the reliability of the sampling process. 
Although scatter plots between output observables and $\beta_M$ are not shown here for the two-dimensional case, as we did in the one-dimensional case in Fig.~\ref{fig:1d_u1_l48_combined}, such plots are helpful to visualize how fluctuations and metastable states affect the sampling process during simulations. 
We note that $\beta_M$ remains an important tool to monitor these effects and assess the reliability of the results. 

\begin{figure}[!t]
\centering
\includegraphics[width=0.60\textwidth]{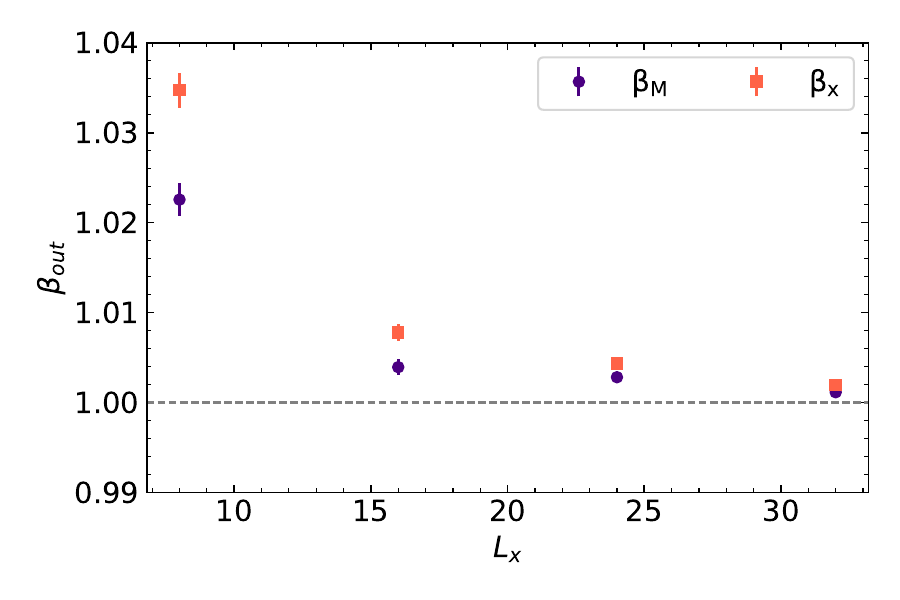}
\includegraphics[width=0.65\textwidth]{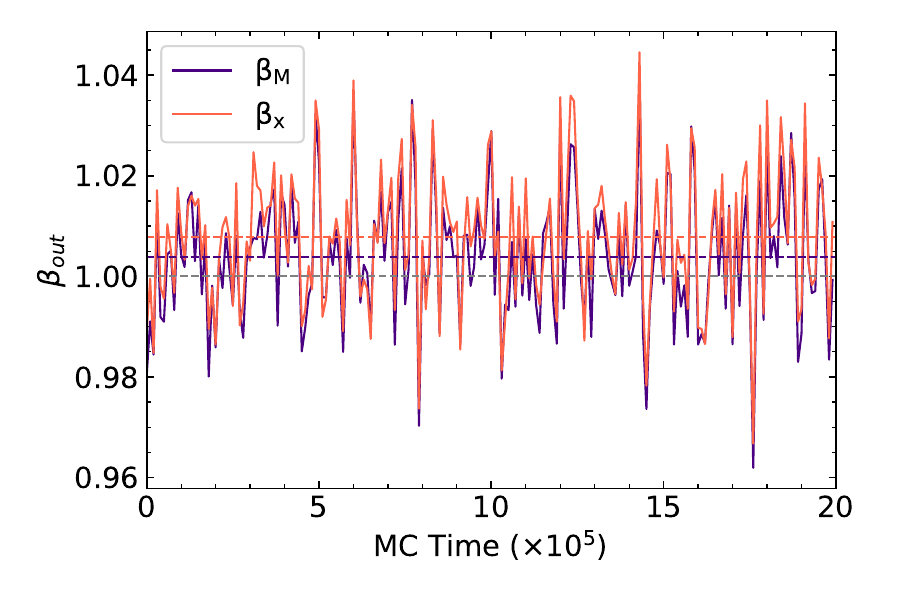}
\caption{\label{fig:beta_vol_2d} (Above) Comparison of the full estimator $\beta_M$ and the directional estimator $\beta_x$ as a function of lattice size $L_x$ for $L_x \times L_x$ lattices at $\beta = 1$. 
(Below) The Monte Carlo time history of both estimators on a $16 \times 16$ lattice illustrates fluctuations and near agreement between the two estimators over simulation time. 
The dotted lines correspond to the average values of the respective observables over the Monte Carlo history.}
\end{figure}

To analyze the dominant contributions to the extracted configurational temperature, we consider a simplified estimator 
\begin{equation}
\beta_x = \left \langle \frac{\sum_{x, y} h_{xx}(x, y)}{\sum_{x, y} \left(g_x(x, y)\right)^2} \right \rangle,
\end{equation} 
constructed solely from the diagonal terms involving derivatives with respect to $x$-oriented links. 
By comparing this with the full estimator $\beta_M$, which includes diagonal and off-diagonal Hessian contributions, we analyze the relative importance of the off-diagonal terms. 
At larger lattice volumes, the diagonal terms dominate, and $\beta_x$ approximates $\beta$. 
However, the off-diagonal terms become more significant for smaller volumes and must be included for a more accurate estimate. 
In Fig.~\ref{fig:beta_vol_2d}, we show the behavior of the full estimator $\beta_M$ and the simplified directional estimator $\beta_x$ for $\beta = 1$ across different lattice sizes. 
This comparison illustrates that the off-diagonal terms contribute very little to the estimator for larger volumes, and $\beta_M$ closely approaches $\beta$. 
In contrast, even the full estimator $\beta_M$ for small volumes deviates from the expected value, indicating that the configurational temperature estimator becomes more reliable in the large-volume limit.

Hence, for larger lattice volumes, one can reliably use the simpler directional estimator as a proxy for the configurational temperature, since it provides an accurate approximation in the large-volume limit where the off-diagonal contributions become negligible. 
In the 4D case considered next, which effectively corresponds to a larger hypervolume than 2D, we omit the off-diagonal terms in the estimator to reduce computational complexity and because their contribution is expected to be negligible in the larger 4D space. 

\section{Four-Dimensional Theory} 

The four-dimensional compact U(1) lattice gauge theory is one of the simplest gauge theories exhibiting a nontrivial phase transition. 
Despite its apparent simplicity, the theory features a rich phase structure, transitioning from a confining phase at strong coupling to a Coulomb-like phase at weak coupling. 
This makes it a valuable testbed for exploring finite-temperature behavior and gauging the performance of our temperature estimator in a controlled but nontrivial setting. 

The compact U(1) lattice gauge theory in four Euclidean dimensions is defined by the Wilson action
\begin{equation}
S = \beta \sum_x \sum_{\mu < \nu} \left[ 1 - \cos\theta_{\mu\nu}(x) \right],
\label{eq:u1_4d_action}
\end{equation}
where $x$ labels the lattice sites and the indices $\mu, \nu \in \{0,1,2, 3\}$ denote the directions in Euclidean spacetime. 
The plaquette angle $\theta_{\mu\nu}(x)$ is constructed as the oriented sum of link variables around the elementary square in the $\mu$-$\nu$ plane:
\begin{equation}
\theta_{\mu \nu}(x) = \theta_\mu(x) + \theta_\nu(x + \hat\mu) - \theta_\mu(x + \hat\nu) - \theta_\nu(x),
\end{equation}
with $\theta_\mu(x) \in (-\pi, \pi]$ being the compact U(1) link variable along the $\mu$-direction at site $x$.

There are $\binom{4}{2} = 6$ such plaquettes per site, corresponding to the planes: $01$, $02$, $03$, $12$, $13$, and $23$. 
The sum over $\mu < \nu$ ensures each plaquette is counted once. 
Each link contributes to six plaquettes, and the local structure of the action allows us to compute its derivatives with respect to individual link angles to obtain local observables such as the gradient and the Hessian. 

For this theory, there is a phase transition around $\beta \approx 1.01$, which has been shown through Monte Carlo simulations in the literature \cite{Arnold:2000hf, Bonati:2013ota, Bhanot:1981zg, Lautrup:1980xr}. 
In this work, our primary objective is not to obtain a high-precision determination of the critical coupling. 
Instead, we aim to evaluate the effectiveness of our proposed temperature estimator in capturing the thermal behavior of the system across a broad range of couplings. 
By applying the estimator to gauge field configurations generated over a wide range of $\beta$, we analyze its robustness and physical sensitivity, particularly its ability to signal transition-like behavior, without relying on traditional order parameters. 
We first investigate the phase structure of this theory by tuning the coupling $\beta$ and studying various gauge-invariant observables that characterize the confinement-deconfinement transition. 
The average plaquette,
\begin{equation}
\langle P \rangle = \frac{1}{6V} \sum_x \sum_{\mu < \nu} \cos\theta_{\mu\nu}(x),
\end{equation} 
serves as a basic diagnostic of gauge field fluctuations and exhibits a sharp increase near the critical coupling. 
We also monitor its susceptibility
\begin{equation}
\chi_P = V \left( \langle P^2 \rangle - \langle P \rangle^2 \right),
\end{equation} 
which enhances sensitivity to critical behavior. 

In addition to these conventional observables, we define the configurational temperature estimator:
\begin{equation}
\hat{\beta} = \frac{\sum_x \sum_\mu h_{\mu \mu}(x)}{\sum_x \sum_\mu \left[ g_\mu(x) \right]^2},
\end{equation} 
where $g_\mu(x)$ and $h_{\mu\mu}(x)$ denote the local gradient and diagonal Hessian of the action with respect to the link variable $\theta_\mu(x)$. 
As discussed in the two-dimensional case, we neglect the contributions from the off-diagonal entries in the Hessian when constructing the estimator. 

The complete configurational-temperature estimator includes both the diagonal and the off-diagonal elements of the Hessian; in the four-dimensional version currently used, we have kept only the diagonal parts. This constitutes an approximation, not an exact simplification of the full Rugh estimator. Our reason for this choice is provided by the finite-volume study described in Section \ref{sec:2d_temp_est}, in which the full estimator and the directional estimator were compared as a function of lattice size. The contribution from the off-diagonal terms of the Hessian decreases as the volume of the lattice increases, whereas the diagonal estimator tends towards the desired value. We thus anticipate that the omitted terms will give a smaller contribution in the larger four-dimensional lattices that are being studied here. Yet their contribution is not exactly zero, and the approximation must therefore be seen as a computationally efficient way of implementing the estimator for large volumes rather than as an exact reformulation of the full estimator.

\begin{figure*}[!t]
\centering
\includegraphics[width=0.60\textwidth]{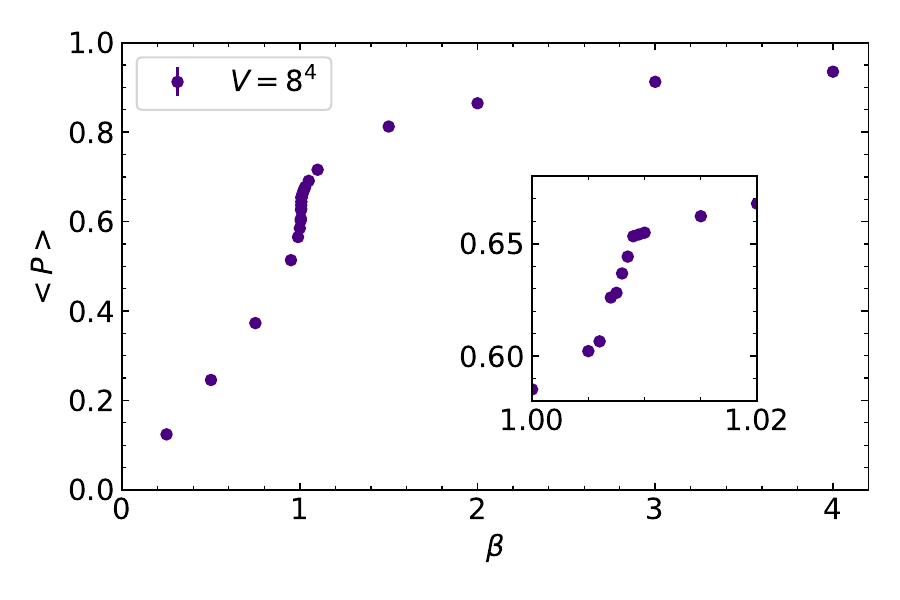}
\includegraphics[width=0.60\textwidth]{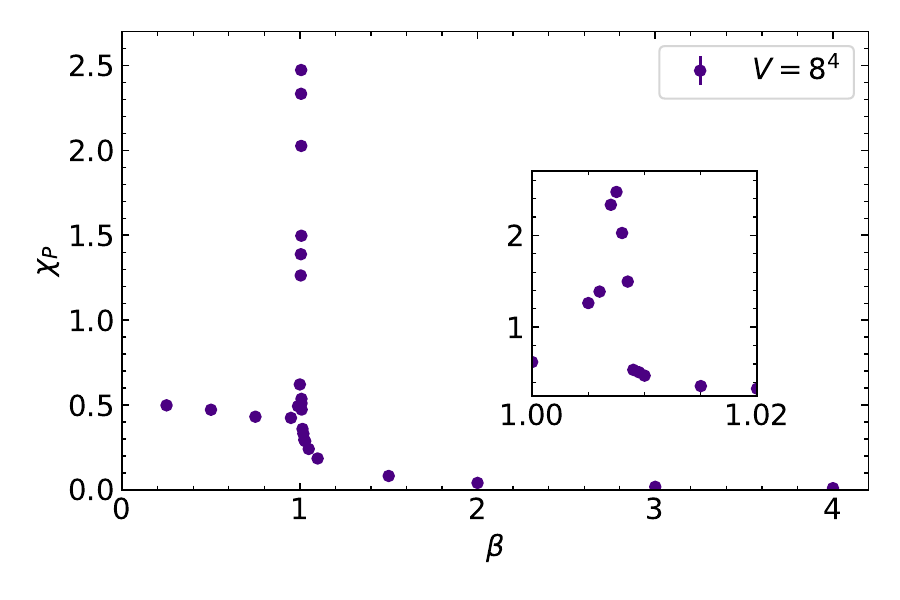}
\includegraphics[width=0.60\textwidth]{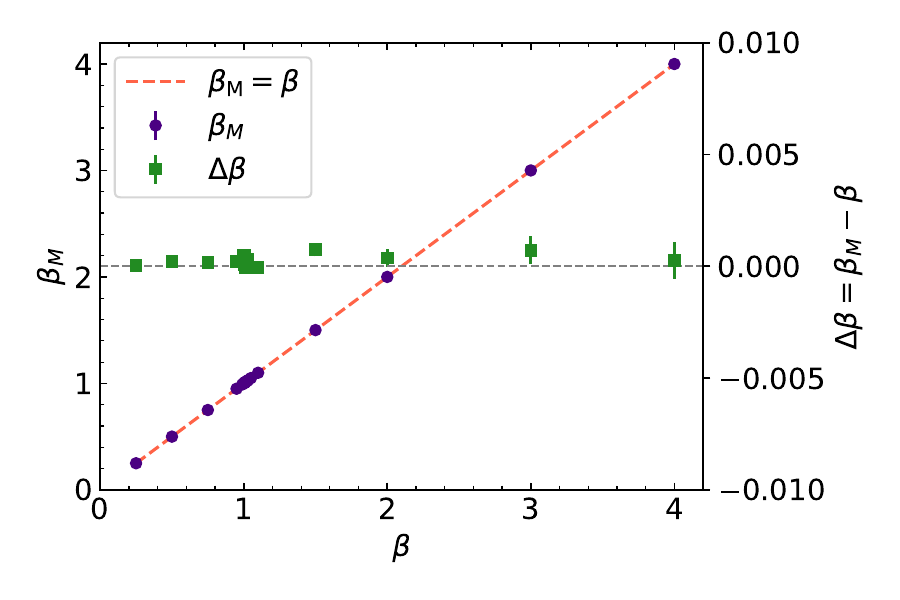} 
\caption{Thermodynamic observables for the four-dimensional U(1) lattice theory on an $8^4$ lattice. The plaquette $\langle P \rangle$ (top), the plaquette susceptibility $\chi_P$ (middle), and the estimated configurational temperature $\beta_M$ (bottom) are plotted against $\beta$.}
\label{fig:4d_u1_8} 
\end{figure*}

\begin{figure*}[ht]
\centering
\includegraphics[width=0.3\textwidth]{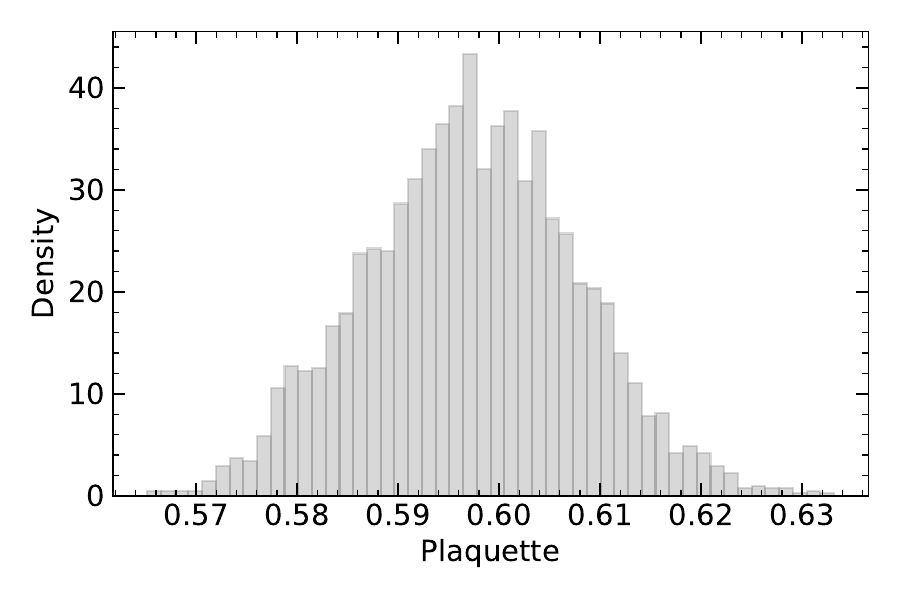}
\includegraphics[width=0.3\textwidth]{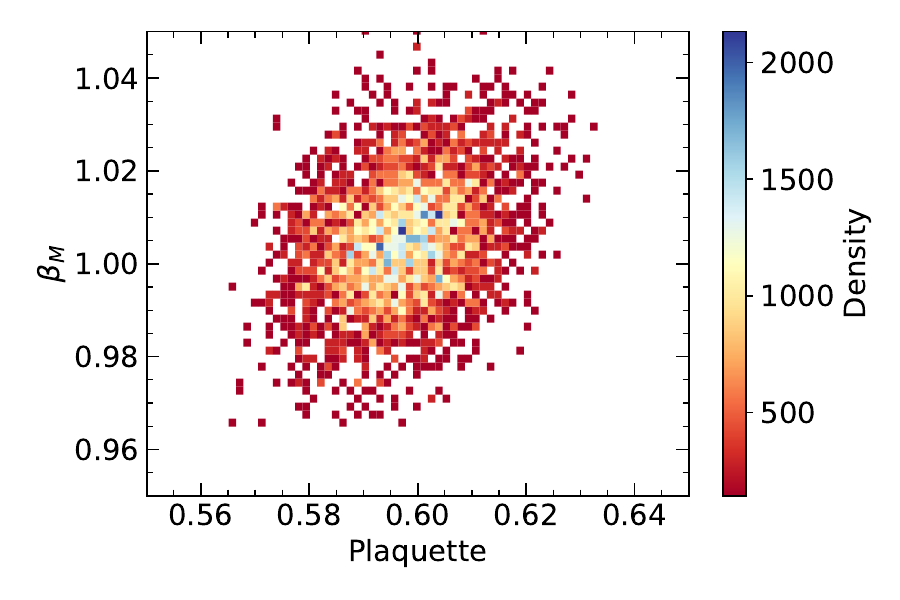}
\includegraphics[width=0.3\textwidth]{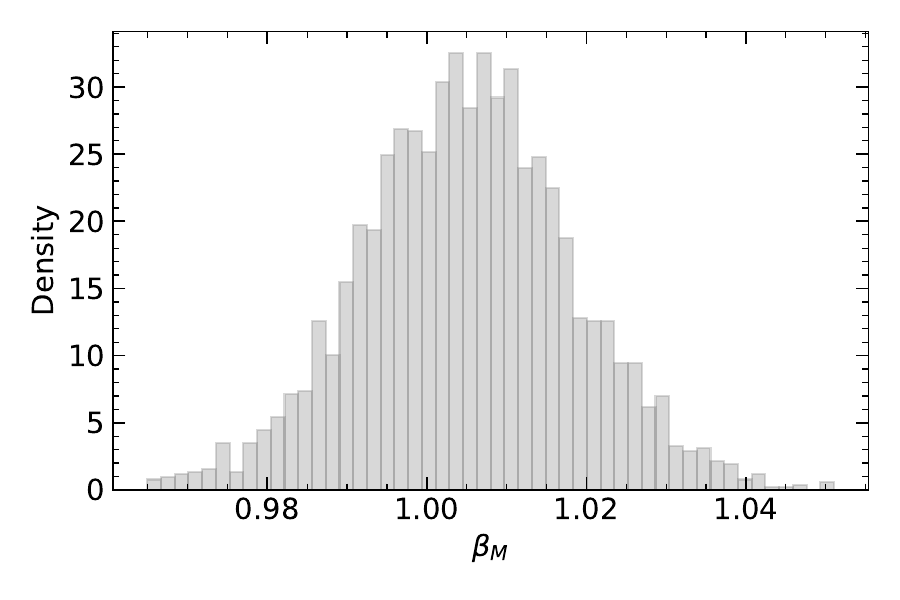}
\includegraphics[width=0.3\textwidth]{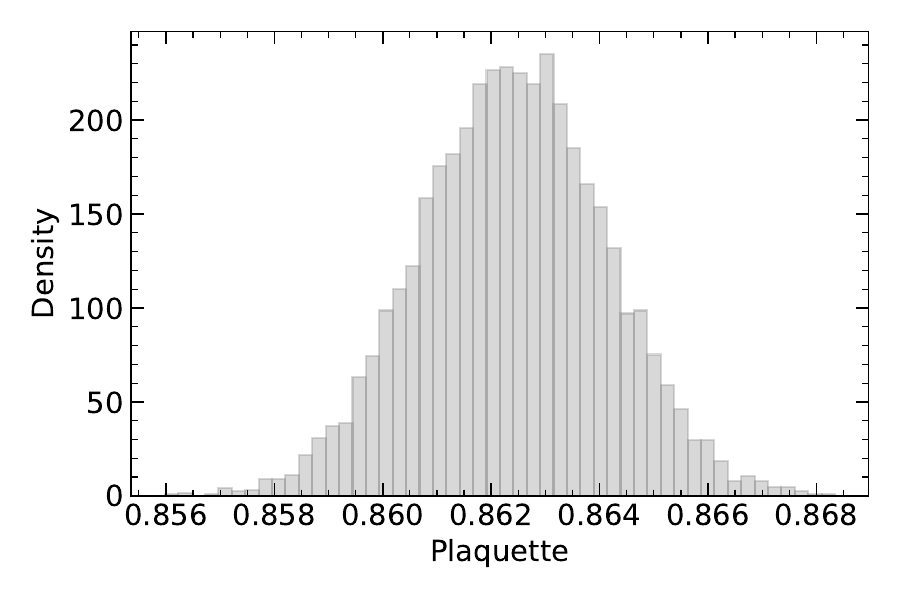}
\includegraphics[width=0.3\textwidth]{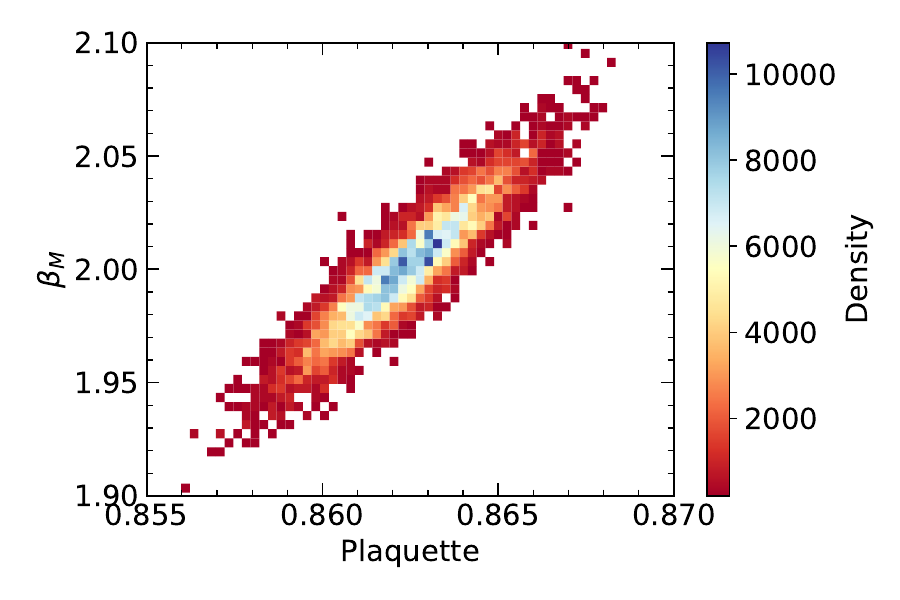}
\includegraphics[width=0.3\textwidth]{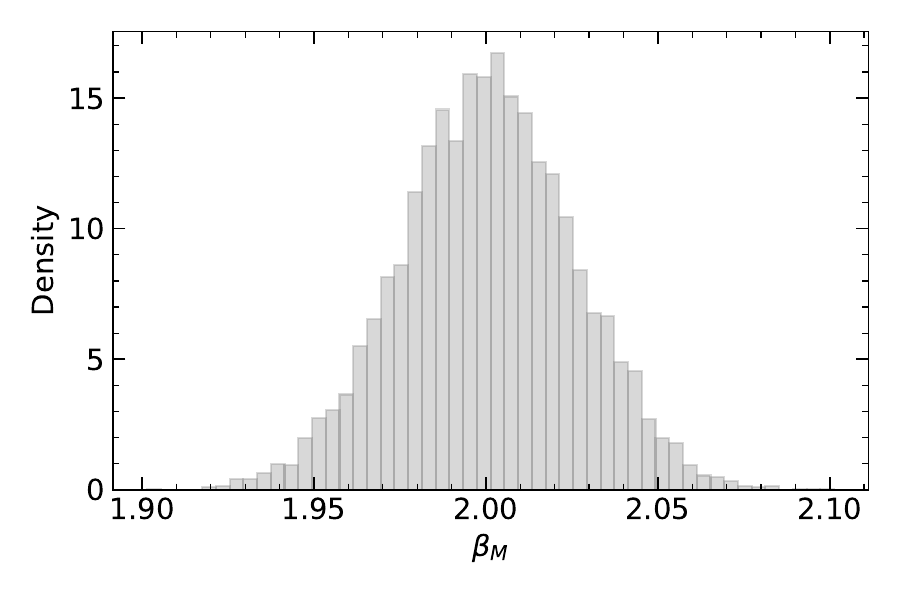}
\caption{Thermodynamic observables for the four-dimensional U(1) lattice theory with $V = 8^4$. 
The figure presents histograms of the plaquette (left), $\beta_M$ (right), and a scatter plot in the form of a heat-map illustrating the relationship between plaquette and $\beta_M$ (middle). 
The top three plots correspond to $\beta = 1.005$, the critical $\beta$, while the bottom three are for $\beta = 2.0$. 
\label{fig:4d_u1_combined}}
\end{figure*}

Fig. \ref{fig:4d_u1_8} presents the thermodynamic observables of the theory on an $8^4$ lattice. 
We show the plots of plaquette $\langle P \rangle$ (left), the plaquette susceptibility $\chi_P$ (middle), and the measured configurational temperature $\beta_M$ against $\beta$ (right). 
For this lattice size, our simulations locate the transition near $\beta \approx 1.0075$. 
Remarkably, even without incorporating the off-diagonal contributions of the Hessian, the measured estimator is close to $\beta$, with high accuracy. 
This agreement reinforces the correctness of the simulations and supports the robustness of the estimator.  
The values corresponding to Fig.~\ref{fig:4d_u1_8} are given in Table~\ref{tab:4d_u1_l8} in Section \ref{sec:table_data}. 
In Fig.~\ref{fig:4d_u1_combined} we show the histograms of the plaquette and $\beta_M$ that exhibit fluctuations around their mean values, reflecting statistical variations from Monte Carlo sampling. 
The central heat-map illustrates the correlation between plaquette and $\beta_M$, capturing how energy responds to fluctuations in the sampling process. 

We present the numerical observations from this four-dimensional case study, highlighting how the estimator reflects simulation behavior.

\begin{itemize}
\item \textit{Thermalization:} 
The estimator can be a valuable indicator for checking simulation thermalization. 
In the early stages of a Monte Carlo run, observables such as the plaquette and $\beta_M$ exhibit similar thermalization behavior, gradually settling into stable distributions, as seen in Fig.~\ref{fig:therm}. 
Even with early statistics, the estimator $\beta_M$ tracks this transition reliably, making it a useful diagnostic tool for probing thermalization in the simulation.

\begin{figure}[ht]
\centering
\includegraphics[width=0.60\textwidth]{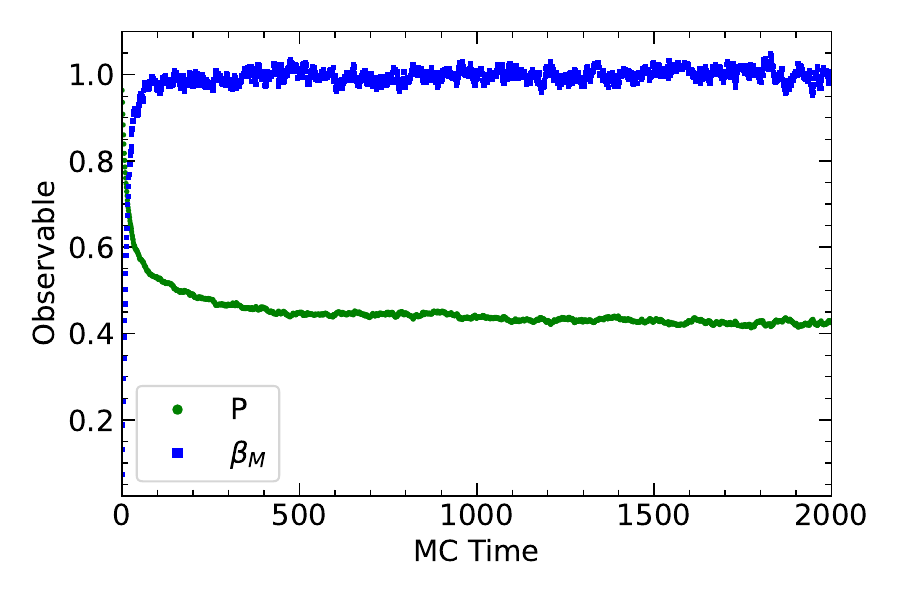}
\caption{The initial part of Monte Carlo time history, including thermalization, of the plaquette and $\beta_M$ on an $8^4$ lattice at $\beta = 1.0$.
\label{fig:therm}}
\end{figure}

\item \textit{Error diagnosis:} 
The estimator can also be used as a diagnostic tool to identify implementation errors in the simulation. 
As a demonstration, we intentionally introduce a flaw in the Metropolis accept-reject step by drawing the random number from a uniform distribution in the range $[0.5, 1.5]$ instead of the correct interval $[0, 1]$. 
This modification alters the acceptance probability, leading to systematic deviations in the simulation outcome. 
In particular, the extracted value of $\beta_M$ from the estimator no longer matches the input value used in the simulation. 
This discrepancy is visible in Fig.~\ref{fig:check}, where we compare the estimator results from the correct and incorrect implementations.

\begin{figure}[ht]
\centering
\includegraphics[width=0.60\textwidth]{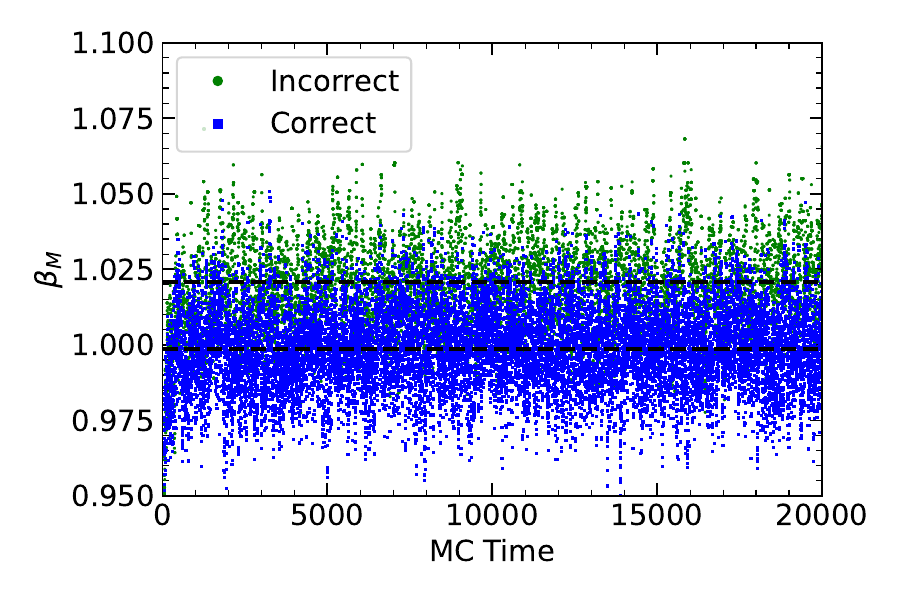}
\caption{Comparison of the extracted $\beta_M$ from simulations using the correct and incorrect Metropolis updates, performed at input $\beta = 1.0$. 
The estimator captures the deviation in the incorrect case, demonstrating its sensitivity to errors in the algorithm.
\label{fig:check}}
\end{figure}

\item \textit{Phase transition order parameter?:} 
While the estimator $\beta_M$ serves well as a diagnostic tool for thermalization and error detection, it is unsuitable as an order parameter for detecting phase transitions. 
In Fig.~\ref{fig:order}, we show the Monte Carlo time histories of the plaquette and $\beta_M$ at the critical coupling $\beta = 1.0075$. 
The plaquette exhibits clear signs of phase coexistence, indicating the presence of a first-order transition, whereas $\beta_M$ remains smooth and does not indicate distinct phases. 
This demonstrates that $\beta_M$ is insensitive to phase separation and is therefore not appropriate as an order parameter for studying critical behavior.

\begin{figure}[ht]
\centering
\includegraphics[width=0.60\textwidth]{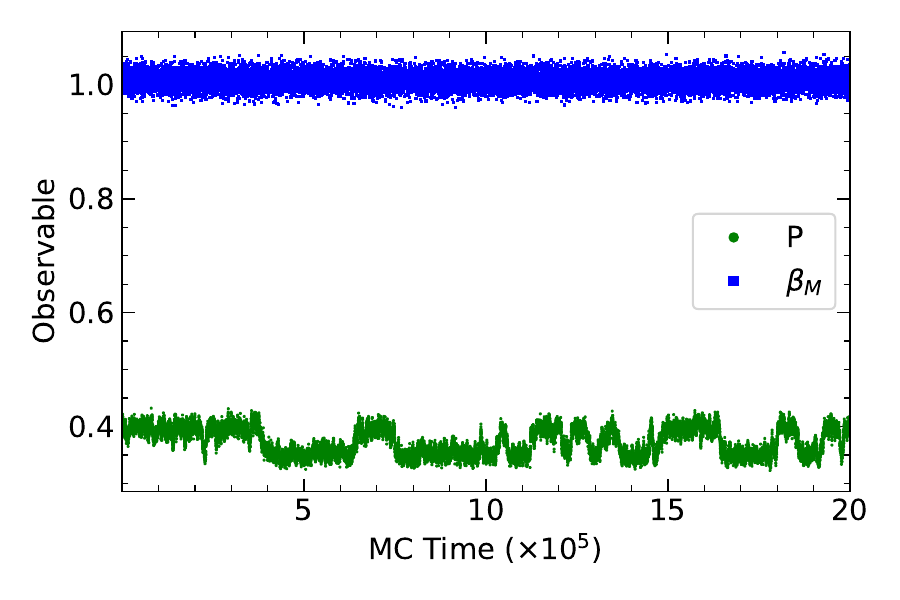}
\caption{Monte Carlo time histories of the plaquette and $\beta_M$ at the critical coupling $\beta = 1.0075$ on an $8^4$ lattice. 
\label{fig:order}}
\end{figure}

\end{itemize}

\section{Conclusions and Future Directions}
\label{sec:conc}

In this work, we build on Rugh’s geometric formulation of temperature in the microcanonical ensemble \cite{PhysRevLett.78.772}, which expresses temperature in terms of phase-space geometry. 
We extend Rugh’s formulation to Euclidean quantum field theories. 
The observable that captures the configurational temperature is made out of the gradient and Hessian of the Euclidean action.
The estimator is local and gauge invariant.
It is independent of momenta and thus suitable for Monte Carlo simulations.

Our investigations across 1D, 2D, and 4D compact U(1) lattice gauge theories demonstrate that the configurational estimator can be used as a sampling consistency check for the theories across various couplings and lattice sizes. 
The configurational temperature estimator closely matched the expected behavior in the analytically tractable 1D and 2D models, validating its stability and sensitivity. 
In the 4D setting, we have a system that undergoes a phase transition. 
Here, the estimator also follows the expected behavior, for various coupling values, across the transition. 

The configurational temperature estimator provides a complementary diagnostic that can reveal subtle violations of equilibrium, such as metastability or poor sampling. 
The standard observables like the average plaquette will eventually reflect the correct sampling distribution upon complete equilibration.
However, they may appear deceptively stable in partially thermalized or poorly sampled ensembles. 
In such cases, the configurational temperature can act as an early indicator of underlying inconsistencies. 
Moreover, we demonstrated that for large lattices, the off-diagonal Hessian contributions to the estimator become negligible.
This enables simplified implementations without significant loss of accuracy. 

Building on the success of the current study, we note that several promising avenues can emerge: $(i.)$ {\it Extension to non-Abelian gauge theories:} Applying the estimator to SU($N$) lattice gauge theories, particularly in QCD-like settings, would test its robustness in more complex, non-Abelian configurations and offer new tools for diagnosing confinement and chiral symmetry breaking; $(ii.)$ {\it Anisotropic and finite-temperature lattices:} The estimator can be leveraged to independently verify sampling anisotropies, especially in systems with nonuniform temporal and spatial lattice spacings, or for simulations at finite temperature and chemical potential; $(iii.)$ {\it Integration into HMC and ML-accelerated algorithms:} Embedding the estimator into HMC or machine learning-guided samplers could offer real-time feedback on sampling accuracy, potentially improving convergence and reducing computational overhead; and $(iv.)$ {\it Investigation of critical slowing down and topological freezing:} By tracking fluctuations around the target value, the estimator could provide new insights into autocorrelation times near criticality or topological sectors that are notoriously hard to sample. 

In finite-temperature QCD, the inverse temperature is controlled by the temporal extent of the lattice: $T = 1/(N_\tau a)$, where  $N_\tau$ is the number of lattice sites in the temporal direction, and $a$ is the lattice spacing. 
Ensuring that simulations indeed sample from the correct thermal distribution is critical, especially near phase transitions such as the deconfinement crossover in QCD. 
The configurational temperature estimator offers a non-perturbative {\it thermometer} that could help verify thermalization in QCD simulations, check for systematic errors, or identify exceptional configurations that may signal systematic issues in large-scale finite-temperature QCD runs.

The QGP studied at RHIC and LHC exists in a regime of extreme temperature and strong coupling, where lattice QCD is the primary non-perturbative tool. 
Observables like Polyakov loops, susceptibilities, and quark condensates are sensitive to thermal fluctuations. 
The configurational temperature estimator could serve as a confidence check for initialization, detect slow thermalization or metastability in ensemble generation.

\newpage

\section{Data Tables for Compact U(1) Lattice Gauge Theories}
\label{sec:table_data}

\begingroup
\renewcommand*{\arraystretch}{1.5}
\begin{table}[ht]
\centering
\begin{tabular}{|p{0.075\textwidth}|p{0.15\textwidth}|p{0.15\textwidth}|p{0.16\textwidth}|}\hline \hline 
~~$\beta$~~ & ~~$E$~~     & ~~$C_v$~~ &~~$\beta_M$~~\\ \hline \hline   
10.00  &  0.05030 (1)  & 0.52069 (354) & 10.00640 (185)\\
 9.00  &  0.05606 (1)  & 0.51860 (353) & 9.00409 (165)\\
 8.00  &  0.06346 (1)  & 0.53036 (357) & 8.00558 (147)\\
 7.00  &  0.07355 (1)  & 0.54866 (363) & 7.01159 (129)\\
 6.00  &  0.08718 (1)  & 0.56794 (369) & 6.00708 (110)\\
 5.00  &  0.10621 (2)  & 0.58282 (374) & 5.00257 (91)\\
 4.00  &  0.13640 (2)  & 0.61667 (385) & 4.00048 (72)\\
 3.50  &  0.15879 (2)  & 0.64027 (392) & 3.50280 (63)\\
 3.00  &  0.18995 (3)  & 0.66721 (400) & 3.00122 (54)\\
 2.50  &  0.23509 (3)  & 0.68122 (404) & 2.49835 (45)\\
 2.00  &  0.30207 (4)  & 0.65918 (398) & 2.00211 (37)\\
 1.50  &  0.40394 (5)  & 0.55666 (366) & 1.49952 (29)\\
 1.00  &  0.55343 (6)  & 0.35419 (292) & 1.00035 (22)\\
 0.75  &  0.64889 (7)  & 0.22962 (235) & 0.75046 (19)\\
 0.50  &  0.75751 (7)  & 0.11398 (165) & 0.49976 (17)\\
 0.25  &  0.87581 (7)  & 0.03050 (86) & 0.25029 (15)\\
\hline \hline
\end{tabular}
\caption{\label{tab:1d_u1_l48}{One-dimensional compact U(1) lattice gauge theory on a 48-site lattice. 
Summary of the thermodynamic observables: coupling $\beta$, energy $E$, specific heat $C_v$, and measured configurational temperature $\beta_M$.}}
\end{table} 
\endgroup

\begingroup
\renewcommand*{\arraystretch}{1.5}
\begin{table}[ht]
\centering
\begin{tabular}{|p{0.075\textwidth}|p{0.15\textwidth}|p{0.16\textwidth}|} \hline \hline 
~~$\beta$~~ & ~~$P$~~   &~~$\beta_M$~~\\ \hline \hline   
    10.00 & 0.94743 (2)& 10.01600 (342) \\
    9.00  &0.94249 (2)& 9.01442 (309) \\
    8.00  &0.93525 (2)& 8.01902 (272) \\
    7.00  &0.92554 (2)& 7.01052 (237) \\
    6.00  &0.91231 (3)& 6.00628 (202) \\
    5.00  &0.89340 (3)& 5.00843 (169) \\
    4.00  &0.86352 (4)& 4.00625 (136) \\
    3.50  &0.84108 (5)& 3.50550 (118) \\
    3.00  &0.80994 (6)& 3.00344 (101) \\
    2.50  &0.76492 (7)& 2.50242 (85) \\
    2.00  &0.69776 (9)& 2.00308 (69) \\
    1.50  &0.59595 (11)& 1.50129 (55) \\
    1.00  &0.44646 (13)& 1.00114 (43) \\
    0.75  &0.35083 (14)& 0.75063 (39) \\
    0.50  &0.24251 (15)& 0.50052 (35) \\
    0.25  &0.12383 (15)& 0.24988 (32) \\
\hline \hline
\end{tabular}
\caption{Two-dimensional compact U(1) lattice gauge theory. 
Summary of the thermodynamic observables: coupling $\beta$, plaquette $P$, and measured configurational temperature $\beta_M$ on a $32 \times 32$ lattice.}
\label{tab:2d_u1_l32_l32}
\end{table} 
\endgroup

\begingroup
\renewcommand*{\arraystretch}{1.5}
\begin{table}[ht]
\centering
\begin{tabular}{|p{0.075\textwidth}|p{0.15\textwidth}|p{0.15\textwidth}|p{0.15\textwidth}|}\hline \hline 
~~$\beta$~~ & ~~$P$~~     & ~~$\chi_P$~~ &~~$\beta_M$~~\\ \hline \hline  
4.0000 & 0.93531 (24) & 0.01000 (0) & 4.00026 (816) \\
3.0000 & 0.91262 (32) & 0.01800 (0) & 3.00072 (618) \\
2.0000 & 0.86461 (49) & 0.04100 (1) & 2.00038 (398) \\
1.5000 & 0.81259 (71) & 0.08200 (2) & 1.50075 (307) \\
1.1000 & 0.71596 (108) & 0.18500 (4) & 1.09994 (217) \\
1.0500 & 0.69117 (122) & 0.24100 (5) & 1.04994 (204) \\
1.0300 & 0.67724 (132) & 0.28700 (6) & 1.03003 (209) \\
1.0250 & 0.67289 (132) & 0.29500 (6) & 1.02531 (207) \\
1.0200 & 0.66779 (140) & 0.33100 (8) & 1.02015 (204) \\
1.0150 & 0.66219 (147) & 0.35900 (8) & 1.01496 (203) \\
1.0100 & 0.65489 (173) & 0.47300 (11) & 1.00991 (204) \\
1.0095 & 0.65411 (177) & 0.51000 (13) & 1.00958 (199) \\
1.0090 & 0.65328 (180) & 0.53600 (14) & 1.00950 (200) \\
1.0085 & 0.64424 (140) & 1.49700 (18) & 1.00854 (93) \\
1.0080 & 0.63677 (159) & 2.02600 (17) & 1.00808 (92) \\
1.0075 & 0.62815 (179) & 2.47300 (17) & 1.00759 (94) \\
1.0070 & 0.62606 (171) & 2.33300 (16) & 1.00709 (93) \\
1.0060 & 0.60657 (132) & 1.38800 (17) & 1.00613 (97) \\
1.0050 & 0.60225 (126) & 1.26300 (17) & 1.00512 (96) \\
1.0000 & 0.58524 (196) & 0.62100 (14) & 1.00049 (210) \\
0.9900 & 0.56540 (170) & 0.49300 (12) & 0.99022 (209) \\
0.9500 & 0.51373 (158) & 0.42400 (10) & 0.95021 (209) \\
0.7500 & 0.37296 (160) & 0.43100 (10) & 0.75016 (200) \\
0.5000 & 0.24579 (166) & 0.47200 (10) & 0.50021 (171) \\
0.2500 & 0.12411 (175) & 0.49800 (11) & 0.25004 (155) \\

\hline \hline
\end{tabular}
\caption{Four-dimensional compact U(1) lattice gauge theory on a $8^4$ lattice. Summary of the thermodynamic observables: coupling $\beta$, plaquette $P$, plaquette susceptibility $\chi_P$, and measured configurational temperature $\beta_M$.}
\label{tab:4d_u1_l8}
\end{table} 
\endgroup

\clearpage\mbox{}\clearpage
\chapter{Summary of Results and Conclusions}
\label{ch:conclusion}

In this thesis, we have carried out a comprehensive numerical investigation of phase transitions in lattice field theories using two complementary approaches. 
The first part, presented in Chapter \ref{ch:tn} and \ref{ch:gxy}, employed tensor network methods and their application to study the generalized XY model. 
The second part, detailed in Chapter \ref{ch:tmp_est} and \ref{ch:u1}, focused on the formulation of the configurational temperature estimator and its validation in compact U(1) lattice gauge theories. 
Although built upon different concepts and ideas, both these studies address the accurate determination of phase structure and the reliability of numerical simulations.

\section{Summary of Results: Generalized XY Model}

In Chapter \ref{ch:gxy}, we investigated the two-dimensional generalized XY (gXY) model using the higher-order tensor renormalization group (HOTRG) method. 
By employing a character expansion of the partition function and controlled truncation in Fourier modes and bond dimensions, we evaluated thermodynamic observables directly in the thermodynamic limit. 

The implementation of impurity tensor techniques, as described in Sec.~\ref{sec:gxy_results}, enabled the computation of physical observables such as the specific heat, magnetization, and magnetic susceptibilities. 
Representative results for these observables are shown in Sec.~\ref{sec:plots_diff_delta}, where distinct signatures of phase transitions are clearly visible. 
A central result of this analysis is the detailed mapping of the phase diagram in the $(\Delta, T)$ plane (see Fig.~\ref{fig:phase_diag}). 
The model exhibits three distinct phases: a ferromagnetic phase characterized by integer vortices, a nematic phase associated with half-integer vortices, and a disordered (paramagnetic) phase. 
These phases are separated by transition lines belonging to different universality classes, including Berezinskii--Kosterlitz--Thouless (BKT), half-BKT, and Ising transitions. 
The identification of these transitions relied crucially on the behavior of susceptibilities and correlation functions, as discussed in Sec.~\ref{sec:gxy_results}. 
We found that the magnetic susceptibility, obtained via finite-field extrapolation to the zero-field limit, provides a more reliable estimator of the critical temperature compared to the peak of the specific heat. 
This is particularly important for BKT-type transitions, where thermodynamic singularities are weak and broadened. 

From our numerical analysis, we identified a multi-critical region around 
\begin{equation} 
\Delta = 0.36(2), \quad T_c \approx 0.716(3), 
\end{equation} 
where multiple transition lines appear to merge. 
The detailed numerical estimates are summarized in the table discussed in Sec.~\ref{sec:gxy_table}. 
The data suggest a sequential merging scenario, in which the half-BKT and Ising transition lines first coincide, then merge with the BKT line at slightly higher values of $\Delta$. 
Furthermore, a systematic comparison between different estimators revealed that the specific heat systematically overestimates the BKT transition temperature, the discrepancy between estimators varies non-monotonically with $\Delta$, and for $\Delta > 0.40$, the system exhibits a single BKT transition consistent with the standard XY limit. 

We also performed detailed convergence checks for the truncation parameters. 
As shown in Sec. \ref{sec:gxy_results}, the results are stable against variations in bond dimension $D$ and Fourier cutoff $m$. 
The use of large effective lattice sizes ensured that finite-size effects are negligible. 

Overall, the results presented in Chapter~\ref{ch:gxy} demonstrate that tensor network methods, particularly GPU-accelerated HOTRG, provide a powerful and accurate framework for studying complex phase diagrams with multiple competing orders. 

\section{Summary of Results: Configurational Temperature Estimator}

In Chapter~\ref{ch:u1}, we introduced a configurational temperature estimator based on Rugh's geometric formulation of statistical mechanics. 
Unlike conventional estimators, this approach determines the sampling consistency of simulations using only configurational information, namely the gradients and Hessians of the Euclidean lattice action. 

The construction of a gauge-invariant estimator, described in Sec.~\ref{sec:derv_temp_est}, allows for direct application to compact U(1) lattice gauge theories. 
The numerical implementation was tested in one, two, and four Euclidean dimensions, with representative results shown in Figs.~\ref{fig:1d_u1_l48}, \ref{fig:plaquette_beta_2d}, and \ref{fig:4d_u1_8}. 
Across all dimensions studied, the measured configurational temperature $\beta_M$ shows excellent agreement with the target value, as summarized in the tables presented in Section~\ref{sec:table_data}. 
This agreement persists even in the vicinity of phase transitions, demonstrating the robustness of the estimator. 

In the four-dimensional case, where the system undergoes a phase transition, the estimator accurately follows its target value across the critical region (see Fig.~\ref{fig:4d_u1_8}). 
This highlights its utility in nontrivial settings where fluctuations are significant. 
An important practical observation is that off-diagonal contributions to the Hessian become negligible for large lattice volumes. 
As discussed in Sec.~\ref{sec:2d_temp_est}, this allows for computational simplifications without compromising accuracy. 

The configurational estimator serves as a diagnostic tool for simulation quality. 
In particular, deviations between $\beta_M$ and $\beta$ can signal slow thermalization, sampling inefficiencies, and algorithmic or numerical errors. 
This makes the estimator particularly valuable for large-scale Monte Carlo simulations, where such issues may otherwise go undetected. 

\section{Conclusions}

The two parts of this thesis address an important aspect of numerical studies of critical phenomena. 
On the one hand, the tensor network approach developed in Chapter~\ref{ch:gxy} enables precise determination of phase boundaries and critical behavior, even in systems with complex competing interactions. 
On the other hand, the configurational temperature estimator introduced in Chapter~\ref{ch:u1} provides an independent consistency check on thermal sampling. 
The two parts of the thesis are scientifically distinct, and the unifying theme is not a single physical model, but rather the problem of reliable numerical investigation of phase transitions. Tensor networks offer a deterministic high-precision route to phase diagrams, while configurational temperature estimators offer an independent diagnostic of sampling quality in stochastic simulations.
Together, these approaches improve both the \emph{accuracy} and \emph{reliability} of numerical simulations in lattice field theory. 
Their combined use is particularly promising for large-scale computations where both precise results and robust diagnostics are essential. 

Several avenues for future work naturally emerge from this study. 
First, the tensor network analysis of the generalized XY model can be extended to models with higher fractionalization ($q \geq 3$), where more intricate topological structures may arise. 
Additionally, coupling the XY model to gauge fields could provide further insights into intertwined order parameters. 
Second, the configurational temperature estimator can be generalized to non-Abelian gauge theories. 
Its application to SU($N$) lattice gauge theories, particularly in the context of finite-temperature QCD, would be of significant interest. 
Another promising direction is the integration of the estimator into Hybrid Monte Carlo (HMC) simulations as a real-time diagnostic tool. 
This could enable early detection of sampling issues and improve the efficiency of large-scale simulations. 

The complex Langevin method (CLM) provides a promising approach to addressing the sign problem in quantum field theories with complex actions. 
However, it is well known that CLM can converge to incorrect results, even in situations where the Langevin evolution appears stable and conventional diagnostics do not signal any pathology. 
This highlights the need for more stringent and direct tests of correctness.

Existing diagnostic tools typically focus on properties of the Langevin dynamics, such as the behavior of the drift term or the validity of Langevin-time evolution identities. 
While these are necessary consistency checks, they do not directly probe whether the generated configurations are sampled according to the correct Boltzmann weight.

A complementary diagnostic can be formulated in terms of the configurational temperature. This estimator tests a specific consistency condition associated with the intended desired weight $e^{-S}$, and therefore provides strong evidence for consistency of the sampled ensemble with the intended distribution with respect to this diagnostic. 

Benchmark studies in one-dimensional $\mathcal{PT}$-symmetric models demonstrate that the configurational temperature reproduces the target value with an accuracy at the level of $0.2\%$--$3\%$ \cite{Joseph:2025fcd}. See also Refs. \cite{Joseph:2025xbn, Joseph:2026xti}.

More importantly, it exhibits a high sensitivity to algorithmic deficiencies, including incorrect noise normalization, finite step-size effects, and incomplete thermalization. 
In such cases, deviations in the configurational temperature are typically more pronounced than those observed using standard drift-based or operator-based diagnostics.

The method depends only on derivatives of the (local) action, and is therefore broadly applicable to a wide class of lattice field theories. 
That said, the computational cost associated with evaluating gradients and Hessians must be carefully managed, particularly in higher-dimensional systems or theories with large numbers of degrees of freedom.

Overall, the configurational temperature provides a powerful and versatile addition to the diagnostic toolkit for complex Langevin simulations. 
It complements existing approaches and offers significant potential for applications ranging from supersymmetric matrix models to lattice QCD at finite density.

Finally, both approaches developed in this thesis are not limited to lattice field theory. 
They can be applied more broadly in condensed-matter systems and statistical mechanics models, where the accurate determination of phase transitions and reliable numerical diagnostics remain central challenges.

\clearpage\mbox{}\clearpage


	
\end{document}